\RequirePackage{fix-cm}

\documentclass[twocolumn,epjc3]{svjour3}  
\smartqed  
\RequirePackage{mathptmx}      
\RequirePackage{bm}            
\RequirePackage{graphicx}
\RequirePackage{makecell}
\RequirePackage{amssymb}
\RequirePackage{amsmath}
\RequirePackage{caption}
\RequirePackage{lineno}
\RequirePackage{booktabs}
\RequirePackage{csquotes}
\RequirePackage{float}
\RequirePackage{comment}
\RequirePackage[version=4]{mhchem}
\RequirePackage{booktabs}
\RequirePackage{url}
\RequirePackage[colorlinks,allcolors=red]{hyperref}
\RequirePackage[multiple]{footmisc}
\RequirePackage{upgreek}
\RequirePackage{mathtools}
\RequirePackage{cuted}
\RequirePackage{relsize}
\RequirePackage{makecell}
\RequirePackage{tablefootnote}
\RequirePackage{siunitx}
\RequirePackage{relsize}
\usepackage[dvipsnames]{xcolor}
\RequirePackage[english]{babel}
\RequirePackage{xspace}
\RequirePackage{multirow}
\usepackage{subcaption} 

\RequirePackage[backend=biber,sorting=none,style=ieee,citestyle=numeric-comp]{biblatex}
\bibliography{bib}
\RequirePackage{etoolbox}
\RequirePackage{authblk}  

\newcommand{\opticalplane}{{\SI{10.5}{\meter^2}}}
\newcommand{\opticalplanes}{{\SI{21}{\meter^2}}}
\newcommand{\tpc}{TPC}
\newcommand{\ds}{DS-20k}

\newcommand{\npduds}{528} 
\newcommand{\ntileds}{8448}
\newcommand{\npcbtotnoa}{10250}
\newcommand{\ntiletotnoa}{10138}

\newcommand{\ntestedtiles}{1507}
\newcommand{\npreproductiontiles}{261}
\newcommand{\nproductiontiles}{1246}
\newcommand{\nwafers}{1400} 
\newcommand{\ndies}{369.600} 
\newcommand{\ntestedpcbs}{1730}
\newcommand{\npreproductionpcbs}{131}
\newcommand{\nproductionpcbs}{1599}

\newcommand{\tileOscilloscope}{Teledyne
HDO6104A}
\newcommand{\tileLaser}{Hamamatsu PLP-10}

\newcommand{\switchingMatrix}{Keithley DAQ6510}
\newcommand{\switchingMatrixsecond}{Keithley DAQ7702}
\newcommand{\tilePSU}{Keysight E3649}
\newcommand{\pcbOscilloscope}{Tektronix MSO64}
\newcommand{\SMU}{Keithley 2450}

\newcommand{\vov}{VoV}
\newcommand{\sipm}{SiPM}
\newcommand{\sipms}{SiPMs}
\newcommand{\sipmarea}{\SI{1}{\centi\meter\squared}}

\newcommand{\diesize}{\SI{11.8}{\milli \meter} $\times$ \SI{7.9}{\milli \meter}}
\newcommand{\vbdsipm}{\SI{27}{\volt}}

\newcommand{\pcb}{PCB}
\newcommand{\pcbs}{PCBs}
\newcommand{\pcbarea}{49.5 $\times$ 49.5 \SI{}{\milli\meter\squared}}
\newcommand{\rgone}{$\mathrm{R}_{\mathrm{g1}}$}
\newcommand{\rgtwo}{$\mathrm{R}_{\mathrm{g2}}$}
\newcommand{\rgthree}{$\mathrm{R}_{\mathrm{g3}}$}
\newcommand{\rbrstar}{$\mathrm{R}_{\mathrm{Br}}^*$}
\newcommand{\tile}{Tile}
\newcommand{\tiles}{Tiles}
\newcommand{\tilesipms}{24}
\newcommand{\vbias}{$\mathrm{V}_{\mathrm{bias}}$}
\newcommand{\tilearea}{\SI{24}{\centi\meter\squared}}
\newcommand{\tilesensarea}{\SI{22.2}{\centi\meter\squared}}
\newcommand{\vbdtile}{\SI{108}{\volt}}

\newcommand{\rdival}{\SI{10}{\mega \ohm}}
\newcommand{\rdivaltot}{\SI{40}{\mega \ohm}}

\newcommand{\cival}{\SI{100}{\nano \farad}}
\newcommand{\rifiltval}{\SI{10}{\kilo \ohm}}
\newcommand{\dividerimpedance}{\SI{40.01}{\mega \ohm}}

\newcommand{\iv}{\mbox{I-V}}
\newcommand{\vbd}{$\mathrm{V}_{\mathrm{bd}}$}
\newcommand{\rd}{$\mathrm{R}_{\mathrm{d}}$}
\newcommand{\ILVon}{$\mathrm{I}_{\mathrm{LV}}^{\mathrm{on}}$}
\newcommand{\ILVoff}{$\mathrm{I}_{\mathrm{LV}}^{\mathrm{off}}$}
\newcommand{\warm}{\SI{300}{\kelvin}}
\newcommand{\cold}{\SI{77}{\kelvin}}
\newcommand{\taurise}{$\tau_\mathrm{rise}$}
\newcommand{\taufall}{$\tau_\mathrm{fall}$}
\newcommand{\gof}{GoF}

\newcommand{\srl}{S.r.l.}
\newcommand{\qaqc}{QA-QC}
\newcommand{\perc}{\%}

\newcommand{\warmcoldpass}{83.5\%}
\newcommand{\warmpasscoldfail}{4.8\%}
\newcommand{\warmfailcoldpass}{5.2\%}
\newcommand{\warmcoldfail}{6.5\%}
\newcommand{\ivwarmfail}{4.3\%}
\newcommand{\pulsefail}{2.9\%}
\newcommand{\ivwarmnoisecoldfail}{2.1\%}
\newcommand{\ivwarmfailoverfailed}{74\%}

\newcommand{\pulsefailoverfailed}{53\%}
\newcommand{\tilesthroughput}{\(32\) Tiles per day}

\newcommand{\Alberta}{Department of Physics, University of Alberta, Edmonton, AB T6G 2R3, Canada}
\newcommand{\APC}{APC, Universit\'e de Paris, CNRS, Astroparticule et Cosmologie, Paris F-75013, France}
\newcommand{\AQLNGS}{INFN Laboratori Nazionali del Gran Sasso, Assergi (AQ) 67100, Italy}
\newcommand{\AQGSSI}{Gran Sasso Science Institute, L'Aquila 67100, Italy}
\newcommand{\AQUni}{
Department of Industrial and Information Engineering and Economics, UniversitÃ  degli Studi dell'Aquila, L'Aquila 67100, Italy}
\newcommand{\AstroCeNT}{AstroCeNT, Nicolaus Copernicus Astronomical Center of the Polish Academy of Sciences, 00-614 Warsaw, Poland}
\newcommand{\Augustana}{Physics Department, Augustana University, Sioux Falls, SD 57197, USA}
\newcommand{\Belgorod}{Radiation Physics Laboratory, Belgorod National Research University, Belgorod 308007, Russia}

\newcommand{\BINP}{Budker Institute of Nuclear Physics, Novosibirsk 630090, Russia}
\newcommand{\Birmingham}{School of Physics and Astronomy, University of Birmingham, Edgbaston, B15 2TT, Birmingham, UK}

\newcommand{\BOINFN}{INFN Bologna, Bologna 40126, Italy}
\newcommand{\BOUniPHY}{Department of Physics and Astronomy, Universit\`a degli Studi di Bologna, Bologna 40126, Italy}
\newcommand{\CAUniCHE}{Department of Mechanical, Chemical, and Materials Engineering, Universit\`a degli Studi, Cagliari 09042, Italy}
\newcommand{\CAUniEEE}{Department of Electrical and Electronic Engineering, Universit\`a degli Studi di Cagliari, Cagliari 09123, Italy}
\newcommand{\CAUniPHY}{Physics Department, Universit\`a degli Studi di Cagliari, Cagliari 09042, Italy}
\newcommand{\CAINFN}{INFN Cagliari, Cagliari 09042, Italy}
\newcommand{\Carleton}{Department of Physics, Carleton University, Ottawa, ON K1S 5B6, Canada}

\newcommand{\Columbia}{Physics Department, Columbia University, New York, NY 10027, USA}
\newcommand{\Chicago}{Department of Physics and Kavli Institute for Cosmological Physics, University of Chicago, Chicago, IL 60637, USA}
\newcommand{\BOCentroFermi}{Museo Storico della Fisica e Centro Studi e Ricerche Enrico Fermi, Roma 00184, Italy}

\newcommand{\CIEMAT}{CIEMAT, Centro de Investigaciones Energ\'eticas, Medioambientales y Tecnol\'ogicas, Madrid 28040, Spain}

\newcommand{\CPPM}{Centre de Physique des Particules de Marseille, Aix Marseille Univ, CNRS/IN2P3, CPPM, Marseille, France}
\newcommand{\CTINFN}{INFN Catania, Catania 95121, Italy}
\newcommand{\CTUNI}{Universit\`a of Catania, Catania 95124, Italy}
\newcommand{\CTLNS}{INFN Laboratori Nazionali del Sud, Catania 95123, Italy}

\newcommand{\ENUniCEE}{Engineering and Architecture Department, Universit\`a di Enna Kore, Enna 94100, Italy}
\newcommand{\ETHZ}{Institute for Particle Physics and Astrophysics, ETH Zurich, Zurich 8093, Switzerland}

\newcommand{\FortLewis}{Department of Physics and Engineering, Fort Lewis College, Durango, CO 81301, USA}
\newcommand{\GEUni}{Physics Department, Universit\`a degli Studi di Genova, Genova 16146, Italy}
\newcommand{\GEINFN}{INFN Genova, Genova 16146, Italy}
\newcommand{\Hawaii}{Department of Physics and Astronomy, University of Hawai'i, Honolulu, HI 96822, USA}
\newcommand{\Houston}{Department of Physics, University of Houston, Houston, TX 77204, USA}
\newcommand{\IHEP}{Institute of High Energy Physics, Chinese Academy of Sciences, Beijing 100049, China}
\newcommand{\INFN}{Istituto Nazionale di Fisica Nucleare, Roma 00186, Italia}

\newcommand{\JINR}{Joint Institute for Nuclear Research, Dubna 141980, Russia}
\newcommand{\Krakow}{M.~Smoluchowski Institute of Physics, Jagiellonian University, 30-348 Krakow, Poland}
\newcommand{\Kurchatov}{National Research Centre Kurchatov Institute, Moscow 123182, Russia}
\newcommand{\Laurentian}{Department of Physics and Astronomy, Laurentian University, Sudbury, ON P3E 2C6, Canada}
\newcommand{\Lancaster}{Physics Department, Lancaster University, Lancaster LA1 4YB, UK}
\newcommand{\Liverpool}{Department of Physics, University of Liverpool, The Oliver Lodge Laboratory, Liverpool L69 7ZE, UK}

\newcommand{\LNLINFN}{INFN Laboratori Nazionali di Legnaro, Legnaro (Padova) 35020, Italy}
\newcommand{\Lodz}{Institute of Applied Radiation Chemistry, Lodz University of Technology, 93-590 Lodz, Poland}

\newcommand{\Manchester}{Department of Physics and Astronomy, The University of Manchester, Manchester M13 9PL, UK}
\newcommand{\MEPhI}{National Research Nuclear University MEPhI, Moscow 115409, Russia}
\newcommand{\MendeleevUniverisity}{Mendeleev University of Chemical Technology, Moscow 125047, Russia}

\newcommand{\MIINFN}{INFN Milano, Milano 20133, Italy}
\newcommand{\MIPoliICA}{Civil and Environmental Engineering Department, Politecnico di Milano, Milano 20133, Italy}
\newcommand{\MIPoliCHE}{Chemistry, Materials and Chemical Engineering Department ``G.~Natta", Politecnico di Milano, Milano 20133, Italy}

\newcommand{\MIUni}{Physics Department, Universit\`a degli Studi di Milano, Milano 20133, Italy}
\newcommand{\MSU}{Skobeltsyn Institute of Nuclear Physics, Lomonosov Moscow State University, Moscow 119234, Russia}
\newcommand{\NAINFN}{INFN Napoli, Napoli 80126, Italy}
\newcommand{\NAUniPHY}{Physics Department, Universit\`a degli Studi ``Federico II'' di Napoli, Napoli 80126, Italy}
\newcommand{\NAUniCHE}{Chemical, Materials, and Industrial Production Engineering Department, Universit\`a degli Studi ``Federico II'' di Napoli, Napoli 80126, Italy}
\newcommand{\NAUniDIST}{Department of Structures for Engineering and Architecture, Universit\`a degli Studi ``Federico II'' di Napoli, Napoli 80126, Italy}
\newcommand{\NAUniPHARM}{Pharmacy Department, Universit\`a degli Studi ``Federico II'' di Napoli, Napoli 80131, Italy}

\newcommand{\Oxford}{University of Oxford, Oxford OX1 2JD, United Kingdom}
\newcommand{\Petersburg}{Saint Petersburg Nuclear Physics Institute, Gatchina 188350, Russia}

\newcommand{\PIINFN}{INFN Pisa, Pisa 56127, Italy}
\newcommand{\PIUniPHY}{Physics Department, Universit\`a degli Studi di Pisa, Pisa 56127, Italy}
\newcommand{\PNNL}{Pacific Northwest National Laboratory, Richland, WA 99352, USA}
\newcommand{\Princeton}{Physics Department, Princeton University, Princeton, NJ 08544, USA}
\newcommand{\Queens}{Department of Physics, Engineering Physics and Astronomy, Queen's University, Kingston, ON K7L 3N6, Canada}
\newcommand{\RHUL}{Department of Physics, Royal Holloway University of London, Egham TW20 0EX, UK}
\newcommand{\RMTreINFN}{INFN Roma Tre, Roma 00146, Italy}
\newcommand{\RMTreUni}{Department of Mathematics and Physics, Roma Tre University, Roma 00146, Italy}
\newcommand{\RMUnoINFN}{INFN Sezione di Roma, Roma 00185, Italy}
\newcommand{\RMUnoUni}{Physics Department, Sapienza Universit\`a di Roma, Roma 00185, Italy}

\newcommand{\SNL}{Savannah River National Laboratory, Jackson, SC 29831, United States}

\newcommand{\SNOLAB}{SNOLAB, Lively, ON P3Y 1N2, Canada}

\newcommand{\STFCInterconnect}{Science \& Technology Facilities Council (STFC), Rutherford Appleton Laboratory, Technology, Harwell Oxford, Didcot OX11 0QX, UK}
\newcommand{\STFCppd}{Science \& Technology Facilities Council (STFC), Rutherford Appleton Laboratory, Particle Physics Department, Harwell Oxford, Didcot OX11 0QX, UK}

\newcommand{\Temple}{Physics Department, Temple University, Philadelphia, PA 19122, USA}
\newcommand{\TNFBK}{Fondazione Bruno Kessler, Povo 38123, Italy}

\newcommand{\TOINFN}{INFN Torino, Torino 10125, Italy}
\newcommand{\TOPoli}{Department of Electronics and Telecommunications, Politecnico di Torino, Torino 10129, Italy}

\newcommand{\TRIUMF}{TRIUMF, 4004 Wesbrook Mall, Vancouver, BC V6T 2A3, Canada}

\newcommand{\UCDavis}{Department of Physics, University of California, Davis, CA 95616, USA}
\newcommand{\UCRiverside}{Department of Physics and Astronomy, University of California, Riverside, CA 92507, USA}

\newcommand{\UCLA}{Physics and Astronomy Department, University of California, Los Angeles, CA 90095, USA}
\newcommand{\UCAS}{University of Chinese Academy of Sciences, Beijing 100049, China}
\newcommand{\UMass}{Amherst Center for Fundamental Interactions and Physics Department, University of Massachusetts, Amherst, MA 01003, USA}

\newcommand{\UniversityofEdinburgh}{School of Physics and Astronomy, University of Edinburgh, Edinburgh EH9 3FD, UK}

\newcommand{\USP}{Instituto de F\'isica, Universidade de S\~ao Paulo, S\~ao Paulo 05508-090, Brazil}
\newcommand{\VTech}{Virginia Tech, Blacksburg, VA 24061, USA}
\newcommand{\Washington}{Center for Experimental Nuclear Physics and Astrophysics, and Department of Physics, University of Washington, Seattle, WA 98195, USA}
\newcommand{\Warwick}{University of Warwick, Department of Physics, Coventry CV47AL, UK}
\newcommand{\WUT}{Institute of Radioelectronics and Multimedia Technology, Faculty of Electronics and Information Technology, Warsaw University of Technology, 00-661 Warsaw, Poland}
\newcommand{\WilliamsCollege}{Williams College, Department of Physics and Astronomy, Williamstown, MA 01267 USA}
\newcommand{\Zaragoza}{Centro de Astropart\'iculas y F\'isica de Altas Energ\'ias, Universidad de Zaragoza, Zaragoza 50009, Spain}

\newcommand{\UniHAM}{Institute of Experimental Physics, University of Hamburg, Luruper Chaussee 149, 22761, Hamburg, Germany}

\journalname{Eur. Phys. J. C}

\makeatletter
\renewcommand{\thanksref}[1]{\nolinebreak\textsuperscript{\ref{#1}}\nolinebreak\checknextarg}
\newcommand{\checknextarg}{\@ifnextchar\bgroup{\nolinebreak\gobblenextarg}{}}
\newcommand{\gobblenextarg}[1]{ \textsuperscript{\nolinebreak\hspace{-4pt}\mbox{\nolinebreak$^,$\nolinebreak\ref{#1}\nolinebreak}\nolinebreak} \@ifnextchar\bgroup{\gobblenextarg}{}}

\begin{document}

\title{Production, Quality Assurance and Quality Control of the \sipm\ \tiles\ for the DarkSide-20k Time Projection Chamber}
 \widowpenalty10000
  \clubpenalty10000

\author{The DarkSide-20k Collaboration$^\text{\normalfont a,1}$}
\thankstext{e1}{e-mail: ds-ed@lists.infn.it}
\institute{See back for author list \label{addr1}}

\date{Received: date / Accepted: date}

\maketitle
\modulolinenumbers[5]

\begin{abstract}
The DarkSide-20k dark matter direct detection experiment will employ a \opticalplanes\ silicon photomultiplier (\sipm) array, instrumenting a dual-phase \SI{50}{tonnes} liquid argon Time Projection Chamber (TPC). 

\sipms\ are arranged into modular photosensors called \emph{Tiles}, each integrating \tilesipms\ \sipms\ onto a printed circuit board (PCB) that provides signal amplification, power distribution, and a single-ended output for simplified readout. \(16\) \tiles\ are further grouped into \emph{Photo-Detector Units} (PDUs).

This paper details the production of the \tiles\ and the quality assurance and quality control (\qaqc) protocol established to ensure their performance and uniformity.

The production and \qaqc\ of the \tiles\ are carried out at Nuova Officina Assergi (NOA), an ISO-6 clean room facility at LNGS. This process includes wafer-level cryogenic characterisation, precision flip-chip bonding, wire bonding, and extensive electrical and optical validation of each \tile.

The overall production yield exceeds \warmcoldpass, matching the requirements of the DarkSide-20k production plan.
These results validate the robustness of the \tile\ design and its suitability for operation in a cryogenic environment.
\end{abstract}

\section{Introduction}
\begin{figure}[tpb]
    \centering
    \begin{subfigure}[t]{\linewidth} 
        \centering
        
        \includegraphics[width=\linewidth]{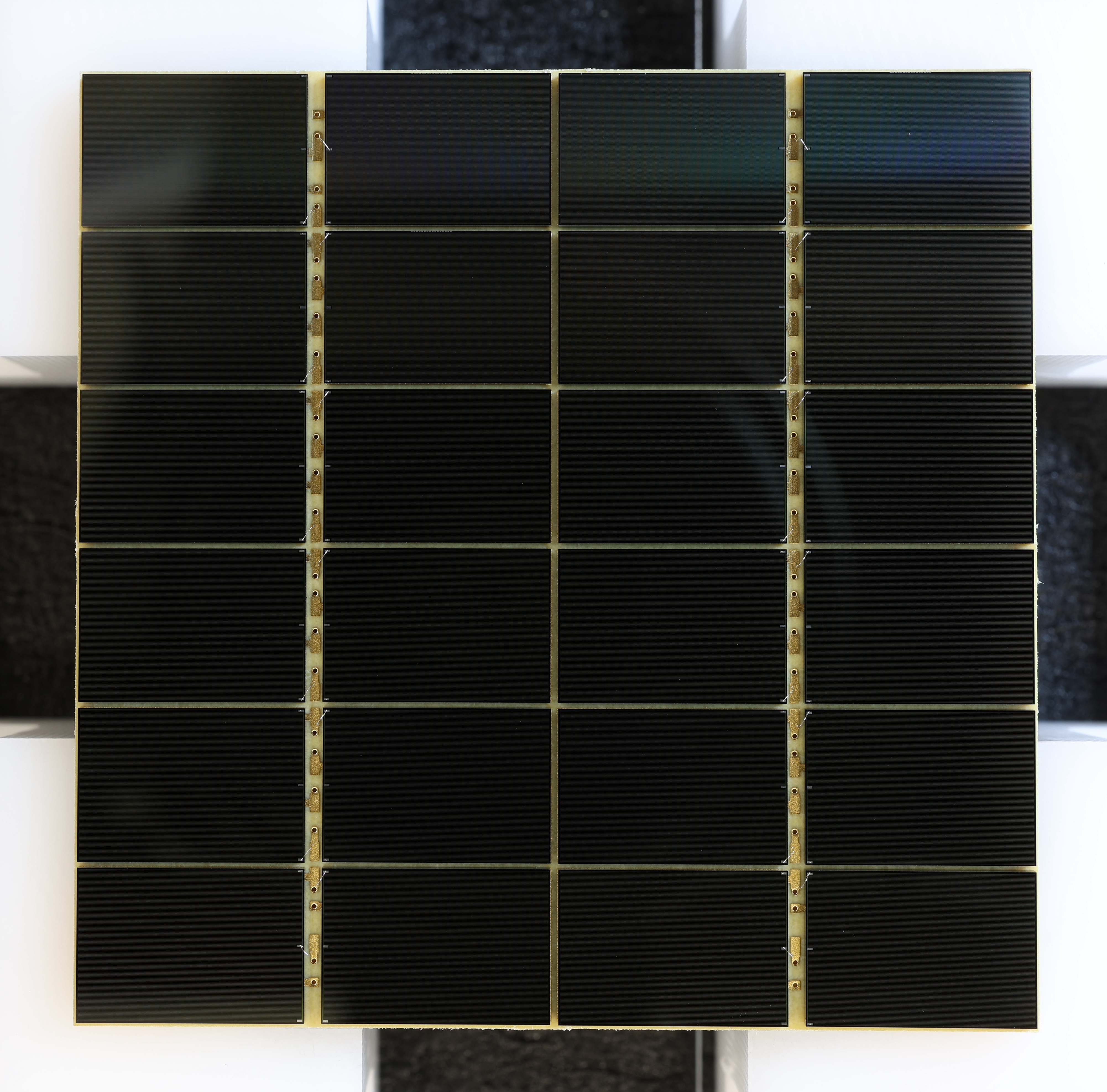}
    \end{subfigure}
    \begin{subfigure}[t]{\linewidth}
        \centering
        \includegraphics[width=\linewidth]{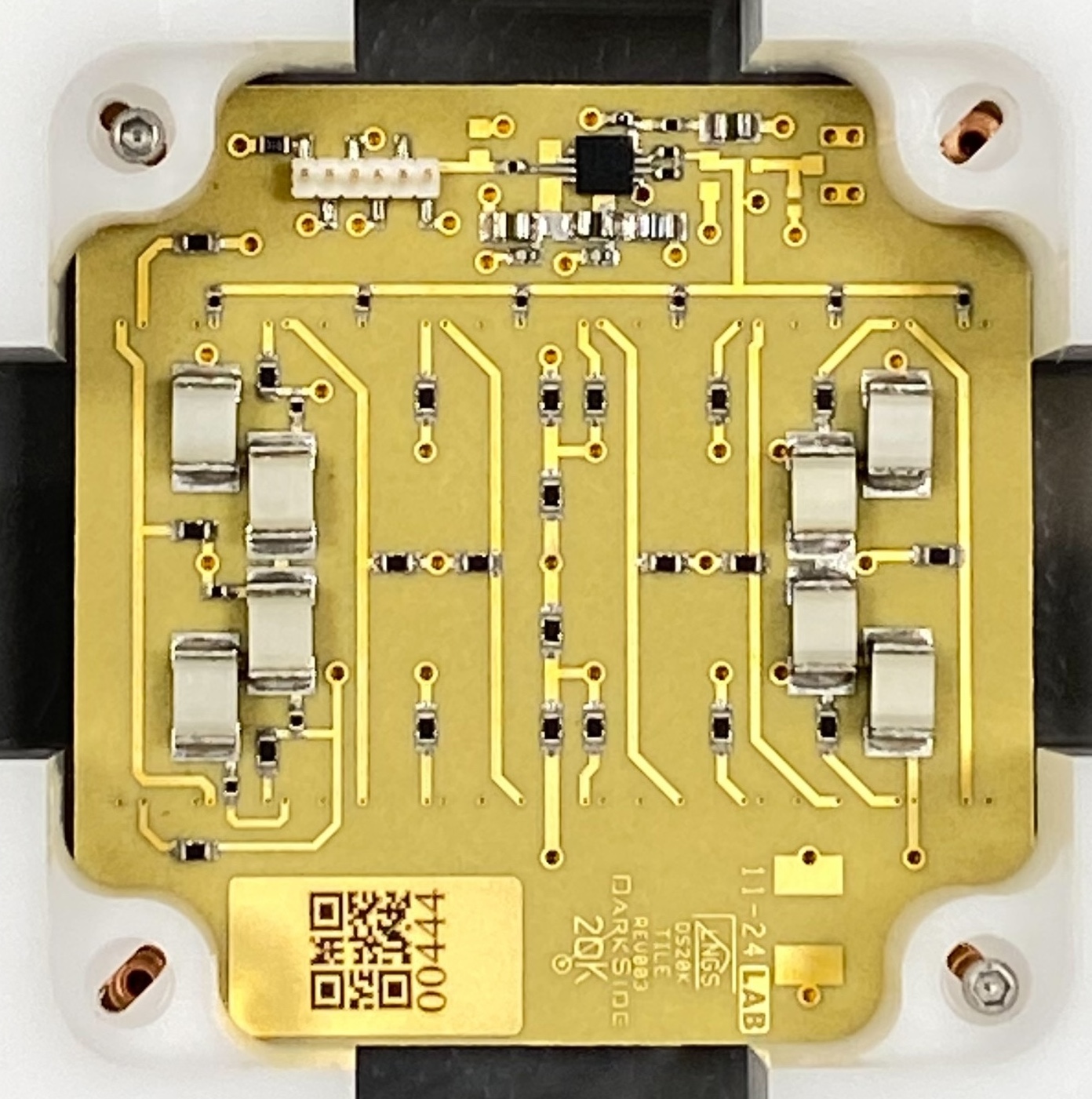}
    \end{subfigure}
    \caption{Picture of a \tile\ showing the \tilesipms\ \sipms, mounted on the top side (Top), and the electronic components, mounted on the bottom side (Bottom). Each \sipm\ is equipped with three aluminium anode pads, of which one is used for the cryogenic \sipm\ assessment and the remaining two are used for the wire bonding. The golden \pcb\ cathode pad is also visible.}
    \label{fig:tile-pictures}
\end{figure}

DarkSide-20k (\ds) is a Time Projection Chamber (\tpc) for dark matter particle searches featuring an active mass of  \SI{50} {tonnes} of liquid argon from underground sources (UAr) currently under construction at Laboratori Nazionali del Gran Sasso (LNGS) in Italy. The collaboration opted to develop with Fondazione Bruno Kessler (FBK) a new generation of silicon photomultiplier (SiPM) especially optimised for operation at cryogenic temperature (\SI{80}{\kelvin}) with reduced noise, and to develop an in-house integration process to produce photo-detectors. These instrument the two large optical planes at the bottom and top of the TPC for a total surface of \SI{21}{\meter^2}.

Among the benefits of utilising \sipms\ over Photo Multiplier Tubes (PMTs) are their low voltage operation, insensitivity to magnetic fields, and compactness that facilitates covering large areas, while minimising empty spots. Additionally, \sipms\ exhibit minimal residual natural radioactivity, which makes them ideal for low-background experiments like \ds~\cite{Aalseth2018}.

At the heart of the \ds\ detector is a dual-phase \tpc\ designed with a vertical electron drift configuration. The \tpc\ is designed as an octagonal prism standing \SI{348}{\centi\meter} tall with an inner diameter of \SI{350}{\centi\meter}. The \tpc\ is filled with low-radioactivity UAr, which serves as the target for Weakly Interacting Massive Particles (WIMPs)~\cite{Agnes2018,Agnes2021,Acerbi2025}. Above the liquid argon phase, a gas layer approximately \SI{7}{\milli \meter} thick is maintained.

The TPC is enclosed in a stainless steel vessel filled with an extra \SI{36}{t} of UAr, which serves as the neutron Inner Veto. 
The stainless steel vessel, the TPC and the Inner Veto are immersed in atmospheric liquid argon housed in a DUNE-like membrane cryostat functioning as the cosmogenic Outer Veto~\cite{Montanari2015}. 

In the liquid argon \tpc, light is produced by two processes: the primary liquid argon scintillation (S1) and gas proportional electroluminescence (S2) from ionisation electrons extracted into the gas layer. In \ds, both S1 and S2 signals will be detected by arrays of \sipms\ mounted on two Optical Planes (OP) positioned on the top and bottom of the \tpc's octagonal barrel. Each OP sits behind an optical window coated with Tetraphenyl butadiene (TPB) and covers an area of approximately \opticalplane. These OPs are equipped with \diesize\ near-UV sensitive, high-density Cryo NUV-HD \sipms\, developed by FBK, further optimised jointly with the DarkSide Collaboration and manufactured by LFoundry \srl~\cite{Aalseth2018, Acerbi2017,Piemonte2016,Gola2019}.

The \sipms\ are arranged into \emph{\tiles}. Every \tile\ functions as a module containing \tilesipms\ \sipms\ bonded to the front side of a printed circuit board (\pcb). The \pcb\ provides common power distribution to all the \tilesipms\ \sipms\ and includes cryogenic low-noise readout electronics, which deliver the total current from the \sipms\ as a single-ended output signal. Therefore, the \tile\ is a compact, functional photodetector module.
A picture of a \tile\ is shown in~\autoref{fig:tile-pictures}.
The Inner Veto and the Outer Veto of \ds\ are also instrumented with \sipm-based photosensors integrated into \tiles.

In \ds, \tiles\ are assembled into \emph{Photo-Detector Units} (PDUs) to further simplify the process of assembling and reading out the OP. A PDU consists of \(16\) \tiles\ mounted on a custom motherboard, where they are aggregated in groups of four to give four readout channels per PDU. All \tiles\ within a single PDU receive a common bias voltage. Each PDU generates four signal outputs, with each output representing the summed signals of each quadrant. The PDU will be the subject of a future paper.

\tiles\ are manufactured within the Nuova Officina Assergi (NOA) facility, located at INFN-LNGS, an ISO-6 clean room packaging space spanning \SI{353}{\square\meter}~\cite{NOAfrontiers}. This facility is equipped with state-of-the-art production machinery for evaluating wafer quality and assembling \sipms\ into \tiles. 
A batch of \(260,000\) \sipms\ derived from \(1400\) LFoundry \srl\ wafers underwent tests at NOA between 2023 and 2025~\cite{acerbi2024qualityassurancequalitycontrol}.
The \sipms\ that successfully meet quality control standards are assembled into \tiles. \ntileds\ \tiles\ among the ones that exhibit good quality are intended for use within the \npduds\ PDUs of the \ds\ \tpc.

This paper describes the production of the \ds\ \tpc\ \tiles\ in NOA and the Quality Assurance and Quality Control (\qaqc) procedures accompanying the production.
The paper is organised as follows: \autoref{sec:tiles} details the \tile's design; \autoref{sec:tile_production} describes the \tile\ production process flow in NOA; \autoref{sec:tile-qaqc} explains the \qaqc\ protocols; \autoref{sec:results} highlights the most relevant results.

\section{The \tile: a compact \sipm-based photodetector}
\label{sec:tiles}

\begin{figure}[tpb]
        \includegraphics[width=\linewidth]{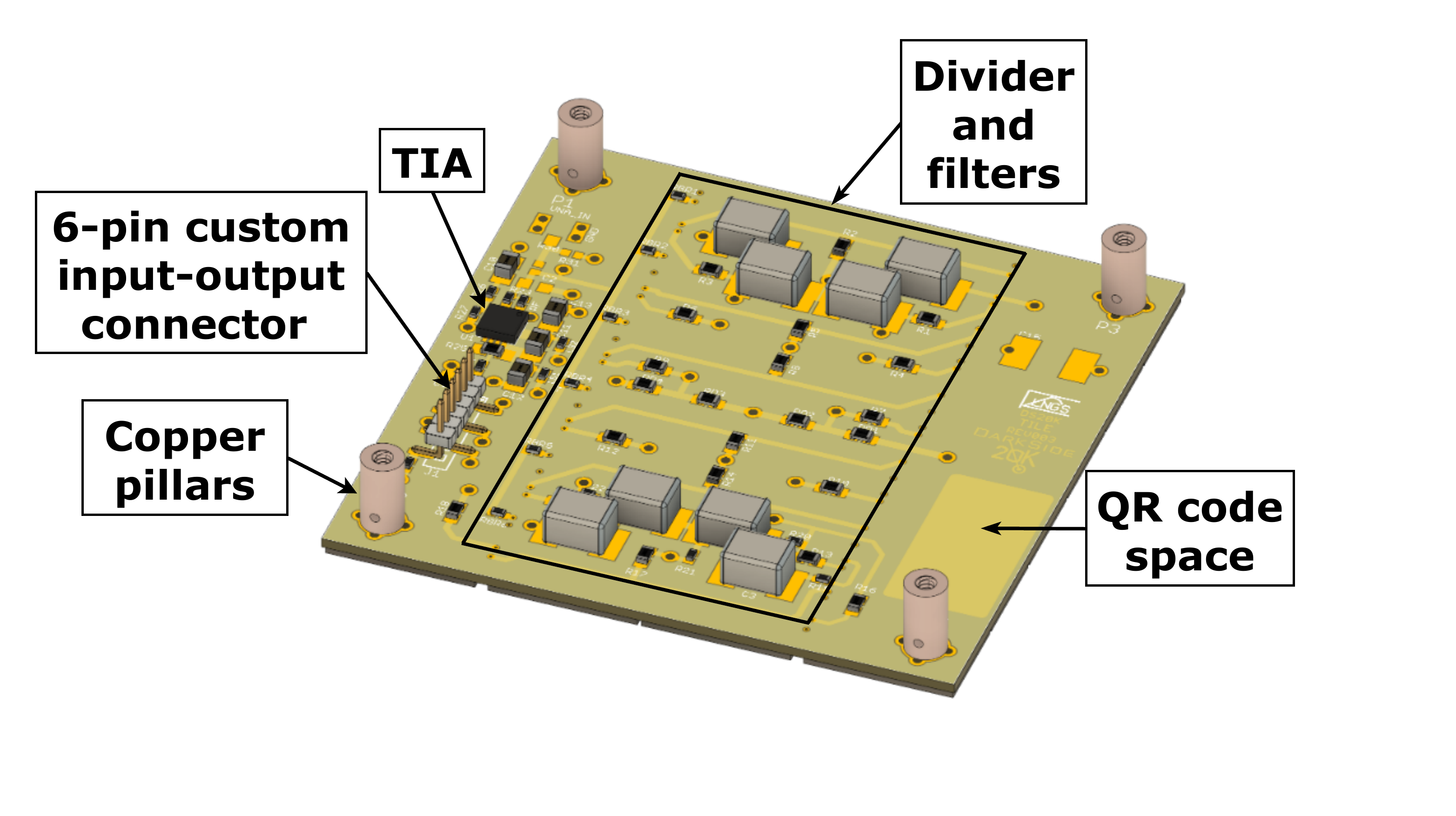}
        \caption{\pcb\ 3D rendering showing the electronic components on the bottom side of the \tile.}
        \label{fig:tile-3D}
\end{figure}

\begin{figure*}[tpb]
    \centering
    \includegraphics[width=\linewidth]{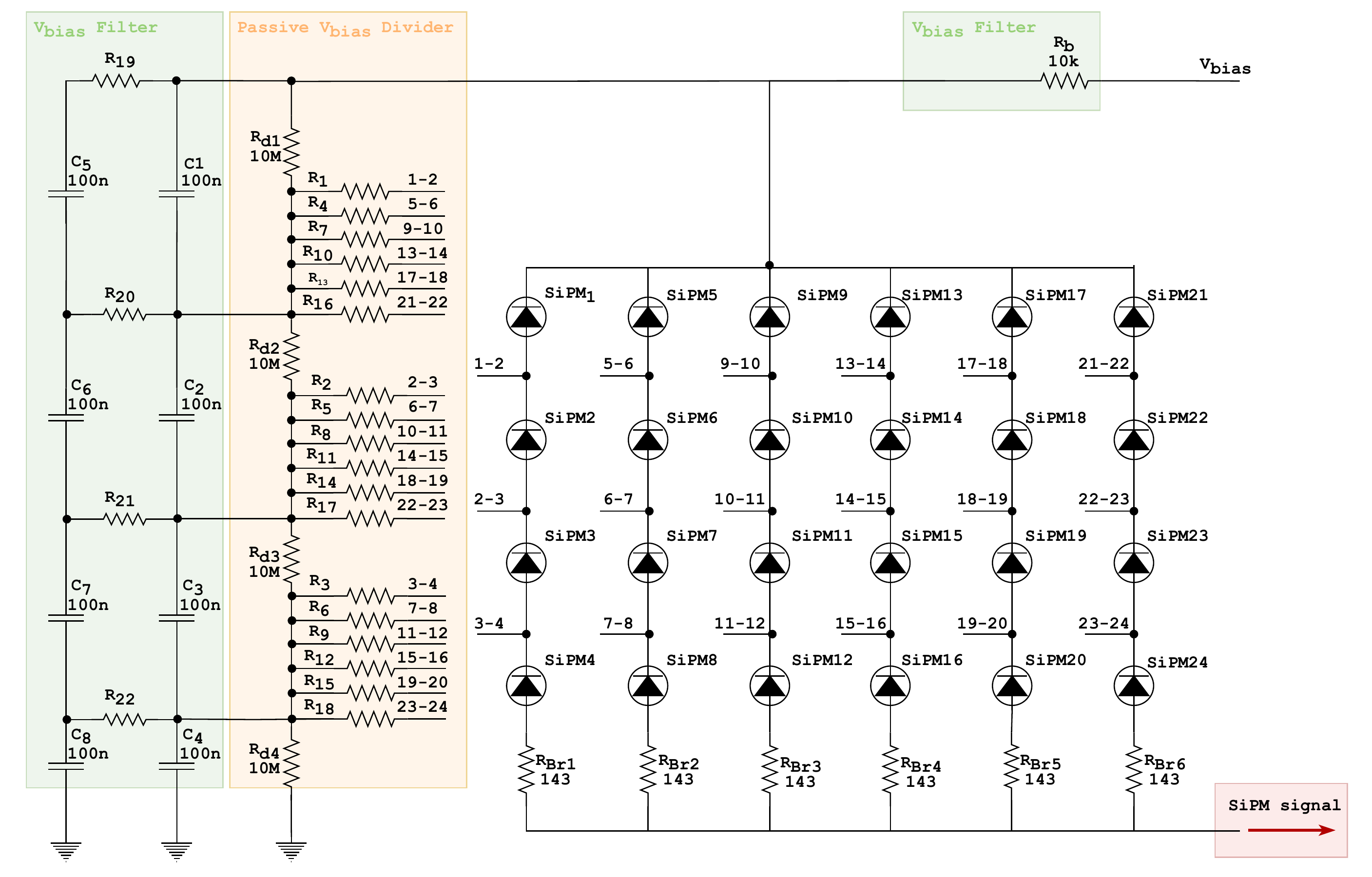}
    \caption{Schematic of the \tile\ electronics: it consists of \tilesipms\ \sipmarea-\sipms\ arranged in a \emph{4s6p} configuration, with \(6\) parallel branches of \(4\) \sipms\ in series. This arrangement reduces capacitance for improved bandwidth while simplifying the readout with a single output signal (\sipm\ signal in the red shade). A precision voltage divider provides 1/4 of the input bias (\vbias) to all the \sipms. The divider is composed of a network of \SI{10}{\mega \ohm} resistors ($\mathrm{R}_{\mathrm{d1}}$ to $\mathrm{R}_{\mathrm{d4}}$, and $\mathrm{R}_1$ to $\mathrm{R}_{19}$, Passive \vbias\ Divider in the orange shade), coupled with \SI{100}{\nano \farad} parallel capacitors ($\mathrm{C}_1$ to $\mathrm{C}_9$, \vbias\ Filter in the green shade) acting as a low-pass filter and charge storage. A \SI{10}{\kilo \ohm} filter resistor is added in series to the divider, resulting in an overall \SI{40.01}{\mega \ohm} impedance. An additional \SI{143}{\ohm} series resistance on each parallel branch ($\mathrm{R}_{\mathrm{Br1}}$ to $\mathrm{R}_{\mathrm{Br6}}$) provides the necessary frequency compensation for the readout amplifier.}
    \label{fig:electronic-scheme}
\end{figure*}

\begin{figure}[tpb]
    \centering
    \includegraphics[width=\linewidth]{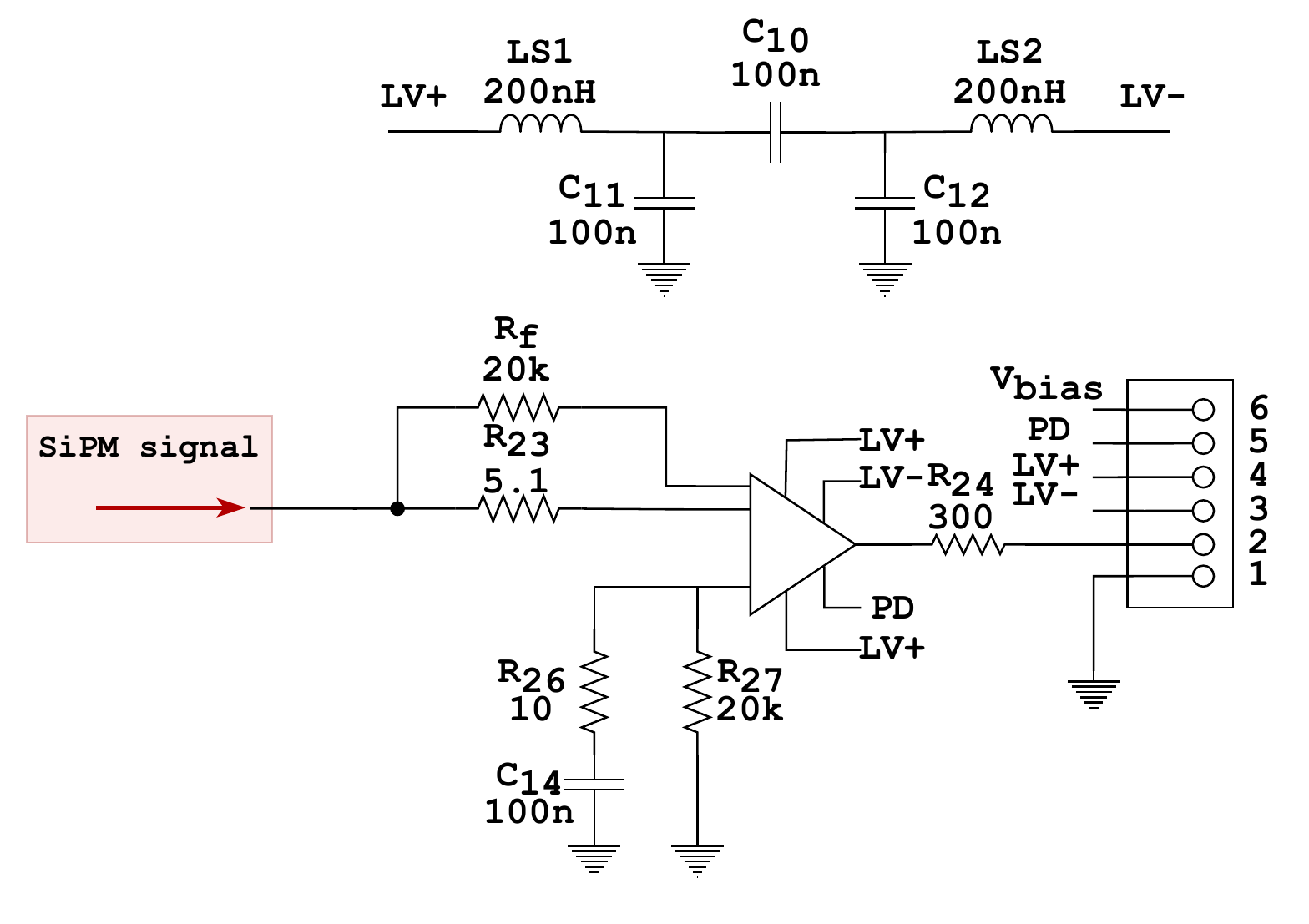}
    \caption{Schematic of the TIA stage with a \SI{20}{\kilo \ohm} feedback resistor ($\mathrm{R}_\mathrm{f}$) setting its gain and the 6-pin custom input-output connector with \SI{1.27}{\milli \meter} pitch. Of these, pin 1 serves as the ground, pin 2 transmits the output signal, pins 3 and 4 supply the low voltage (LV- and LV+), pin 5 manages the \tile\ power down (PD), and pin 6 supplies the \sipm\ bias voltage input (\vbias).}
    \label{fig:tia-scheme}
\end{figure}

Silicon photomultipliers (\sipms) are single-photon detectors composed of arrays of closely arranged Single Photon Avalanche Diodes, which operate above the breakdown voltage, \vbd, to initiate self-sustaining charge avalanches upon photon absorption~\cite{Gallina2019Avalanche}.

The distinct challenge of extracting signals from large areas of \sipms\ arises primarily from their capacitance, which is on the order of \SI{50}{\pico\farad/\square{\milli\meter}}. A single \sipm\ covering a \sipmarea\ area exceeds the nanofarad range.
In the \ds\ detector, the TPC's optical planes have a surface area around \opticalplane.
Such a large surface requires a careful design of the readout electronics and the connection configuration of \sipms.

The solution implemented in \ds\ involves arranging the \tilesipms\ \sipms\ within a \tile\ composed of \(6\) parallel branches, each containing \(4\) \sipms\ connected in series~\cite{DIncecco2018}. This arrangement, hereafter known as \emph{4s6p} (where \emph{s} stands for series and \emph{p} stands for parallel), helps decrease the total capacitance of each series branch, thus improving the bandwidth, at the cost of reducing the output current by a factor of \(4\).
The surface of a \tile\ is \tilearea\ and within it, \tilesensarea\ is the total sensitive area.

The \emph{4s6p} \sipm\ matrix is read out by a single trans-impedance amplifier (TIA).
The \tile\ outputs a single-ended signal that represents the cumulative photo-currents of all \sipms.
A 3D illustration of the \tile\ is depicted in~\autoref{fig:tile-3D}.
The schematics of the \tile\ electronics are illustrated in \autoref{fig:electronic-scheme} and \autoref{fig:tia-scheme}.

In the \tiles, the \sipm\ matrix is attached to the front surface of a \pcb\ with dimensions \pcbarea.
Each \pcb\ comes equipped with a custom 6-pin input-output connector with a \SI{1.27}{\milli \meter} pitch. Of these, two pins supply the low voltage (LV- and LV+) for the operational amplifier, one serves as the ground, another is designated for the \sipm\ bias voltage, one pin manages power down (PD), and the last one transmits the output signal (refer to \autoref{fig:tia-scheme}).
The \pcb's back side accommodates the connector, the low-noise cryogenic readout electronics, and delivers the input voltages to the various components.

A precision voltage divider ensures an even distribution of bias voltage by providing 1/4 of the input bias to all the \sipms.
This passive voltage divider is based on \rdival\ resistors ($\mathrm{R}_{\mathrm{d1}}$ to $\mathrm{R}_{\mathrm{d4}}$ in~\autoref{fig:electronic-scheme}). The divider is also coupled with \cival\ parallel capacitors to act as a low-pass filter of cut-off frequency about \SI{300}{\hertz}. A decoupling capacitor acts as a charge storage device, providing energy when required by the \sipm\ and making it less susceptible to short-term changes in the working point. 
High-precision electronics components (0.5\perc\ resistors, 5\perc\ capacitors) ensure the balance of the bias circuit through the series of \sipms.
The \rdival\ divider resistors result in an equivalent \rdivaltot\ divider resistance. A \rifiltval\ filter resistor ($\mathrm{R}_\mathrm{b}$ in~\autoref{fig:electronic-scheme}) is added in series to the voltage divider, resulting in a \dividerimpedance\ overall impedance.
In the \emph{4s6p} network, an additional \SI{143}{\ohm} series resistance on each parallel branch ($\mathrm{R}_{\mathrm{Br1}}$ to $\mathrm{R}_{\mathrm{Br6}}$ in~\autoref{fig:electronic-scheme}) provides the necessary frequency compensation for the readout amplifier. The resistor introduces a low-pass filter effect, which reduces the gain of the amplifier at higher frequencies and adds a pole in the frequency response of the amplifier, dampening the peak in the noise gain~\cite{DIncecco2018}.

The \tile\ configuration provides a single positive current signal to the amplification stage. 
Early developments in \tile\ readout indicated that charge pre-amplifiers are not suitable for such a large input capacitance, suggesting that a TIA could be a viable alternative~\cite{Aalseth2018}.
For this reason, the DarkSide Collaboration has selected available commercial SiGe operational amplifiers whose performance peaks at \SI{77}{\kelvin}~\cite{DIncecco2018,aggr_out}. \footnote{An alternative technology is implemented for the \tiles\ of the Inner and Outer Vetoes, where the amplification stage is achieved through a custom ASIC~\cite{Kugathasan2020}.}
Reading out the whole \tile\ surface with a single amplifier significantly reduces the readout complexity~\cite{aggr_out}.

The TIA stage is based on a high-speed, ultra-low noise LMH6629 amplifier, made with SiGe technology by Texas Instruments~\cite{tia}, that converts the photocurrent from the \sipms\ to a voltage output with a gain set by the \SI{20}{\kilo \ohm} feedback resistor.
The TIA is supplied by \SI{5}{\volt} (\SI{\pm 2.5}{\volt} with respect to the local middle ground point, refer to~\autoref{fig:tia-scheme}). 
The TIA stage drains an idle current of \SI{15}{\milli\ampere} at room temperature and \SI{11}{\milli\ampere} at cryogenic temperature, which means that each \tile\ has a total power consumption of \SI{55}{\milli\watt} in liquid argon.
Power dissipation in a cryogenic environment is critical since it can cause bubbling when immersed in liquid argon~\cite{Acerbi2025}.
The single-ended output of the TIA is a negative voltage pulse due to the amplifier's inverting configuration.
It is possible to power the TIA on and off using the PD input.

The \ds\ detector needs to meet stringent radiopurity requirements to maximise its sensitivity to very rare particle dark matter signals~\cite{LightDM2024, Pesudo2023}. Therefore, significant effort was dedicated to selecting materials that minimise the radioactivity of the \tile\ while preserving its functionality. The primary background contributions originate from the \pcb\ itself and the Polyethylene Naphthalate (PEN) capacitors mounted on its backside ($\mathrm{C}_1$ to $\mathrm{C}_8$ in~\autoref{fig:electronic-scheme}).

The \pcb\ is made with Arlon 55NT by Arlon Electronic material~\cite{arlon}, an epoxy laminate and pre-preg system, reinforced with nonwoven aramid, meeting the requirements of IPC-4101/55. 
The main \pcb\ stack-up structure is based on two \SI{0.48}{\milli\meter} Arlon laminate sheets between two \SI{17}{\micro\meter} metal layers of copper and separated by two \SI{75}{\micro\meter} pre-preg sheets.

This choice of material efficiently incorporates compatibility with lead-free processing to prevent unintended radioactive contamination from lead isotopes, such as \ce{^{214}Pb}, utilising a high-temperature multifunctional epoxy resin. The resin features a Thermal Expansion Coefficient ranging from \SIrange{6}{9}{ppm \per \kelvin} compatible with that of silicon to avoid mechanical stress and offers excellent dimensional stability owing to the nonwoven aramid reinforcement. Moreover, \pcbs\ with polymeric reinforcement are intrinsically cleaner compared to traditional glass-reinforced laminates. 

The bare \pcbs\ and the individual electronic components were characterised using high-purity broad-energy germanium (BEGe) detectors to quantify radioactive contamination from the uranium and thorium decay chains, as well as gamma-emitting isotopes such as \ce{^{40}K}, \ce{^{60}Co}, and \ce{^{137}Cs}~\cite{Franchini2024,Santone2024}.
The measured activity is below the requirements set for \ds.

\section{\tiles\ production workflow}
\label{sec:tile_production}


\begin{figure*}[tpb]
    \centering
    \includegraphics[width=0.75\linewidth]{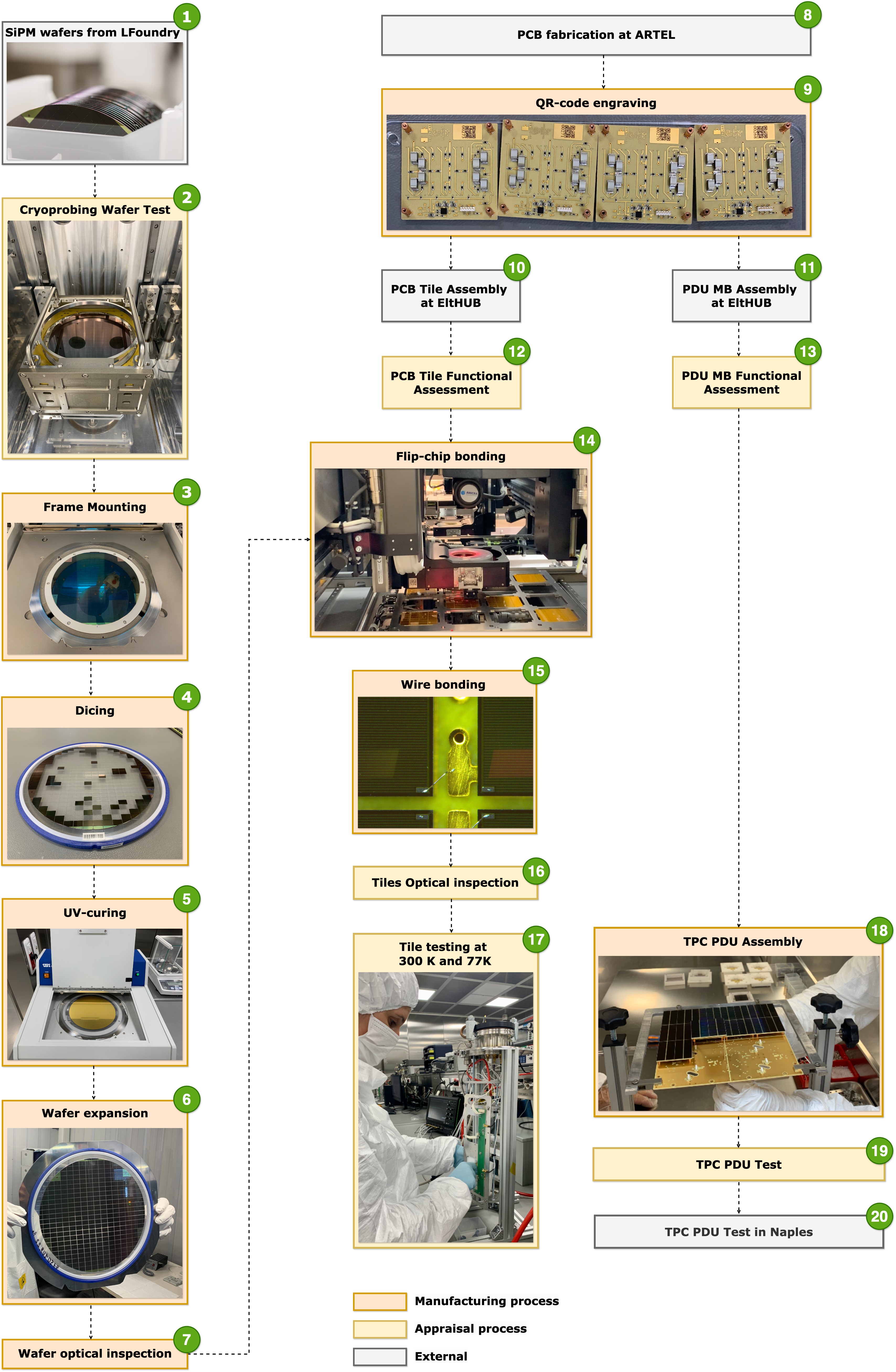}
    \caption{\tile\ Production process flow in NOA. Orange boxes refer to manufacturing processes, yellow boxes to appraisal processes and grey boxes to processes performed by external companies. Wafers from LFoundry \srl\ (1) are characterised at cryogenic temperatures using a high-precision, semi-automated probe system (cryoprobing, 3). Faulty \sipms\ are discarded. Next, wafers are mounted on a metal frame with Blue tape (4) and diced into \(268\) dice (5). The tape is UV-cured (6) to reduce its adhesiveness and expanded (7) for \sipm\ pick up. After wafer optical inspection (8), \sipms\ are bonded on a \pcb\ substrate via thermal compression (14). PDU Motherboards (MBs) and \tile\ \pcbs\ from ARTEL \srl\ (2) are engraved with a QR code (9), assembled with electrical components (10 and 11), and tested before bonding (12 and 13). \sipms\ are wire bonded to \pcb\ to connect anode and cathode terminals of two consecutive \sipms\ within the same \tile\ branch (15). Fully mounted \tiles\ are inspected (16) and tested at room temperature and in liquid nitrogen with the \tile\ Testing Setup (17). Finally, \tiles\ that pass the quality criteria are used to assemble PDUs (18), which are tested at room temperature (19) before being shipped to Naples for a long-term test in liquid nitrogen (20).}
    \label{fig:tile-production-process-flow}
\end{figure*}

\begin{figure}[tpb]
    \centering
    \includegraphics[width=\linewidth]{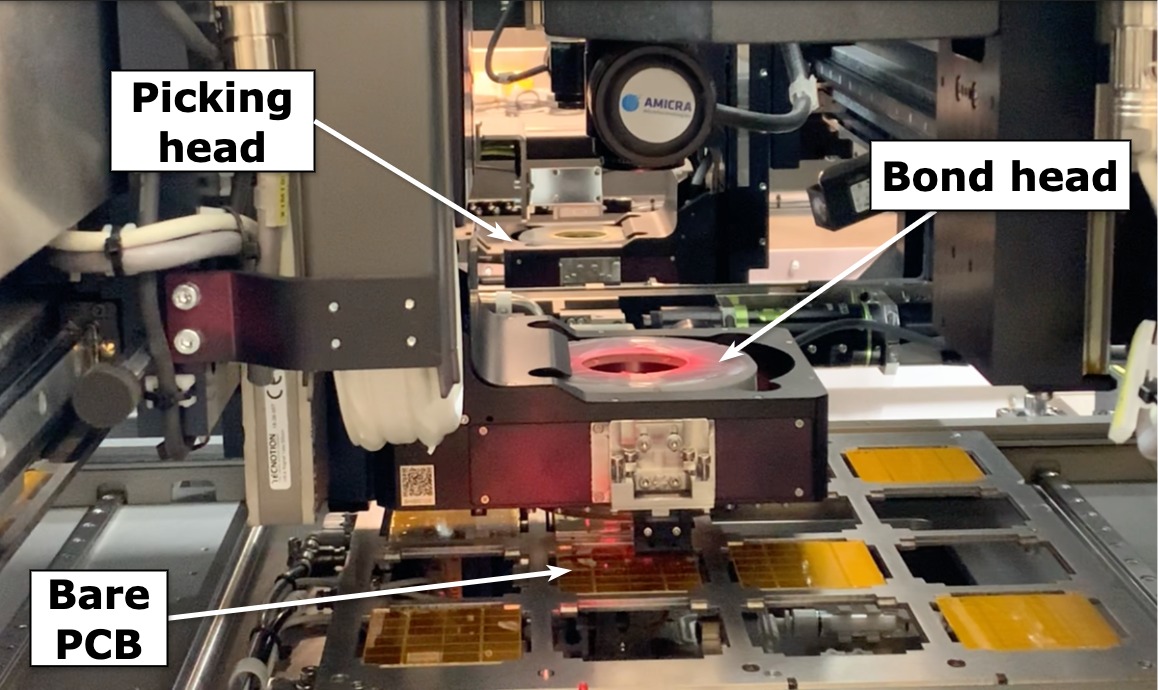}
    \caption{\sipm\ assembly on the \pcbs\ performed using an AMICRA NOVA PLUS flip chip bonder. The custom \pcb\ frame holder, developed to accommodate up to \(16\) \pcbs\ simultaneously, is visible at the bottom. }
    \label{fig:flip-chip}
\end{figure}

\begin{figure}[tpb]
    \centering
    \includegraphics[width=\linewidth]{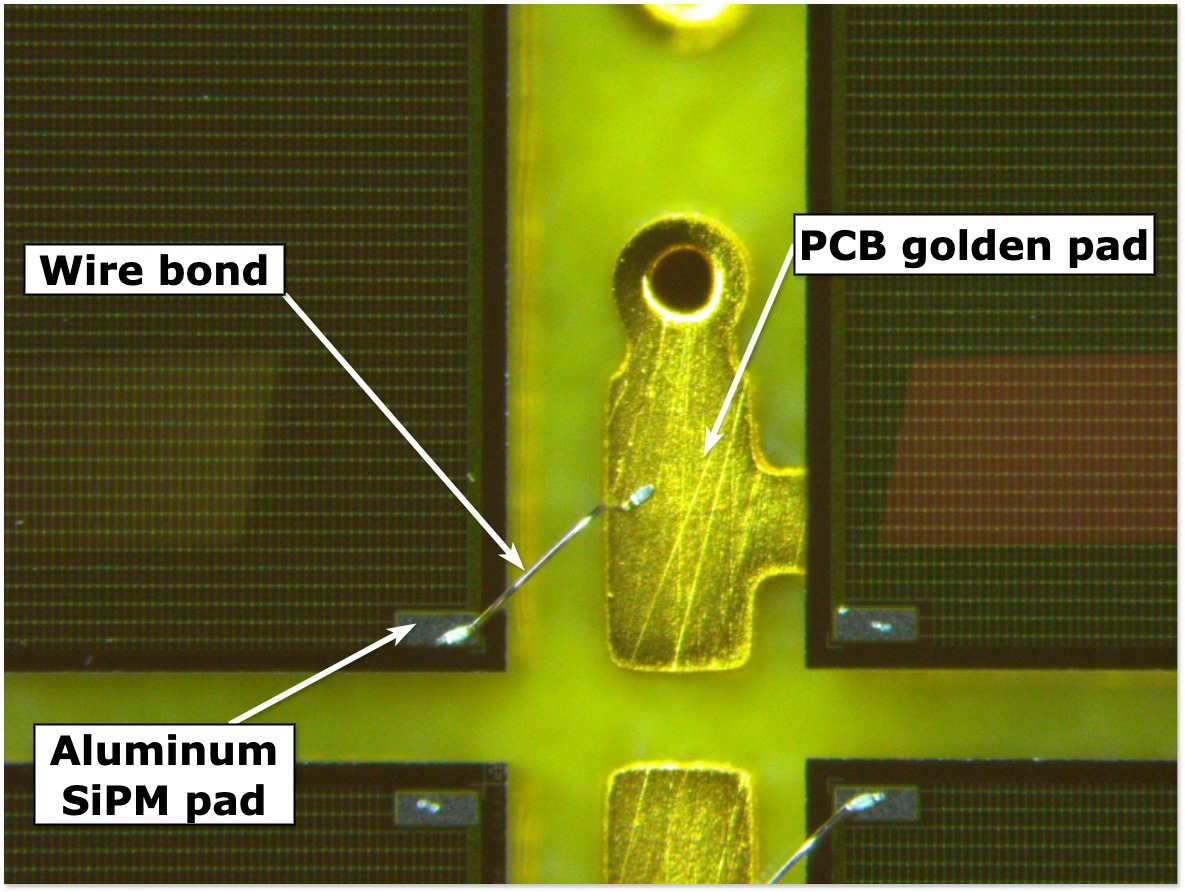}
    \caption{Wire bonding: a low-resistance electrical contact is built between the \sipm\ anode and the \pcb. The contact is made using a \SI{25}{\micro\meter} thick aluminium conductive wire, through a wedge bonding.}
    \label{fig:wire-bonding}
\end{figure}


The production workflow of the \ds\ \tpc\ photoelectronics in NOA follows the block diagram illustrated in \autoref{fig:tile-production-process-flow}. Orange boxes refer to manufacturing processes, yellow boxes to appraisal processes and grey boxes to processes performed by external companies.
This paper focuses on processes \(3\) to \(7\), \(12\), and \(14\) to \(17\). Processes 1 and 2 are detailed in the technical paper~\cite{acerbi2024qualityassurancequalitycontrol}, while processes \(13\), \(18\), \(19\), and \(20\) will be described in other publications of the collaboration.

All wafers for probing are supplied by LFoundry \srl\ (Avezzano, AQ, Italy) ~\cite{Organtini2020}, in lots, and stored in the NOA clean room (process 1 in \autoref{fig:tile-production-process-flow}).

The wafer characterisation is detailed in the \ds\ technical paper~\cite{acerbi2024qualityassurancequalitycontrol}. Here we only report the most relevant aspects.
The wafer characterisation is carried out using a high-precision, semi-automated cryogenic probe system, the Cascade FormFactor PAC200~\cite{FormFactor}, which performs automatic electrical testing at the wafer level at a temperature of \cold\ (process 2 ``cryoprobing wafer test" in \autoref{fig:tile-production-process-flow}).

The cryogenic characterisation of each \sipm\ includes measurements of the breakdown voltage, the quenching resistance, and the leakage current as a function of the bias voltage. The results of these tests are compiled into a wafer map, which is used to asses the performance of individual \sipms. Devices that meet the specified requirements are identified as validated \sipms, while those failing to comply with the performance criteria are excluded from deployment in the \tiles. The acceptance criteria for the \sipms\ are detailed in the technical \ds\ paper~\cite{acerbi2024qualityassurancequalitycontrol}.

The actual throughput of the wafer characterization is \(4\) wafers per day, with the machine in use for \(10-11\) hours per day, with a rate of about \(0.35\) wafers per hour, including the wafer loading, cooling, measurement, warm up, and unloading time~\cite{acerbi2024qualityassurancequalitycontrol}.

After the characterization, each wafer is mounted on a metal frame and secured with Blue tape~\cite{bluetape}, needed to hold the wafer during the dicing process and to retain the dice after they are cut (process 3 ``frame mounting" in \autoref{fig:tile-production-process-flow}). The wafer is then diced into \(268\) \diesize\ dice  (4 ``dicing" in \autoref{fig:tile-production-process-flow}) using a dicing blade with diamond particles and cooled with deionised water. After dicing, the Blue tape undergoes UV curing to reduce its adhesiveness. It is mechanically expanded to increase the spacing between the \sipms, facilitating their pick-up during the die bonding process (processes 5 ``UV curing" and process 6 ``Wafer expansion" in \autoref{fig:tile-production-process-flow}).

An optical inspection using a microscope is performed to identify surface damages that occurred during wafer handling and dicing, as well as manufacturing defects that were not detected during the cryoprobe test (process 7 ``Wafer Optical Inspection" in \autoref{fig:tile-production-process-flow}).
Production wafers have proven to be devoid of large-scale defects; only a small number exhibited issues before the cryoprobe test. Wafer defects include surface scratches caused by the fabrication process, discoloured metal lines, black pads impacting wire bonding connectivity, and missing metallisation among the \sipm\ SPADs. Occasionally, surface features such as edge chipping of the \sipm\ are observed following the wafer dicing.

The operator has access to a list of wafer damages accompanied by pictures to facilitate the identification of defects.
This additional down-selection process reduces the failure rate due to \tiles\ that exhibit anomalous current versus voltage (\iv) characteristic curves.
The \sipms\ that do not pass the optical inspection are excluded from the subsequent flip-chip bonding process. The
percentage of discarded \sipms\ in the optical inspection is around 3\perc.

The \pcb\ substrates were manufactured by ARTEL \srl\ \cite{artel} (Pieve al Toppo, AR, Italy) and stored in the NOA clean room  (process 8 ``PCB fabrication at ARTEL" in \autoref{fig:tile-production-process-flow}). Each \pcb\ substrate then undergoes a QR code engraving process through a laser engraver LUXFIBER30-E JOKE~\cite{laserengraver}  (process 9 ``QR Code Engraving" in \autoref{fig:tile-production-process-flow}).
With the QR code, each \pcb\ can be uniquely identified and tracked during all the stages of the production.
The time needed to engrave a packet of \pcbs\ (\(20\) pieces) is about half an hour, including the unpacking and repacking of the boards. An engraving rate of about \(1000\) \pcbs\ per week is
achieved by a full-time operator.

The bare \pcbs\ are then sent to EltHUB \srl \cite{elthub} (Carsoli, AQ, Italy) for the electronic component population  (process 10 ``PCB Tile assembly at EltHUB" in \autoref{fig:tile-production-process-flow}).
The \pcbs\ with assembled electronics components are cleaned with demineralised water and sent back to NOA to be electrically assessed, as described in \autoref{sec:pcb_test} (process 12 ``PCB Tile Electrical Assessment" in~\autoref{fig:tile-production-process-flow}).

The \sipms\ are integrated on the \pcbs\ by using an AMICRA NOVA PLUS flip chip bonder (shown in \autoref{fig:flip-chip}). This machine is modular,  tailored for micro-assembly applications, and features a placement accuracy of a few \SI{}{\micro \meter}~\cite{amicra}, using a thermocompression bonding technique  (process 14 ``Flip-Chip Bonding" in \autoref{fig:tile-production-process-flow}). This packaging solution incorporates indium solder bump bondings, offering exceptionally robust bonds. The bonding tools of the machine have been customised to attach the large size \sipm\ dice using an indium NC-SMQ80 solder paste (\SIrange{18}{21}{\milli \gram} per \tile, where the range balances insufficient solder paste that causes die detachment and excess paste which may overflow from the \sipms). A custom \pcb\ frame holder has been developed to accommodate up to \(16\) \pcbs\ simultaneously, optimising the assembly throughput~\cite{NOAfrontiers}.

At the start of the process, the silicon wafer is mounted on a custom metal frame and loaded into the machine. Meanwhile, the indium paste is dispensed onto the top surface of each of the \pcbs\ following a predefined dot pattern.
Only \sipms\ that have been validated by both the cryoprobing (process 2) and optical inspection (process 7) are selected for integration. 

\sipms\ are bonded individually onto the tile \pcbs.
Each \pcb\ is first heated to \SI{80}{\celsius} on a chuck holder. Two bond heads, maintained at \SI{135}{\celsius}, then execute the pick-and-place operation, precisely positioning and bonding the \sipms\ with an accuracy of few \SI{}{\micro \meter}. This process ensures a low-resistance electrical contact between the \sipm\ backside (cathode) and the \pcb.
In NOA, the Flip-Chip Bonder machine is able to assemble up to \(32\) complete \tiles\ in \(8\) hours of operation.

Subsequently, a wire bonding process is performed using the HESSE Bondjet BJ855 machine~\cite{wirebonder}  (process 15 ``Wire Bonding" in \autoref{fig:tile-production-process-flow}). This process builds an electrical contact with negligible resistance between the \sipm\ anode and \pcb. The contact is made using a \SI{25}{\micro\meter} thick aluminium conductive wire, through a wedge bonding, that joins the anode pad of the \sipm\ with the contact pad of the tile \pcb, as displayed in \autoref{fig:wire-bonding}. 

The wire bonding process must establish a reliable electrical connection, resilient to temperature variations due to cryogenic operations of the devices, between the anode and cathode terminals of two consecutive \sipms\ within the same branch of a \tile.
Upon the completion of this process, two distinct substrates are interconnected, specifically, the aluminium pad of a \sipm\ and the golden pad of the \pcb\ (refer to \autoref{fig:tile-pictures}), on which the cathode of the next \sipm\ in the series is soldered.
This contact is achieved through a wedge bonding process.
Although all the SiPMs have three anode
pads, the \ds\ \tile\ design only uses one wire between one of these pads and the corresponding cathode terminal.
The time needed to wire bond a batch of \(16\) \tiles\ is about 1 hour.

Fully equipped \tiles\ undergo an optical inspection  (process 16 ``Tile Optical Inspection" in \autoref{fig:tile-production-process-flow}) to verify the correct placement and the quality of the wire bonds, as well as the lack of defects.
Defects in the \tile\ primarily arise during the packaging stage. Dust particles might settle on the wafer and get picked up by the bonding tools during die placement. These particles adhere to the \sipm\ surface during the thermocompression process, becoming permanently fixed.

The \tiles\ that pass the optical inspection are subsequently mounted on the \tile\ test setup where they undergo the quality control procedure detailed in \autoref{subsed:tile_testing_procedure} (process 17 ``Tile Testing at room temperature and in liquid nitrogen" in~\autoref{fig:tile-production-process-flow}).

The \tiles\ that pass the quality control procedure are used to assemble PDUs (process 18 ``TPC PDU Assembly" in \autoref{fig:tile-production-process-flow}). These PDUs then undergo a room temperature test in NOA (process 19 ``TPC PDU Functional Assessment", before being shipped to the Naples Photosensor Test Facility PTF, at Università di Napoli Federico II and INFN, NA, Italy)~\cite{Balmforth2023}. There, the PDUs undergo an electrical assessment at room temperature, followed by a long-term test in liquid nitrogen (process 20 ``TPC PDU test in Napoli" in \autoref{fig:tile-production-process-flow}).

A production database maintains a record of all components employed in assembling \tiles, their progression through each stage, and measurements taken at every assembly phase. It tracks the location and shipment of each part.
The database is a PostgreSQL~\cite{PostgreSQL}, hosted at the INFN-CNAF (Italy) infrastructure and having multiple levels of backup and replication~\cite{Franchini2025}.
The database is accessible from the outside world only by using a password-protected API to retrieve and insert data~\cite{ds20kdb}. In the production phase, various users are granted distinct access levels, depending on their particular roles in assembly or testing.
A web interface provides simplified access to the database contents.

\section{\tile\ Quality Assurance and Quality Control in NOA}
\label{sec:tile-qaqc}

The \tile\ performance is crucial for ensuring the reliability of light collection in the \ds\ \tpc\ and the uniformity of the optical response across the OP.
Spatial uniformity enhances energy, position, and particle-type reconstruction accuracy, particularly for pulse shape discrimination~\cite{Agnes2016, Agnes2018}. Moreover, the \tiles\ must sustain operation in a cryogenic environment without performance degradation.

Furthermore, maintaining a uniform response across the \tiles\ is essential when grouping \tiles\ in PDUs. The \tiles\ within a PDU are operated at a common bias voltage. Therefore, all of the \tiles\ must be able to operate at the maximum overvoltage to avoid the need for reducing the bias voltage or disabling individual \tiles.

The Quality Assurance and Quality Control (\qaqc) process is designed to ensure high-quality \tiles\ with uniform response, suitable for installation on the PDUs.
Specifically, the \qaqc\ ensures uniformity and reliability of the \tiles\ by identifying defects early in production, optimising the manufacturing process, and verifying the operation in liquid nitrogen.

The quality assurance protocol guarantees that \sipms\ within \tiles\ are sourced from the same wafer.
Assembling a \tile\ with \sipms\ from the same wafer significantly reduces \sipm-to-\sipm\ variability. Cryoprobing wafer tests (process ``Cryoprobing wafer test", 2 in~\autoref{fig:tile-production-process-flow}) have indeed shown that most variations of key performance parameters occur between different wafer lots~\cite{acerbi2024qualityassurancequalitycontrol}.
Specifically, \sipms\ within the same wafer exhibit breakdown voltages with a variance of \SI{0.011}{\square \volt} and quenching resistors with a variance of \SI{0.019}{\square \mega \ohm}~\cite{acerbi2024qualityassurancequalitycontrol}.
The variability present within a wafer is due to imperfections in the manufacturing processes across the wafer's surface.

Selecting \sipms\ that meet the acceptance criteria, from the same wafer, guarantees a narrow range of breakdown voltages (with nominal value \vbdsipm\ at \cold) and quenching resistors. This helps prevent gain discrepancies when operating the \sipms\ at a common bias within a \tile, thereby enhancing the uniformity of the signal amplitude and pulse shape across the entire \tile\ surface~\cite{acerbi2024qualityassurancequalitycontrol}.

The risk of superficial defects on the \sipms\ induced by the die bonding is mitigated by continuous monitoring of the process stability and regular maintenance of the die bonding tools (process 14 in \autoref{fig:tile-production-process-flow}).
The constant monitoring of the wire bonding parameters allows us to evaluate the reliability of the wire bonding process (process 15 in \autoref{fig:tile-production-process-flow}). In addition, the quality and long-term stability of the process under various temperature conditions are evaluated through regular destructive tests of a set of wires that are bonded individually onto the golden metalisation of a \pcb\ and the aluminium pads on two opposite \sipms.

The quality assurance protocol also includes an initial electrical assessment of the \pcb\ before mounting the \sipms, as detailed in~\autoref{sec:pcb_test} (process 12 in \autoref{fig:tile-production-process-flow}). 
Fully mounted \tiles\ undergo an optical inspection (process 13 in \autoref{fig:tile-production-process-flow}). Subsequently, the quality of each \tile\ is assessed both at room temperature and in liquid nitrogen (process 13 in \autoref{fig:tile-production-process-flow}). The tests evaluate both the electrical conformity and the response to pulsed light, as detailed in~\autoref{subsed:tile_testing_procedure}.

Two custom LabVIEW applications were developed for the \pcb\ electrical assessment and for the \tile\ quality control.
The applications manage instrument interfaces, assist users through the test setup, automate the data acquisition and quality control, and archive the data in the production database.

\subsection{\pcb\ electrical assessment}
\label{sec:pcb_test}

\begin{figure}[tpb]
    \centering
    \includegraphics[width=\linewidth]{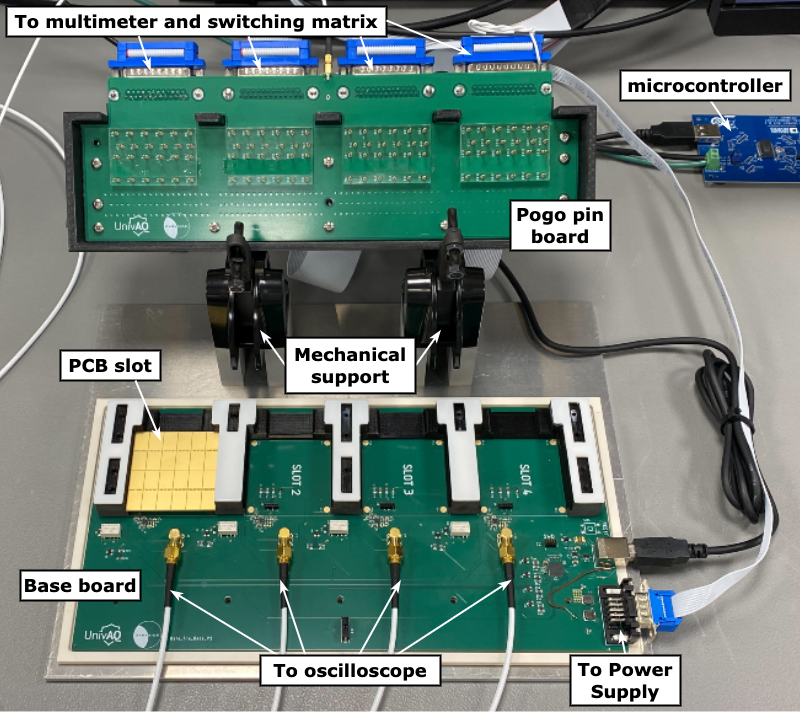}
    \caption{\pcb\ electrical assessment setup. The testing setup is composed of a base with \(4\) connectors hosting \(4\) \pcbs. A lid with a pogo pin board is placed on the \pcbs\ to measure the passive components of the voltage divider.}
    \label{fig:pcb-setup-picture}
\end{figure}

\begin{figure}[tpb]
    \centering
    \includegraphics[width=\linewidth]{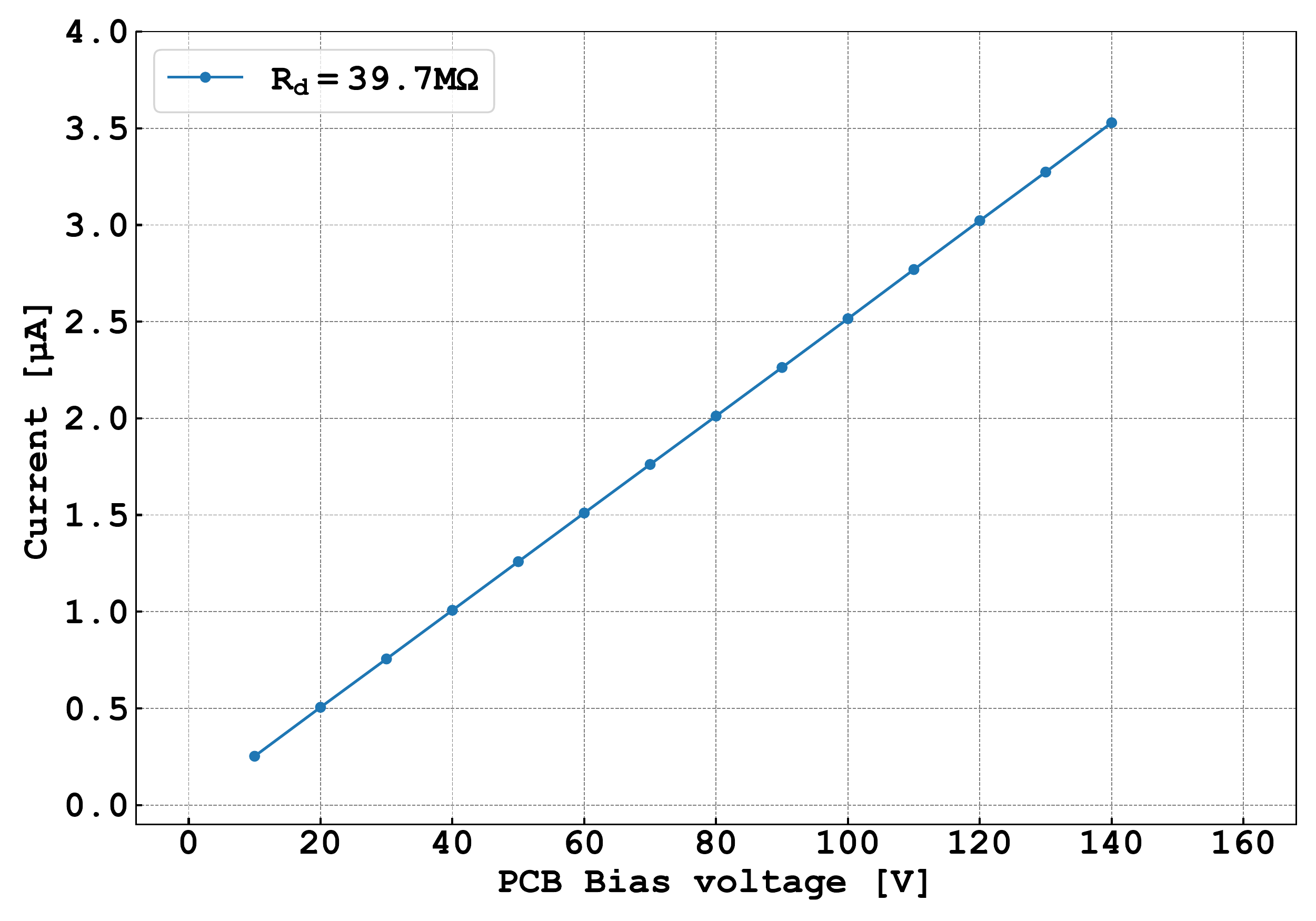}
    \includegraphics[width=\linewidth]{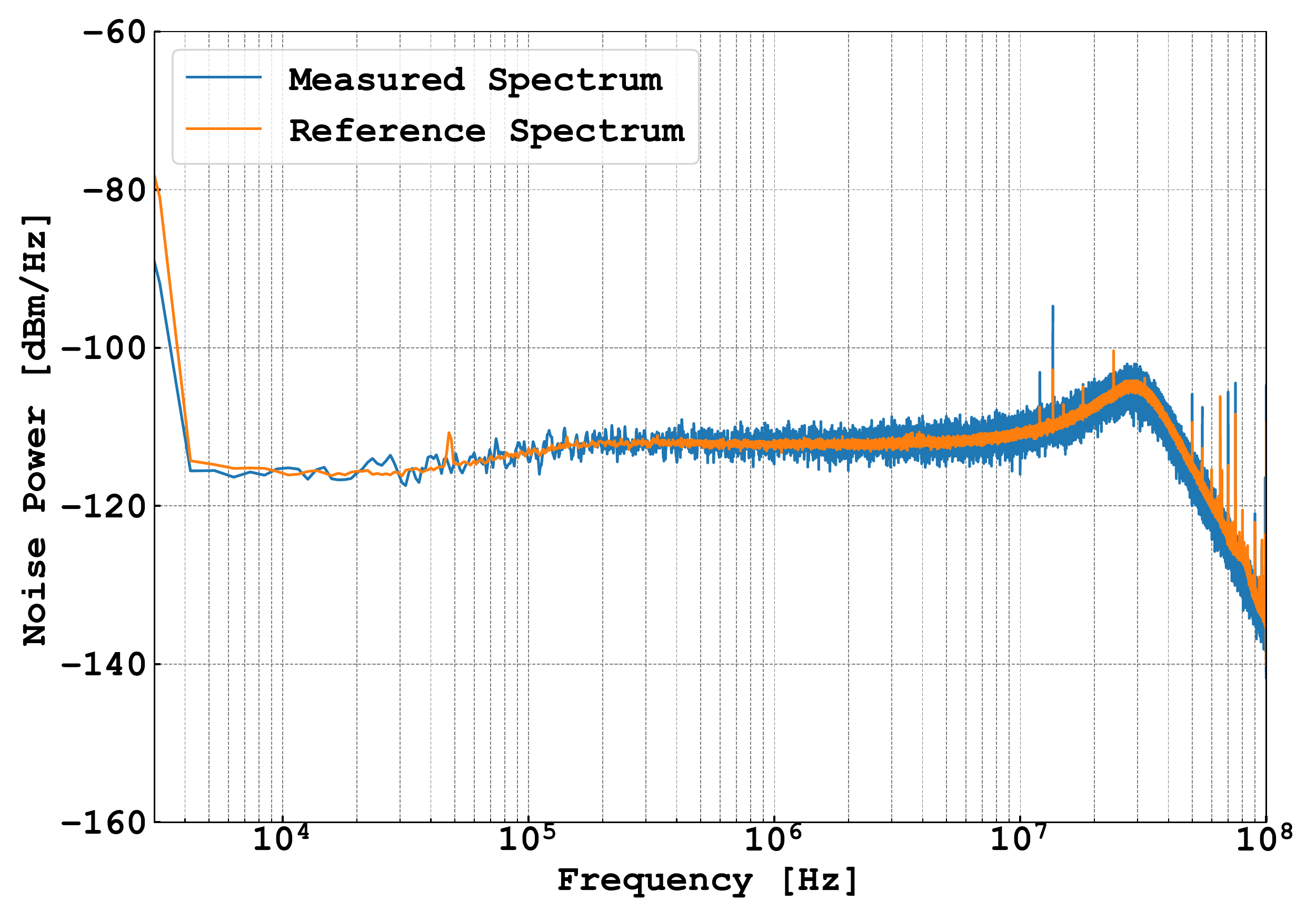}
    \caption{\pcb\ quality assessment. (Top) \iv\ curve, measured to check the linearity of the bias network within its working range and to extract its equivalent resistance. The divider resistance \rd\ is the reciprocal of the slope of the \iv\ curve. (Bottom) \pcb\ noise spectrum compared to the reference spectrum.}
    \label{fig:pcb-test}
\end{figure}

\begin{table}[tbp]
    \centering
    \begin{tabular}{c|c|c}
         Parameter   & Min. & Max.  \\
         \hline
         \hline
         \rd\ (\iv) & \SI{39.4}{\mega\ohm} & \SI{40.1}{\mega\ohm} \\
         \hline
         \hline
         DC offset  & \SI{-150}{\milli\volt} & \SI{150}{\milli\volt} \\
         \hline
         Noise RMS  & - & \SI{6.2}{\milli\volt} \\
         \hline
         Spectrum integral  & - & \SI{900}{\nano\watt} \\
         \hline
         GoF  & - & 5 \\
         \hline
         \hline
         \rgone\  & \SI{14}{\mega\ohm} & \SI{14.4}{\mega\ohm} \\
         \hline
         \rgtwo\  & \SI{23.7}{\mega\ohm} & \SI{24.2}{\mega\ohm} \\
         \hline
         \rgthree\  & \SI{33}{\mega\ohm} & \SI{33.5}{\mega\ohm} \\
         \hline
         \rbrstar\ & \SI{286.3}{\ohm} & \SI{287.5}{\ohm} \\
         \hline
         \rd\ (pogo pin) &  \SI{39.6}{\mega\ohm} & \SI{40.1}{\mega\ohm} \\
         \hline
    \end{tabular}
    \caption{Quality parameters for the \pcb\ test. Min and Max are the boundaries of the specification range. \rd\ (\iv) is the equivalent divider resistance computed as the reciprocal of the slope of the \iv curve. The DC offset and the noise RMS are measured with the oscilloscope in the time domain. 
    Spectrum integral is the integral of the Fast Fourier Transform (FFT) expressed in \SI{}{\nano\watt}, and \gof\ represents the Goodness of Fit in a least squares comparison of the FFT with a reference shape. \rd\ (pogo pin) is the divider equivalent resistance directly measured with the SMU at \SI{40}{\volt} bias. \rgone, \rgtwo, and \rgthree\ are the equivalent resistances of multiple combinations of the divider components, detailed in the text. \rbrstar\ is the value of the combinations of the resistances  $\mathrm{R}_{\mathrm{Br1}}$ to  $\mathrm{R}_{\mathrm{Br6}}$ in each \sipm\ branch. Refer to \autoref{sec:pcb_test} for more information.}
    \label{tab:pcb-params}
\end{table}

A PCB testing station (\autoref{fig:pcb-setup-picture}) was developed to execute electrical tests and to characterise the \pcbs\ from EltHUB after the electrical components assembly (process 12 ``\pcb\ \tile\ functional assessment" in~\autoref{fig:tile-production-process-flow}). The setup includes three structures: the ``base board", the ``pogo pin board" and a mechanical support. 

The ``base board" accommodates up to \(4\) \pcbs\ alongside a microcontroller. The latter is used to drive the PD inputs of the \pcbs\ in order to switch on and off their TIAs. 
The \pcbs\ are manually connected to SMD 6-pin connectors by the user.
The single-ended output of the \pcbs\ is connected to an oscilloscope (\pcbOscilloscope) via \SI{50}{\ohm} SMA coaxial cables to measure the TIA DC offset and to acquire the noise power spectrum.
A DB25 connector connects the board to a \SMU\ Source Meter Unit (SMU).

The ``pogo pin board" is used to apply the appropriate voltage to the \pcbs\ divider resistances, thereby preventing them from remaining in a floating state. This board is equipped with \(24\) spring-loaded contacts (pogo pins) per \pcb\ and \(24\) multi-pin connectors for the external Switching Multimeter (\switchingMatrix\ equipped with a \switchingMatrixsecond).

A mechanical support is needed to apply the appropriate force on the pogo pin board (\SI{60}{\gram} for each pin) and to hold and stabilise the ``base board" and the pogo ``pin board".

The duration of a full electrical test is about \SI{15}{\minute} for \(4\) \pcbs.
The first part of the test consists of measuring the characteristic current-voltage curve with the SMU.
The measurement of the \iv\ curve of the \pcb\ is performed to check the linearity of the bias network within its working range and to extract its equivalent bias resistance through a linear regression. The divider resistance \rd\ is computed as the reciprocal of the slope of the \iv\ curve. An example of an \iv\ curve is shown in~\autoref{fig:pcb-test} (Top).

The second part of the test is the assessment of the noise characteristics using the oscilloscope. The noise characterisation is performed at \SI{40}{\volt} of bias voltage.
For each \pcb\ under test, the oscilloscope returns the measure of the DC offset, the noise RMS, and the noise Fast Fourier Transform (FFT).
The typical shape of a noise spectrum is shown in~\autoref{fig:pcb-test} (Bottom).
For each of the spectra, an assessment is carried out in the frequency range between \SI{100}{\kilo\hertz} and \SI{9.6}{\mega \hertz}. In this range, we compute the integral of the spectrum and a least square comparison with respect to a reference noise curve, hereby called Goodness of Fit (\gof). A detailed description of the \gof\ computation is given in \autoref{sec:noise}.

The quality checks on the noise features (refer to \autoref{tab:pcb-params}) ensure that the \pcb\ has DC offset and noise levels within specifications and that the spectrum is compatible with that of a \pcb\ in which all components have their assigned values within the expected tolerance and are assembled correctly.

In the third part of the \pcb\ test, the discrete components mounted on the \pcb\ are evaluated by measuring the current draw at specific points of the circuit when the bias network is supplied with \SI{40}{\volt}, and by measuring a set of equivalent resistances in the \pcb\ voltage divider using the custom pogo pin board.

The \SI{10}{\mega \ohm} network of the voltage divider (``Passive \vbias\ Divider" in the orange shaded area of \autoref{fig:electronic-scheme}) is assessed by measuring some equivalent resistances using the Switching Multimeter and the SMU. 
The equivalent resistances $\mathrm{R}_{\mathrm{g1}}$,  $\mathrm{R}_{\mathrm{g2}}$, and $\mathrm{R}_{\mathrm{g3}}$ under test are defined as 

\begin{align}
    \mathrm{R}_{\mathrm{g1}} &= \mathrm{R}_{\mathrm{d1}} + \left(\mathrm{R}_{\mathrm{(1,7,13)}}//\mathrm{R}_{\mathrm{(4,10,16)}}//\left(\mathrm{R}_{\mathrm{d2}}+\mathrm{R}_{\mathrm{d3}}+\mathrm{R}_{\mathrm{d4}}\right)\right)  \, ,\\
    \mathrm{R}_{\mathrm{g2}}  &= \mathrm{R}_{\mathrm{d1}} + \mathrm{R}_{\mathrm{d2}} + \left(\mathrm{R}_{\mathrm{(2,8,14)}}//\mathrm{R}_{\mathrm{(5,11,17)}}//\left(\mathrm{R}_{\mathrm{d3}} + \mathrm{R}_{\mathrm{d4}}\right)\right) \, ,\\
    \mathrm{R}_{\mathrm{g3}} &= \mathrm{R}_{\mathrm{d1}} + \mathrm{R}_{\mathrm{d2}} + \mathrm{R}_{\mathrm{d3}} + \left(\mathrm{R}_{\mathrm{(3,9,15)}}//\mathrm{R}_{\mathrm{(6,12,28)}}//\mathrm{R}_{\mathrm{d4}}\right) \, , 
    \label{eq:pcb_test_conf}
\end{align}
where the ``+" and ``//" symbols indicate series and parallel combinations, respectively, and the subscripts with parenthesis denote which resistor is used in each specific measure (for instance, $\mathrm{R}_{\mathrm{g1}}$ is measured three times, once with the parallel of  $\mathrm{R}_{\mathrm{1}}$ and $\mathrm{R}_{\mathrm{4}}$, then $\mathrm{R}_{\mathrm{7}}$ and $\mathrm{R}_{\mathrm{10}}$ and finally $\mathrm{R}_{\mathrm{13}}$ and $\mathrm{R}_{\mathrm{16}}$).
The nominal values of $\mathrm{R}_{\mathrm{g1}}$, $\mathrm{R}_{\mathrm{g2}}$ and $\mathrm{R}_{\mathrm{g3}}$ are \SI{14.3}{\mega \ohm}, \SI{24}{\mega \ohm}, and \SI{33.3}{\mega \ohm}, respectively, with a tolerance of about \SI{0.5}{\mega \ohm}.

The Switching Multimeter is then used to asses the resistors $\mathrm{R}_{\mathrm{Br1}}$ to $\mathrm{R}_{\mathrm{Br6}}$ in each parallel branch of the \emph{4s6p} network (refer to \autoref{fig:electronic-scheme}).  
Since a direct measurement of these resistors is not possible, an indirect measure of five equivalent resistors is performed.
This process involves fixing one resistor in the positive pole and measuring its series resistance with combinations of the remaining five resistors. These resistors are sequentially grounded using the internal generator of the \switchingMatrix, configured in multimeter mode. The combinations of the five resistors are:

\begin{align}
\mathrm{R}_{\mathrm{Br}}^{(1+2)} &= \mathrm{R}_{\mathrm{Br1}} + \mathrm{R}_{\mathrm{Br2}}\\
\mathrm{R}_{\mathrm{Br}}^{(1+3)} &= \mathrm{R}_{\mathrm{Br1}} + (\mathrm{R}_{\mathrm{Br2}}//\mathrm{R}_{\mathrm{Br3}})\\
\mathrm{R}_{\mathrm{Br}}^{(1+4)} &= \mathrm{R}_{\mathrm{Br1}} + (\mathrm{R}_{\mathrm{Br2}}//\mathrm{R}_{\mathrm{Br3}}//\mathrm{R}_{\mathrm{Br4}})\\
\mathrm{R}_{\mathrm{Br}}^{(1+5)} &= \mathrm{R}_{\mathrm{Br1}} + (\mathrm{R}_{\mathrm{Br2}}//\mathrm{R}_{\mathrm{Br3}}//\mathrm{R}_{\mathrm{Br4}}//\mathrm{R}_{\mathrm{Br5}})\\
\mathrm{R}_{\mathrm{Br}}^{(1+6)} &= \mathrm{R}_{\mathrm{Br1}} + (\mathrm{R}_{\mathrm{Br2}}//\mathrm{R}_{\mathrm{Br3}}//\mathrm{R}_{\mathrm{Br4}}//\mathrm{R}_{\mathrm{Br5}}//\mathrm{R}_{\mathrm{Br6}})
\end{align}
The nominal value of each of the $\mathrm{R}_{\mathrm{Br}}^{*}$ is \SI{143}{\ohm} with a tolerance of about \SI{1}{\ohm}.

Finally, the divider resistance \rd\ is measured by employing the pogo pin alongside the SMU resistance measurement mode, at an input bias voltage of \SI{40}{\volt}. This approach might yield a result that deviates slightly from the slope of the \iv\ curve, as it is sensitive to the \SI{10}{\mega \ohm} resistors $\mathrm{R}_{\mathrm{1}}$ to $\mathrm{R}_{\mathrm{16}}$ in \autoref{fig:electronic-scheme}.

The \pcb\ quality control relies on the parameters listed in~\autoref{tab:pcb-params}.
The ranges were initially set according to the tolerances, then fine-tuned to exclude the outliers in the measured distributions.
A \pcb\ is considered accepted when all the quality parameters are within the specifications.
The divider resistance \rd\ is evaluated with two different methods because of its substantial influence on \tile's performance.

\subsection{\tile\ quality control}
\label{subsed:tile_testing_procedure}

The quality control on the \tiles\ happens during the appraisal process 17 (``\tile\ testing at \warm\ and \cold\ in~\autoref{fig:tile-production-process-flow}). The test is divided into two stages: an initial assessment at room temperature to identify potential electronic failures, followed by a second test in liquid nitrogen (\cold) to evaluate the \tile\ performance. The electrical behaviour of a \tile\ is expected to remain consistent when immersed in liquid argon, which has a boiling point of approximately \SI{87}{K}.

The test starts with an initial assessment of the current drawn at zero bias voltage to check the electrical connections (\autoref{sec:lv_curr_ass}). The test proceeds with a measure of the \iv\ curve in reverse bias, detailed in~\autoref{sec:iv}, which allows the extraction of the breakdown voltage and measurement of the \tile\ voltage divider resistance.
Subsequently, the noise RMS and the noise power spectrum are acquired with an oscilloscope, as detailed in \autoref{sec:noise}.
The test in liquid nitrogen includes a pulse counting measurement with a laser source. This process entails collecting pulse waveforms to evaluate their adherence to our quality criteria, as detailed in~\autoref{sec:pulse}.

The total time needed to complete one full test, hence four \tiles, is about \(1\) hour, including the time needed to warm up the \tiles\ after the test in liquid nitrogen.

\subsubsection{Hardware Setup}
\label{subsubsec:hardware_setup}

\begin{figure}[tbp]
    \centering
    \includegraphics[width=0.99\linewidth]{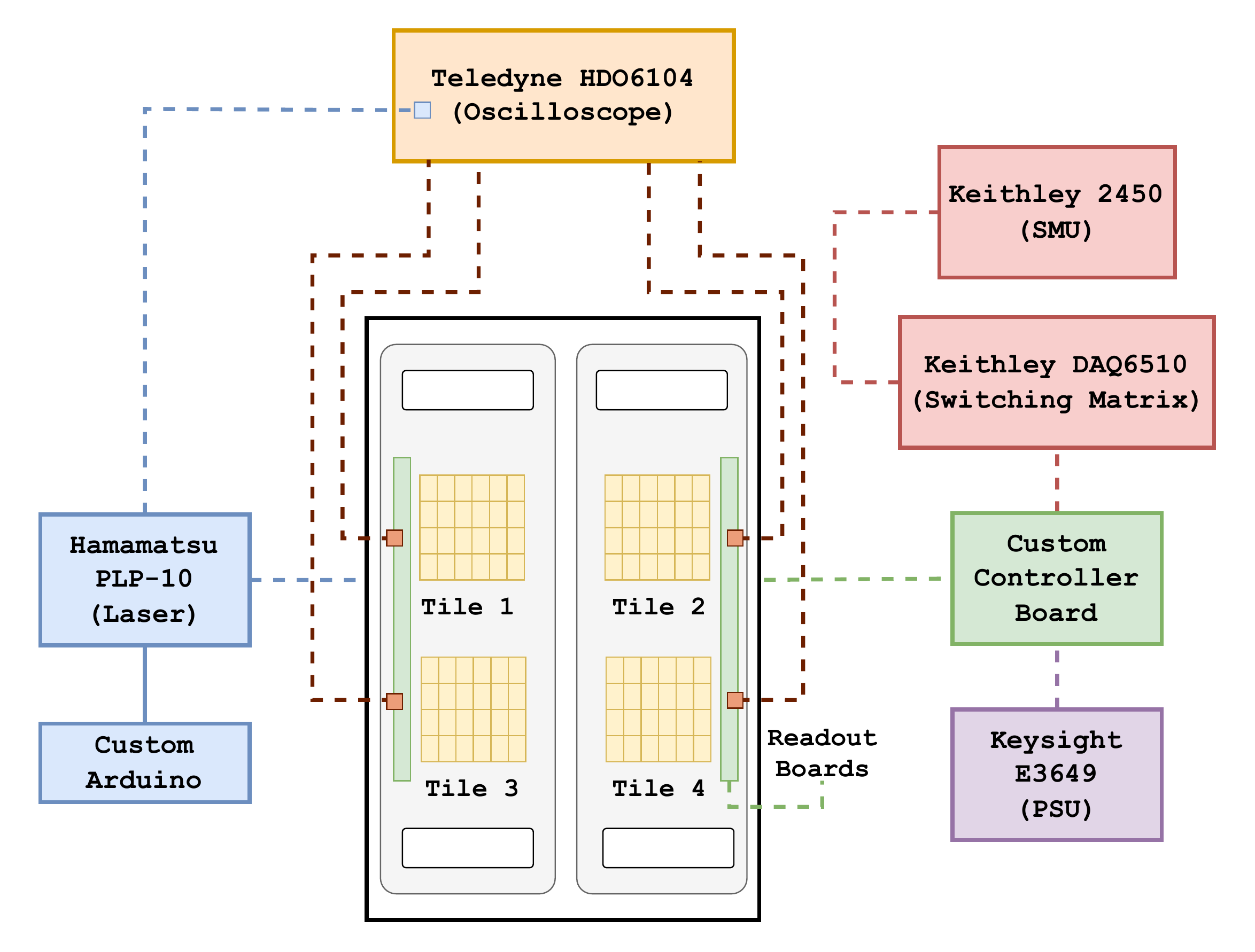}
    \caption{Schematic representation of the \tile\ Testing Setup. The black box represents the dewar, where up to four \tiles\ can be mounted in pairs on two mechanical holders. Signals from the \tiles\ are acquired through a \tileOscilloscope\ oscilloscope. \tiles\  are illuminated by an optical fibre connected to an external laser source (\tileLaser), which is triggered by a custom Arduino. A custom controller board housing an onboard microcontroller drives the power down of the \tiles\ and triggers temperature checks. A Power Supply Unit (PSU, \tilePSU) provides the Low Voltage (LV) to the electrical part of the \tile. The Source Meter Unit (SMU, \SMU) generates the bias voltage, and a Switching Matrix (\switchingMatrix) provides it to the \tile\ under test.}
    \label{fig:Tile-setup-scheme}
\end{figure}

\begin{figure*}[tpb]
    \centering
    \includegraphics[width=0.71\linewidth]{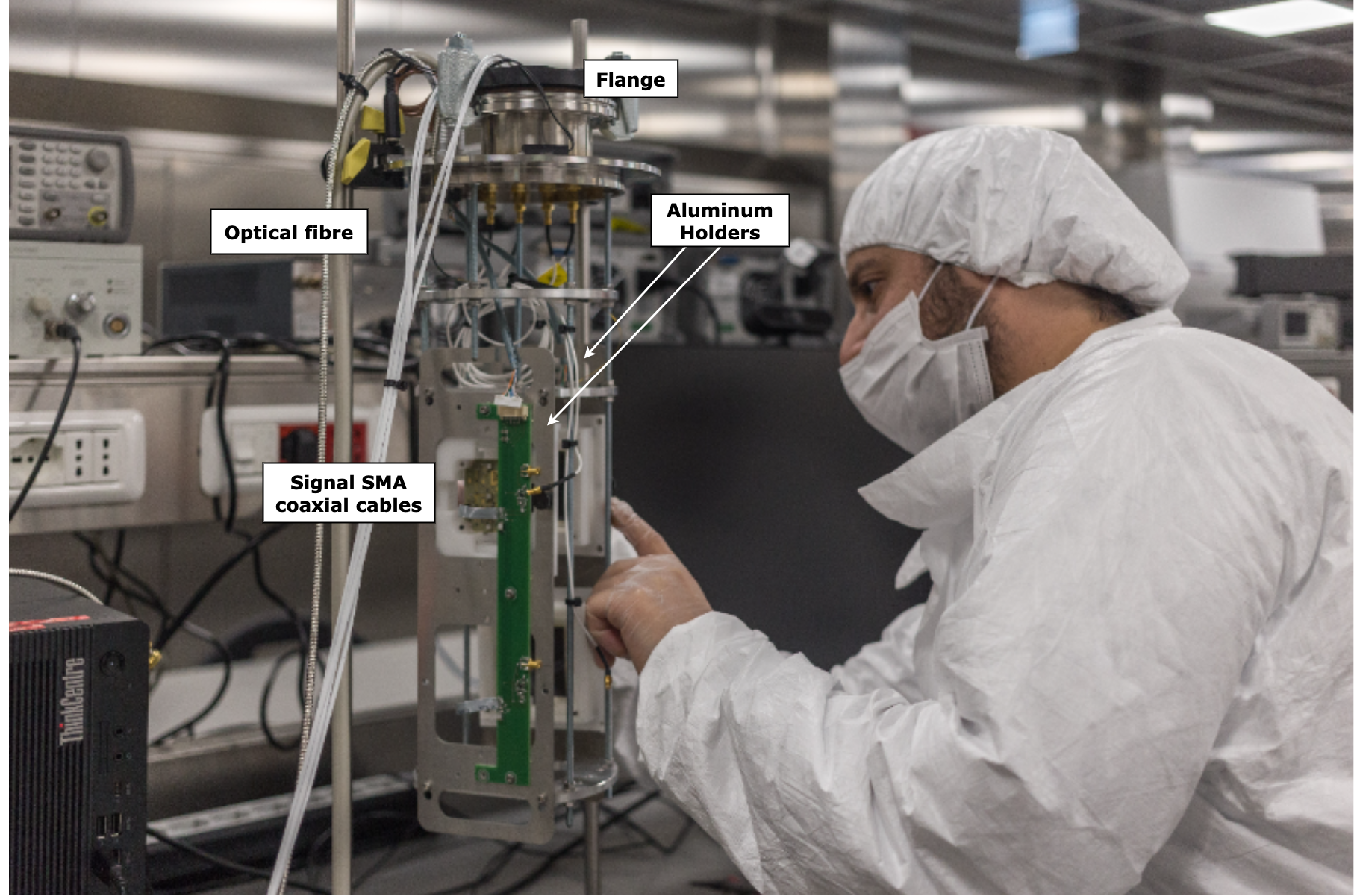}
    \includegraphics[width=0.28\linewidth]{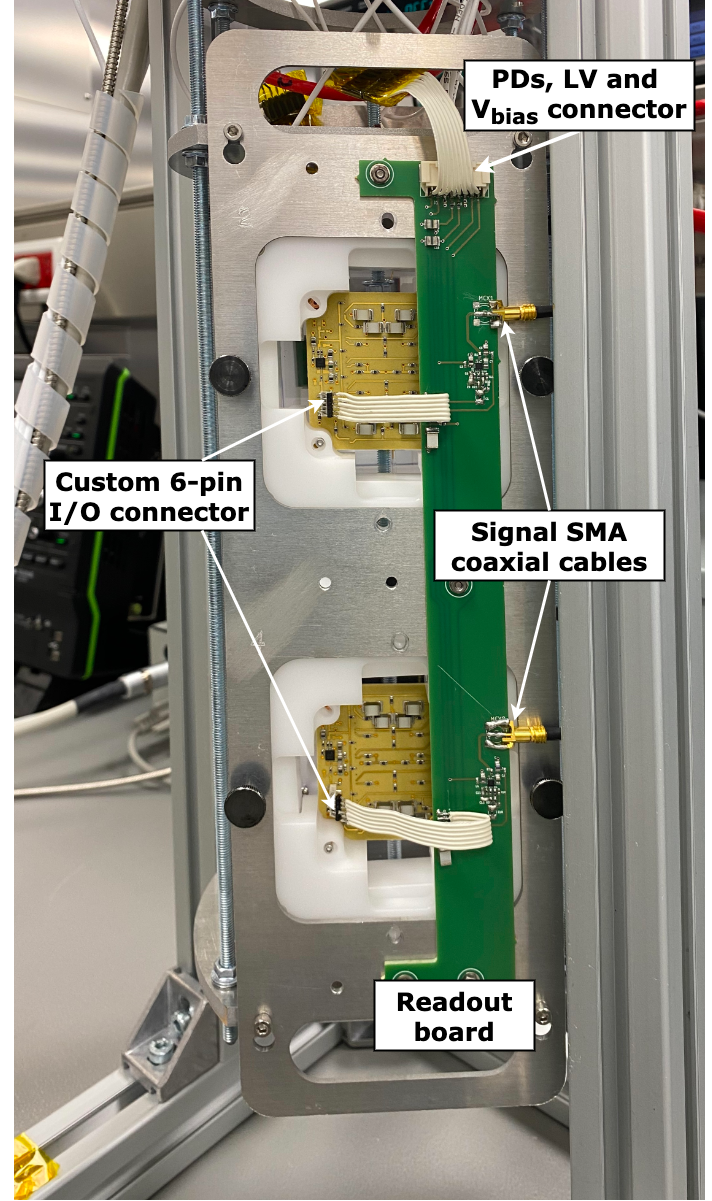}
    \caption{Mechanical structure of the \tile\ Testing Setup. Four \tiles\ are mounted in pairs on two aluminium holders secured to the main frame. The \tiles' output signals are amplified by custom readout boards and connected to an oscilloscope via SMA coaxial cables exiting from the flange. The latter is also equipped with an optical fibre connected to a laser and illuminating the four \tiles.}
    \label{fig:Tile-setup-picture}
\end{figure*}

\begin{figure}[tpb]
    \centering
    \includegraphics[width=\linewidth]{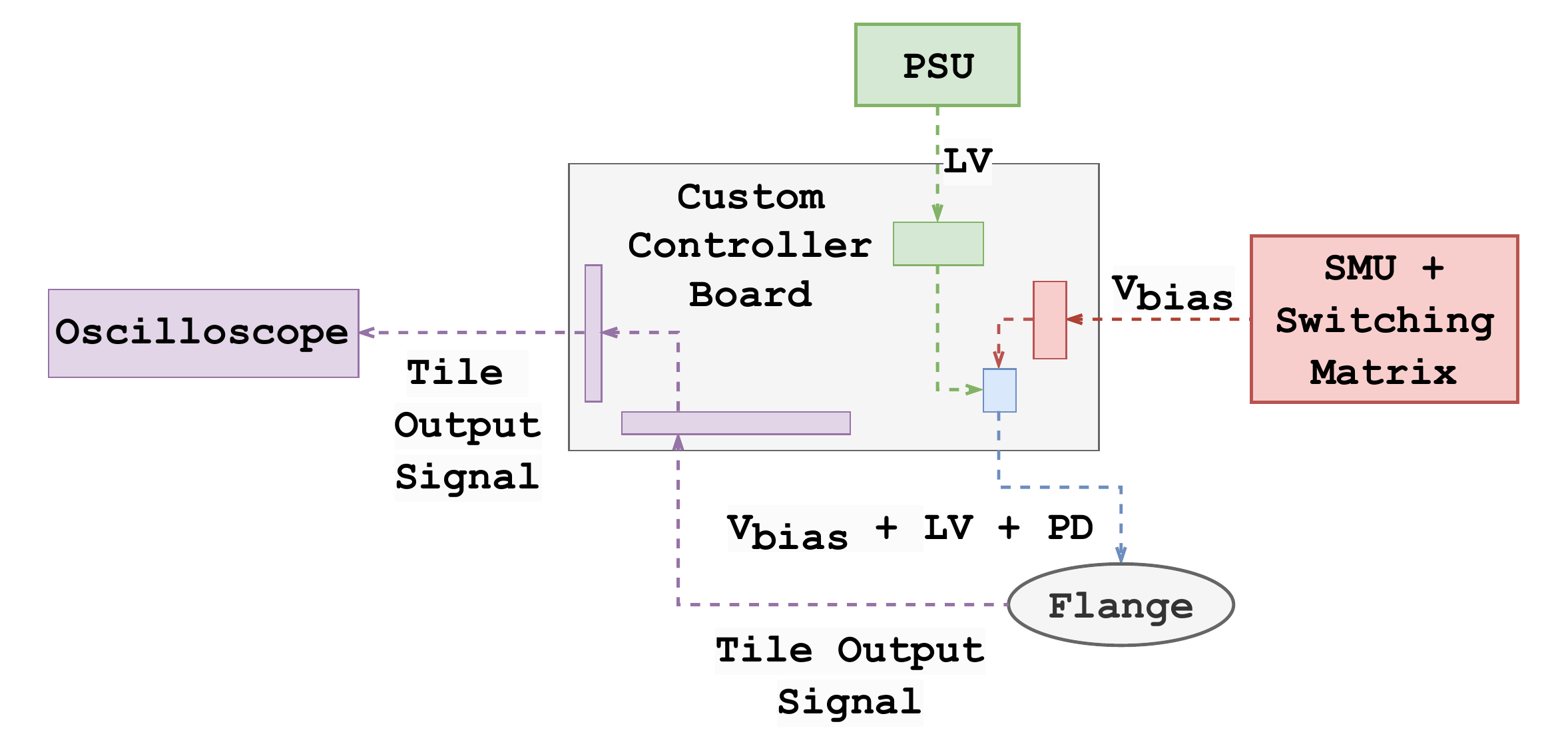}
    \caption{Schematic representation of the bias voltage \vbias\ and Low Voltage (LV) paths through the custom controller board.}
    \label{fig:hv-lv}
\end{figure}


The \tiles\ are tested using a custom setup which can carry up to four units. A schematic representation of the \tile\ testing setup is shown in~\autoref{fig:Tile-setup-scheme}. Two identical tile test setups are available in NOA, to increase the \tiles\ production throughput. 

Before each measurement, the four \tiles\ under test are first tightly screwed on a plastic support and covered with an acrylic layer and then mounted in pairs on two aluminium holders, in turn, secured to the main frame (\autoref{fig:Tile-setup-picture}).

The mechanical structure is positioned inside a dewar, equipped with a sealed flange featuring a venting copper tube and electrical and optical feedthroughs. Four coaxial SMA connections are employed for the output signal transmission and connected to the \tiles\ via custom readout boards.
The two readout boards serve as an impedance adapter between the output of the \tiles\ and the oscilloscope, also providing a tenfold amplification of the signals via an onboard operational amplifier.
The \tile\ output signals are then acquired with an oscilloscope (\tileOscilloscope).

Additionally, a hermetically-sealed optical fibre feedthrough is integrated to allow the \tiles\ to be illuminated by an external laser source (\tileLaser), triggered by a custom Arduino~\cite{arduino}. 
The fibre is terminated with a cylindrical plastic diffuser and placed between the four \tiles\ to improve the uniformity of the illumination.
The setup also includes a custom controller board housing an onboard microcontroller (Adafruit Feather M0, SAMD21 \cite{adafruit}), which triggers temperature checks and drives the power down of the TIA (refer to \autoref{fig:tia-scheme}), which is needed to turn on and off each \tile\ individually.
The custom controller board is also responsible for distributing the utilities to the \tiles. A Power Supply Unit (PSU, \tilePSU) provides the Low Voltage (LV) to the \pcb\ (\autoref{fig:tia-scheme}), while the SMU (\SMU) generates the bias voltage (\vbias, refer to \autoref{fig:electronic-scheme}) needed to bias the \sipms. A \switchingMatrixsecond\ Switching Matrix is used to deliver \vbias\ to the \tile\ under test. \autoref{fig:hv-lv} shows a schematic representation of the \vbias\ and LV paths through the controller board.
To perform the test in liquid nitrogen, the mechanical structure is slowly inserted into the dewar to avoid abrupt temperature changes.
After the test in liquid nitrogen, the \tiles\ are moved inside a cylinder, where they are thermalised using a flow of heated gaseous nitrogen to prevent condensation.

\subsubsection{LV current assessment}
\label{sec:lv_curr_ass}
\begin{table}[tpb]
    \centering
    \begin{tabular}{c|c|c|c}
         Parameter & Temperature & Min. & Max.  \\
         \hline
         \hline
         \multirow{2}{*}{\ILVoff} & \warm & \SI{60}{\milli\ampere} & \SI{71}{\milli\ampere} \\ 
         &\cold  & \SI{35}{\milli\ampere} & \SI{52}{\milli \ampere}\\
         \hline
         \multirow{2}{*}{\ILVon}& \warm &\SI{74}{\milli \ampere} & \SI{87}{\milli \ampere}\\ 
         &\cold&\SI{4}{\milli \ampere} &\SI{62}{\milli \ampere}\\
    \end{tabular}
    \caption{Quality assurance requirements for the Low Voltage (LV) current drawn with the \tile\ powered down (\ILVoff) and turned on (\ILVon).
    The criteria have been tuned to identify a poor connection of the \tile\ with the measurement setup.
    }
    \label{tab:lv}
\end{table}

At the beginning of each test, the current consumption of the \tile\ is measured under nominal LV conditions ($\pm$ \SI{2.5}{\volt}) without any bias voltage.
Specifically, we asses the LV current supplied by the Power Supply Unit. When the \tile\ is not biased, the current (\ILVoff) is mostly absorbed by the amplifying readout boards (refer to~\autoref{subsubsec:hardware_setup}); when the \tile\ is powered, the current (\ILVon) is predominantly drawn by the TIA.
Typical average values include \SI{66}{\milli \ampere} for the readout boards and \SI{15}{\milli \ampere} for the TIA at room temperature, whereas in liquid nitrogen, these values drop to \SI{39}{\milli \ampere} and \SI{11}{\milli \ampere}, respectively. \autoref{tab:lv} details the current consumption criteria.
If these specifications are not satisfied, it is usually due to a poor connection of the \tile\ within the measurement setup.

\subsubsection{\iv\ curve quality control}
\label{sec:iv}

\begin{table}[tpb]
    \centering
    \begin{tabular}{c|c|c|c}
         Parameter & Temperature & Min. & Max.  \\
         \hline
         \hline
         \multirow{2}{*}{\vbd}& \warm & \SI{130.9}{\volt} & \SI{133}{\volt} \\ 
         &\SI{77}{K}& \SI{107}{\volt} & \SI{110.1}{\volt}\\
         \hline
         \multirow{2}{*}{\rd}& \warm  &\SI{38}{\mega \ohm} & \SI{39.5}{\mega \ohm}\\ 
         &\SI{77}{K}&\SI{40.2}{\mega \ohm} &\SI{41.5}{\mega \ohm}\\
    \end{tabular}
    \caption{Quality parameters for the \iv\ curves and requirements at room temperature (\warm) and in liquid nitrogen (\cold). The parameter \vbd\ is the breakdown voltage, and \rd\ is the equivalent resistance of the voltage divider.}
    \label{tab:iv-params}
\end{table}

\begin{figure*}[tpb]
    \centering
    \begin{subfigure}{0.49\textwidth} 
        \centering
        \includegraphics[width=\linewidth]{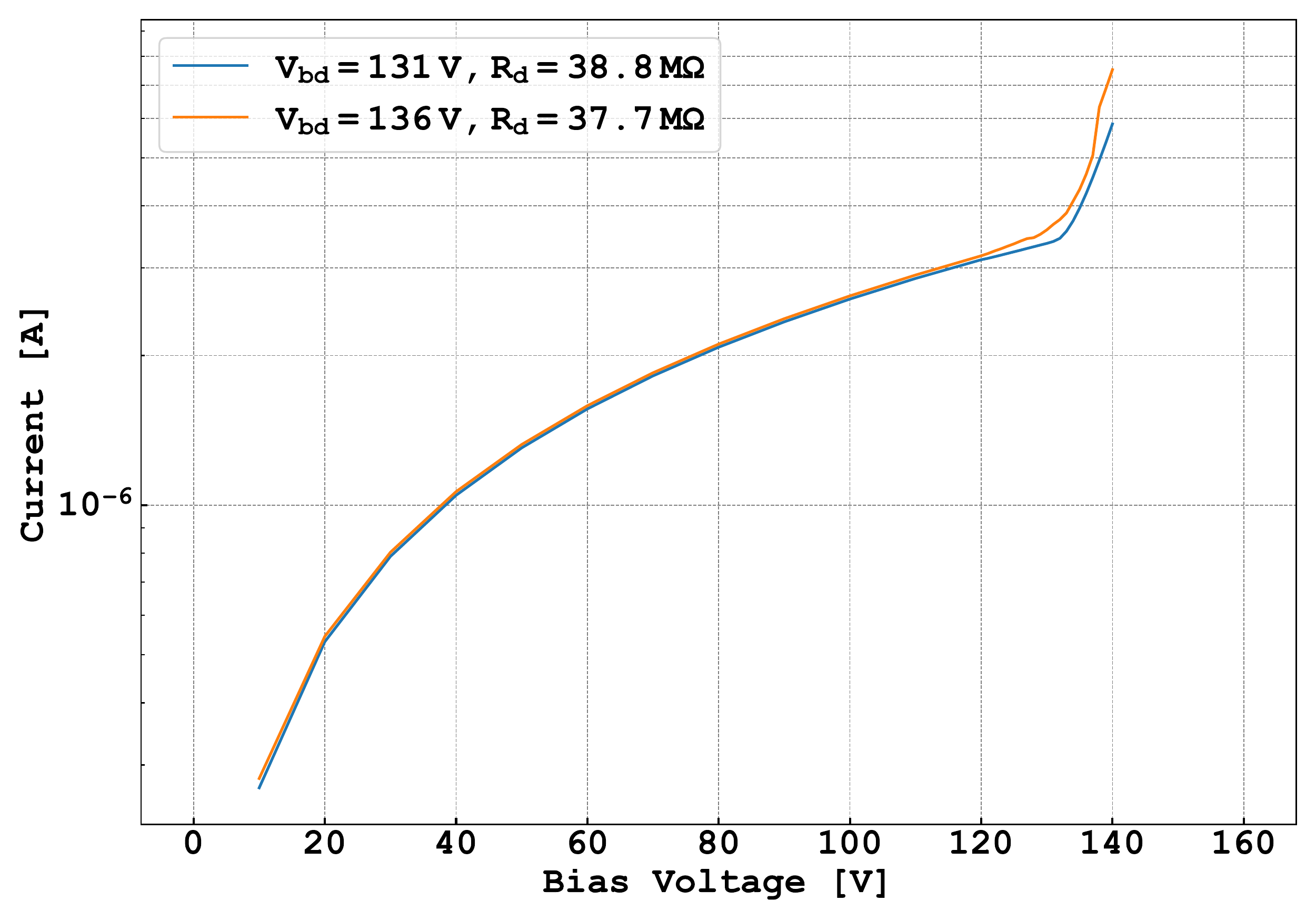}
    \end{subfigure}
    \begin{subfigure}{0.49\textwidth}
        \centering
        \includegraphics[width=\linewidth]{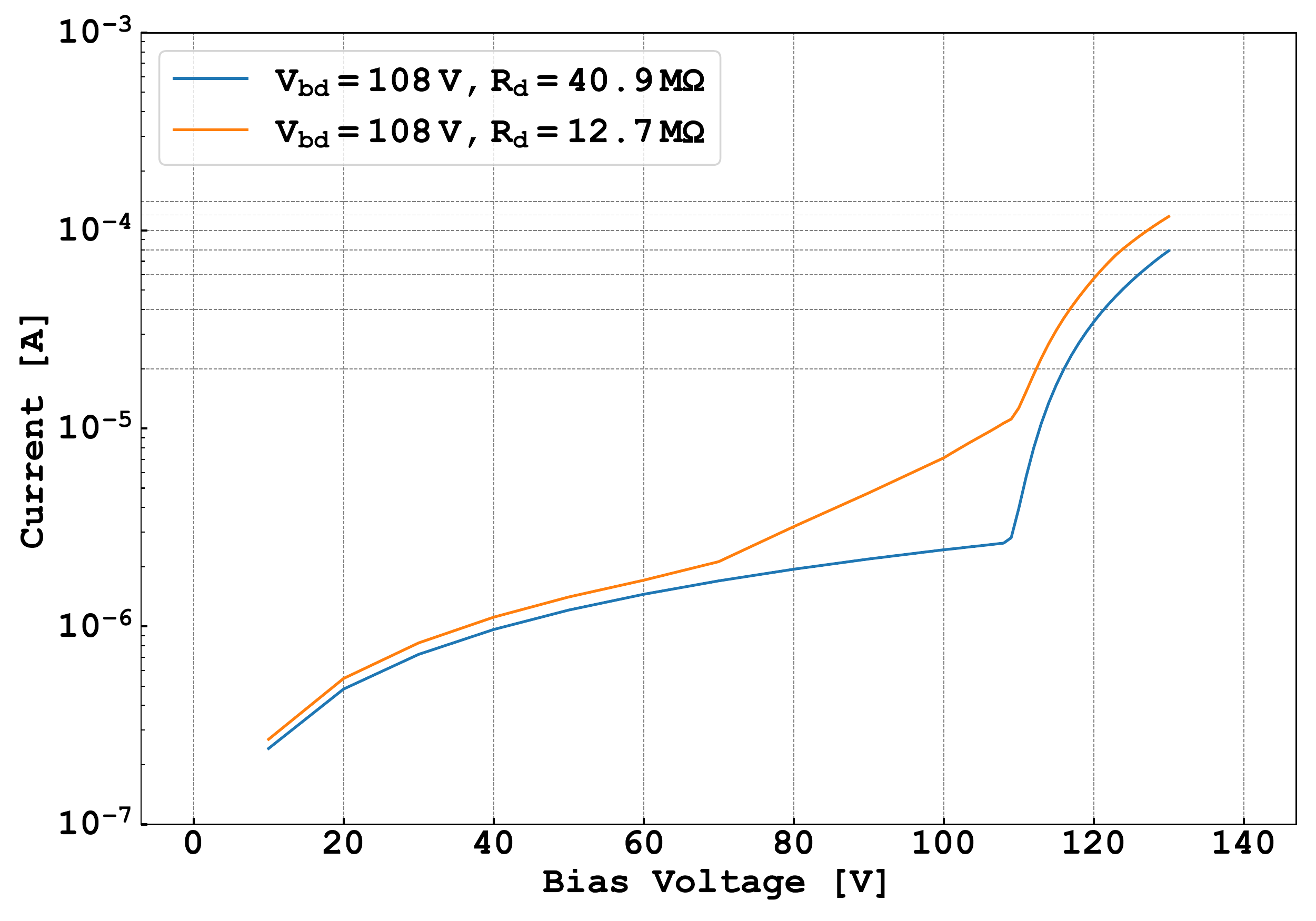}
    \end{subfigure}
    \caption{\iv\ curves examples at room temperature (Left) and in liquid nitrogen (Right). The blue curves belong to \tiles\ that passed the quality control, indicating that the tile breakdown voltage \vbd\ and the divider resistance \rd\ are within the acceptance ranges. Conversely, the orange curves belong to \tiles\ with quality parameters that deviate from the requirements.}
    \label{fig:iv-curves}
\end{figure*}


The \iv\ of each \tile\ in reverse-bias mode is obtained using a SMU \SMU. The measure is acquired both at room temperature and in liquid nitrogen.
During the liquid nitrogen measurement, a laser source at \SI{16}{\mega \hertz} and maximum intensity is illuminating the \tiles. Such a high light flux is needed to induce a detectable current, of the order of tens of \SI{}{\micro \ampere}, in the divider. 

The voltage sweep is performed with a \SI{10}{\volt} step where the linear region of the \iv\ curve is expected. Here the current flows through the \tile\ voltage divider. On the other hand, a step of \SI{1}{\volt} is used for the region where the breakdown is foreseen. 
At room temperature, the former is \SIrange{10}{120}{\volt} and the latter is \SIrange{121}{140}{\volt}, while in liquid nitrogen the ranges are respectively \SIrange{10}{100}{\volt} and \SIrange{100}{120}{\volt} (the nominal breakdown voltage of a \sipm\ at \cold\ is \vbdsipm, corresponding to \vbdtile\ for a \tile). 
A one-second delay is introduced between the voltage being applied and the current measurement. The current is measured five times for each voltage, with the median value recorded to mitigate the impact of fluctuations and outliers when adjusting the bias.

The parameters estimated from the \iv\ curve to asses the electrical functioning of the \tile\ are the \tile\ breakdown voltage \vbd\ and the equivalent resistance of the bias voltage divider, \rd.  

The \tile\ breakdown voltage \vbd\ is the turning point of the \iv\ curve in the reverse bias region. The nominal value of \vbd\ for each \tile\ is \SI{131}{V} at room temperature and \SI{108}{V} in liquid nitrogen. We define the ``turning point" of the \iv\ curve as the voltage corresponding to the maximum of the second derivative of the logarithm of the current,

\begin{equation}
\centering
\frac{ \mathrm{d}^2 \, { (\ln \mathrm{I}) } }{ \mathrm{d} \, {\mathrm{V}^2} }\bigg|_{\mathrm{V}=\mathrm{V}_{\mathrm{bd}}} = \text{max} \, .
\label{eq:2log}
\end{equation}

The derivatives are computed via finite difference methods. The first derivative at point $n$ is obtained by subtracting the value at point $n$ from that at point $n+1$. Similarly, the second derivative follows the same calculation process. Subsequently, the algorithm identifies the maximum value in the second derivatives array, represented by the index $n$. Consequently, \vbd\ is estimated as the voltage corresponding to index $n+1$.

This second derivative method is distinct from the first derivative approach previously employed in the wafer quality assurance procedure for estimating the breakdown voltage of the \sipms~\cite{acerbi2024qualityassurancequalitycontrol}. For the individual \sipms, \vbd\ is determined as the voltage at which the first derivative of the current reached its peak. Because the \tiles\ incorporate a TIA stage, the post-breakdown curve does not rise as sharply as it does in the \sipms. As a result, using the first derivative becomes less reliable for identifying the transition point due to the less pronounced peak in its curve. Conversely, the second derivative offers a more accurate indication of the transition point.

The current flowing prior to breakdown is the one that passes through the \tile\ voltage divider.
The nominal value of its equivalent resistance \rd\ is about \SI{40}{\mega \ohm}, corresponding to the series combination of the four \SI{10}{\mega \ohm} resistors that make up the \tile\ voltage divider (refer to~\autoref{fig:electronic-scheme}).
The parameter \rd\ is estimated from the \iv\ curve as the reciprocal of the slope determined by a linear regression conducted in the linear region.

This parameter is primarily employed to detect suboptimal \iv\ curves, such as those exhibiting another, albeit milder, deviation from linearity significantly below \vbd\ (for instance, approximately \SI{70}{\volt} in \autoref{fig:iv-curves}).
This phenomenon, known as \emph{double slope} or \emph{double breakdown}, leads to an estimate of \rd\ that is significantly lower than our acceptance range.

\autoref{tab:iv-params} lists the quality requirements for the parameters \vbd\ and \rd. The requirement on \vbd\ is established to accept \tiles\ populated with working \sipm\ with nominal breakdown voltage. The acceptance range on \rd\ has been fine-tuned to exclude \tiles\ which exhibit the double slope feature.

\autoref{fig:iv-curves} shows some examples of \iv\ curves measured at room temperature and in liquid nitrogen. Each plot includes the results of the quality test. 
In both figures, the blue curves belong to \tiles\ that passed the quality control on the \iv\ curve, indicating that both \vbd\ and \rd\ are within the specifications of~\autoref{tab:iv-params}. Conversely, the orange curves belong to \tiles\ that deviate from the quality specifications.

\subsubsection{Noise quality control}
\label{sec:noise}

\begin{table}[ht!]
    \centering
    \begin{tabular}{c|c|c|c}
         Parameter & Temperature & Requirement  \\
         \hline
         \hline
         \multirow{2}{*}{DC offset}&\warm& $<$ \SI{180}{\milli \volt}\\
         &\cold& $<$ \SI{150}{\milli \volt}\\
         \hline
         \multirow{2}{*}{Noise RMS}&\warm& $<$ \SI{9}{\milli \volt}\\
         &\cold& $<$ \SI{4.1}{\milli \volt}\\
         \hline
         \multirow{2}{*}{Spectrum integral}& \warm  & $\in$ (\SI{2.0}{\micro \watt}, \SI{2.5}{\micro \watt})\\
         &\cold& $\in$ (\SI{0.32}{\micro \watt}, \SI{0.5}{\micro \watt} ) \\
         \hline
         \multirow{2}{*}{Spectrum \gof}&\warm& $<$ 3  \\
         &\cold& $<$ 4  \\
         \\
    \end{tabular}
    \caption{Quality parameters on the \tile\ noise at room temperature (\warm) and in liquid nitrogen (\cold). The absolute value of the DC offset and the noise RMS measured by the oscilloscope in the time domain, 
    Spectrum integral is integral of the power spectrum expressed in $\mu$W, and \gof\ is the Goodness of Fit.}
    \label{tab:noise-params}
\end{table}

\begin{figure*}[tpb]
    \centering
    \begin{subfigure}{0.49\textwidth} 
        \centering
        \includegraphics[width=\linewidth]{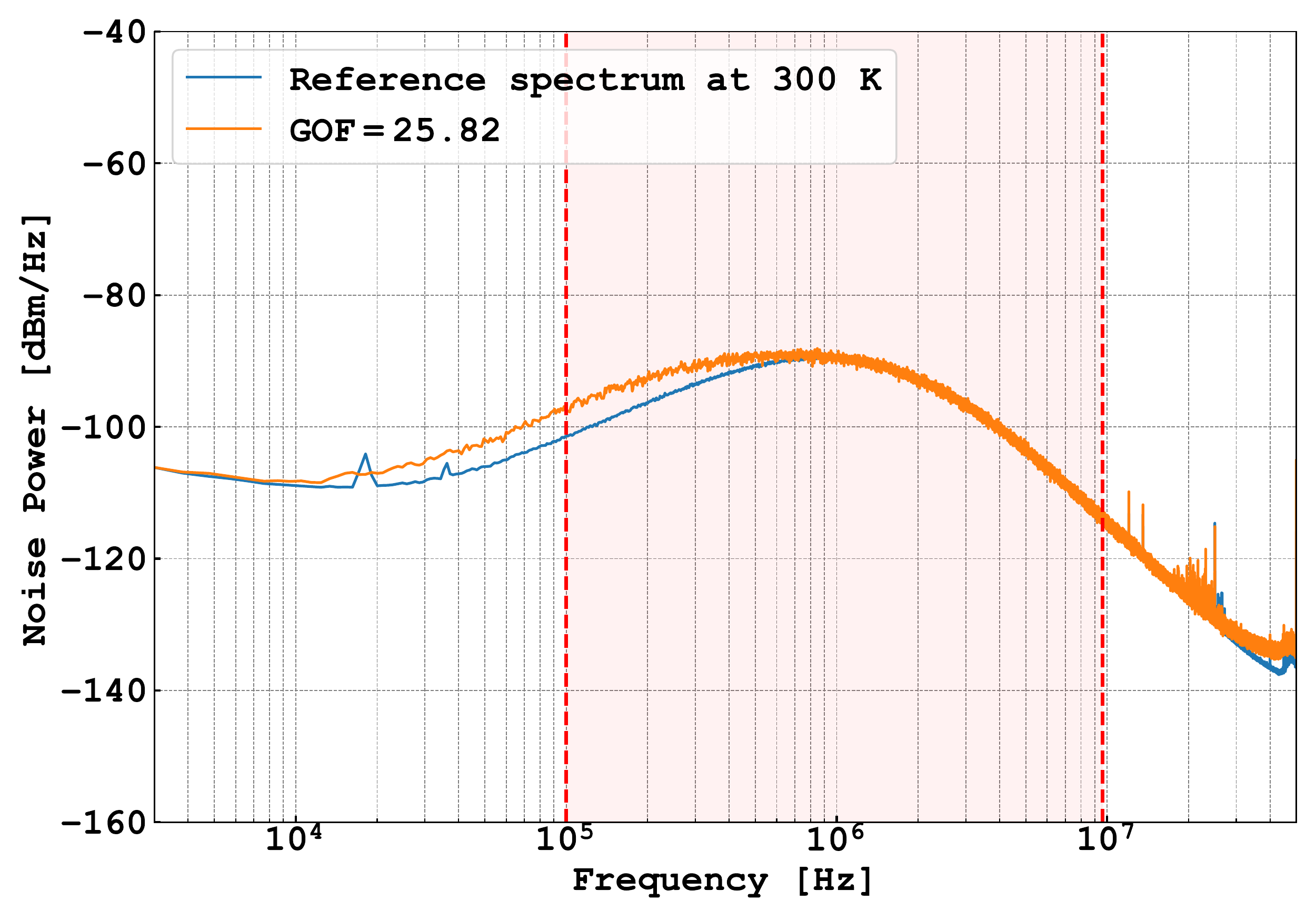}
    \end{subfigure}
    \begin{subfigure}{0.49\textwidth}
        \centering
        \includegraphics[width=\linewidth]{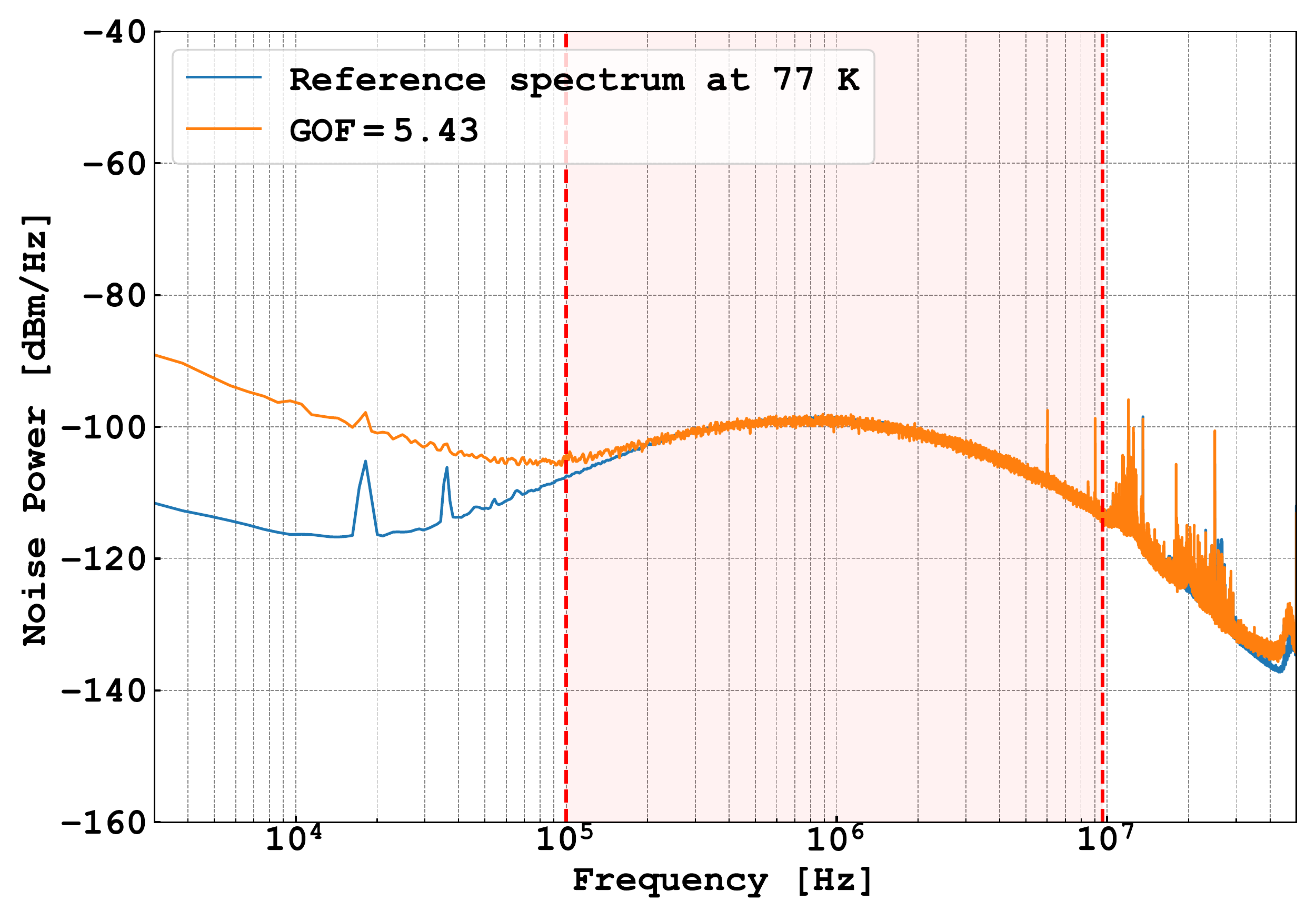}
    \end{subfigure}
    \caption{Noise spectra examples at room temperature (Left) and in liquid nitrogen (Right). The blue curves represent the reference noise spectra. Conversely, the orange curves represent \tiles\ with at least one quality parameter deviating from these requirements. The shaded area indicates the region where the Goodness of Fit parameter is evaluated (\SI{100}{\kilo \hertz} to \SI{9.6}{\mega \hertz}).
    In this case, the Goodness of Fit parameter (\gof) of the orange curve does not meet the specifications.}
    \label{fig:noise-spectrum}
\end{figure*}

\begin{figure}[tpb]
\centering
\includegraphics[width=\linewidth]{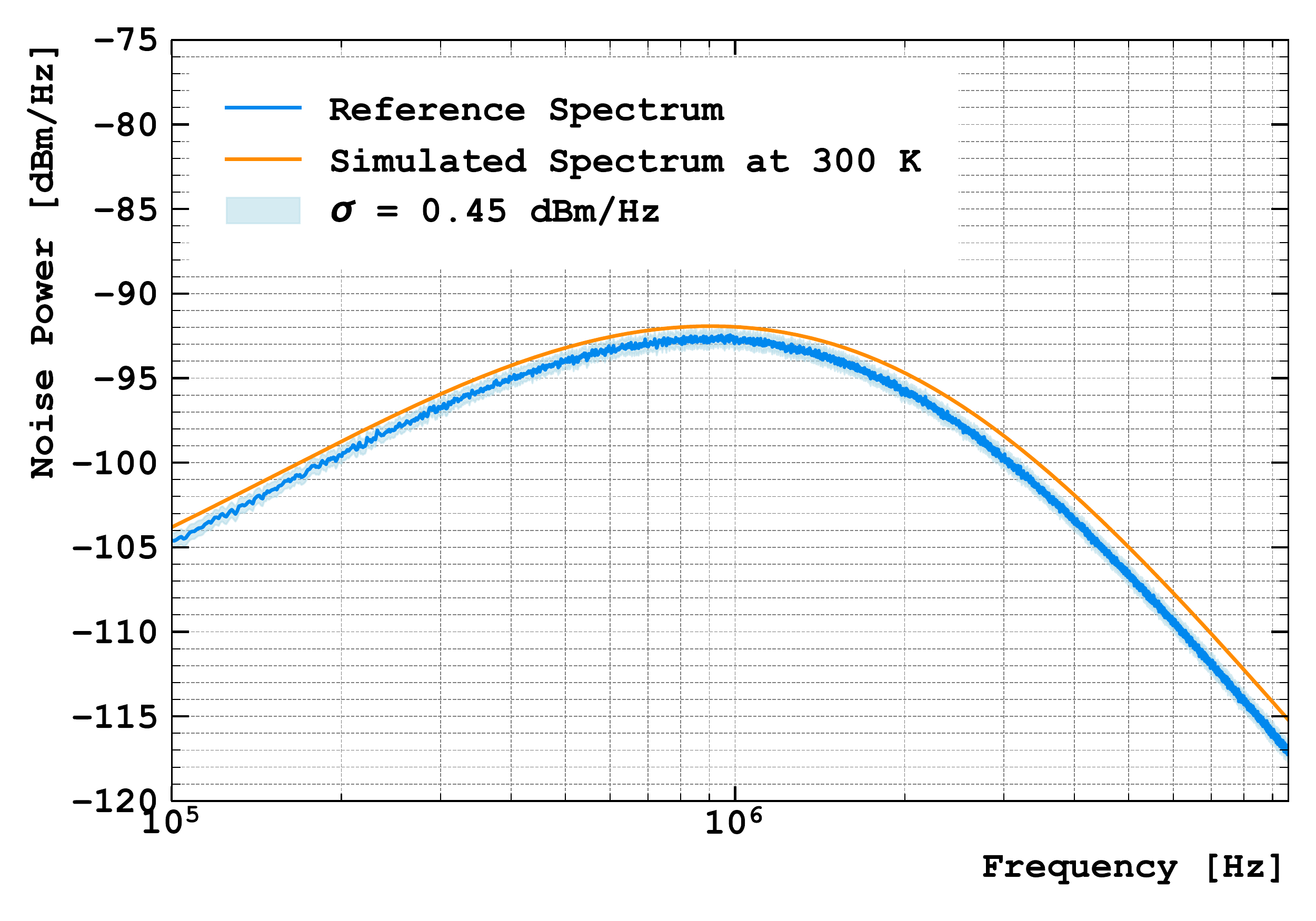}
    \caption{Noise spectrum of the \tile\ simulated with the TINA-TI software (orange) compared to the noise spectrum of a reference \tile\ at room temperature (blue). The shaded region displays the error band including the statistical uncertainty.}
    \label{fig:noise-simulation}
\end{figure}

The noise root mean square (RMS) and the noise power spectrum for each \tile\ under test are measured using a \tileOscilloscope\ oscilloscope. These measures occur both at room temperature and in liquid nitrogen, at \SI{40}{V} bias voltage, significantly below the \tile\ breakdown voltage, so that the avalanche generation in the \sipms\ is suppressed.

In \emph{Spectrum} mode, the oscilloscope directly measures the power spectrum. We collect the average power spectrum across \(100\) waveform triggers, each having a duration of \SI{0.2}{\milli \second}, sampled at \SI{250}{MSample/s} and constrained by a bandwidth limit of \SI{200}{\mega \hertz}. The oscilloscope returns the power spectrum in dBm, with a resolution bandwidth of about \SI{2}{\kilo \hertz}.
Examples of noise spectra are shown in~\autoref{fig:noise-spectrum}. 
For the same set of data, the oscilloscope also returns the noise RMS and the DC offset.

The analysis of the noise spectrum is conducted within the frequency range spanning from \SI{100}{\kilo \hertz} to \SI{9.6}{\mega \hertz}, as this is the range of interest for the \tile's output signals.
Above this range, high-frequency spikes due to electromagnetic pick-up from the signal cables pollute the analysis outcome.
Below this range, the rising spectrum at low frequency observed in some \tiles\ is attributed to burst noise linked to production defects in the TIA and can be solved by replacing the chip.

The spectra are then normalised by converting the output spectrum of the oscilloscope from dBm to \SI{}{\milli \watt}.
The integral of the spectrum (in units of Watt) is computed in the frequency range \SI{100}{\kilo \hertz} to \SI{9.6}{\mega \hertz}. A low value of the spectrum integral is typically due to detached branches in the \sipm\ bias network.

A \gof\ parameter is introduced to quantify the similarity to a noise spectrum of a \tile, where each component is assembled correctly and meets its designated value within the acceptable tolerance.
The \gof\ is defined as:
\begin{equation}
\mbox{\gof} = \frac 1 {N -1 } \sum_{i=1}^N \frac{(y_i \, - \, k \bar Y_i)^2}{\sigma ^2} \, ,
\label{eq:gof}
\end{equation}
Here, $y$ is the value of the spectrum under evaluation, $\bar Y$ is the value of the reference spectrum, the subscript $i$ refers to the $i$-th of the $N$ points in the set, $\sigma$ is the statistical uncertainty on the reference, and $k$ is a global scale factor to parametrise small deviations to the reference. The \gof\ is minimised with respect to $k$.

The \gof\ is computed for a set of $N = 7$ frequencies chosen to have a flat distribution in logarithmic space (i.e. \SI{100}{\kilo \hertz}, \SI{200}{\kilo \hertz}, \SI{400}{\kilo \hertz}, $\dots$, \SI{6.4}{\mega \hertz}). This choice guarantees that deviations in the frequency range between \SI{100}{\kilo \hertz} and \SI{1}{\mega \hertz} count as much as deviations in the frequency range between \SI{1}{\mega \hertz} and \SI{10}{\mega \hertz}.

The reference spectrum shape is computed as an average among \(16\) \tiles\ of good quality in the early stages of production.
The scale factor $k$ is introduced to account for small differences in environment noise levels between measurements.
Typically, $k$ differs from unity by less than one per cent.
The statistical uncertainty $\sigma$ was evaluated as the standard deviation of a flat sub-range of the spectrum values around the peak and corresponds to \SI{0.45}{dBm/\hertz}.

The experimental \gof\ distribution is close to a reduced $\chi^2$ distribution with $N-1$ degrees of freedom, with a distinct peak around 1.
A higher value of this parameter is often a hint of defects in the \tile\ electronics, such as the detachment of one or more of the branches in the \sipm\ bias network, typically resulting from faulty or missing wire bonds.

As a cross-check on the reference shape at room temperature, the \tile\ noise spectrum was simulated with the TINA-TI software \cite{tinati} by reproducing the \tile\ circuitry. The simulated output spectrum is compared to the acquired data. The input parameters of the simulation are the \tile\ circuit elements discussed in~\autoref{sec:tiles} and the parameters included in the \sipm\ equivalent electrical model, namely a collection of microcells, each of them modelled as an avalanche photodiode operated in Geiger mode and passively quenched through a series resistor. The model used here is derived from~\cite{Corsi2006, Corsi2007}. 

The simulation reproduces the shape of the measured spectrum within \(2\) \perc, as shown in~\autoref{fig:noise-simulation} in the range of interest for the noise spectrum quality assurance.

The simulation cannot be performed at cryogenic temperature, as the software does not include the cryogenic characterisation of the TIA down to \cold. However, a simplified calculation can be performed to obtain the expected maximum of the spectrum in liquid nitrogen at \SI{1}{\mega \hertz}, as detailed in~\cite{DIncecco2018}. This can be achieved by summing the following contributions: (i) the voltage noise, which is the sum of the intrinsic voltage noise of the TIA and the Johnson-Nyquist voltage noise of the resistive components of the circuit ($\sim$~\SI{-171}{dBm}), (ii) the asymptotic noise gain ($\sim$~\SI{50}{dBm})  and (iii) the amplification factor given by the readout boards ($\sim$~\SI{20}{dBm}). Thus, the expected maximum in liquid nitrogen is \SI{-101}{dBm/\hertz}, in excellent agreement with the measurements, as shown in \autoref{fig:noise-spectrum}.

The quality parameters related to the noise measurements are the DC offset, the noise RMS, the integral of the power spectrum and the \gof.
\autoref{tab:noise-params} lists the quality requirements on these parameters. The spectrum requirements were checked by testing a \tile\ where one of the branches in the \sipm\ bias network was removed on purpose.

\subsubsection{Pulse quality control}
\label{sec:pulse}


\begin{table*}[tpb]
    \centering
    \begin{tabular}{c|c|c| c|c}
         Parameter & \tile\ Bias Voltage (V)& \sipm\ Overvoltage (VoV) & Minimum & Maximum  \\
         \hline
         \hline
         SPE mean (amplitude) &136& 7  & \SI{41}{\milli \volt} &  \SI{50}{\milli \volt} \\
         \hline   
         SPE mean (amplitude) &144&9  &  \SI{53}{\milli \volt}& \SI{65}{\milli \volt}\\
         \hline 
         SPE mean (charge) &136&7  &  \SI{16}{\nano \volt \second}& \SI{28}{\nano \volt \second}\\
         \hline  
         SPE mean (charge) &144&9  &  \SI{20}{\nano \volt \second}& \SI{34}{\nano \volt \second}\\
         \hline 
         SPE resolution (amplitude)&136&7  &  -& \SI{0.11}{}\\
         \hline
         SPE resolution (amplitude)&144& 9  & -& \SI{0.09}{}\\
         \hline
         SZR (amplitude) &136& 7  &  13.5& -\\
         \hline
         SZR (amplitude) &144& 9  &18.5& -\\
         \hline
         \taurise & 136&7&\SI{70}{\nano \second}&\SI{100}{\nano \second}\\
         \hline
         \taurise & 144&9&\SI{70}{\nano \second}&\SI{100}{\nano \second}\\
         \hline
         \taufall &136&7&\SI{250}{\nano \second}&\SI{390}{\nano \second}\\
         \hline
         \taufall &144&9&\SI{270}{\nano \second}&\SI{390}{\nano \second}\\
         \\
    \end{tabular}
    \caption{Quality parameters for the pulse assessment. The overvoltage refers to a nominal \sipm\ with \vbdsipm\ breakdown voltage. Single PhotoElectron (SPE) mean, resolution, and Single-to-Zero Ratio (SZR) are estimated from the corresponding finger plot. Refer to \autoref{sec:pulse} for the detailed definition and computation.}
    \label{tab:pulses-params}
\end{table*}

\begin{figure*}[tbp]
    \centering
    \includegraphics[width=\linewidth]{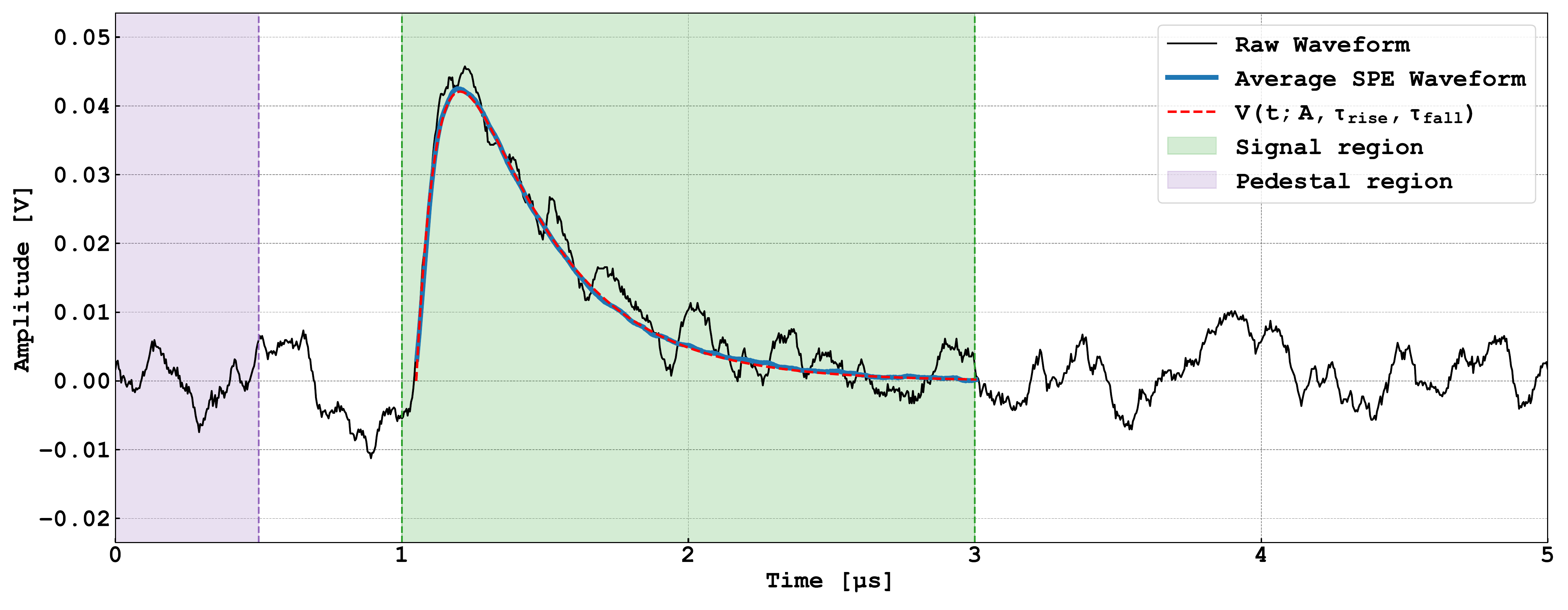}
    \caption{Example of a raw pulse waveform in the Single PhotoElectron (SPE) regime, acquired at \SI{7}{\vov}. The amplitude and charge (integral) of the pulse are evaluated in the \emph{Signal region} (shaded in green). The region where the pedestal is computed (shaded in violet) is in the first \SI{0.5}{\micro \second}. The blue line is the average SPE waveform. The red dashed line is the best fit to the average SPE waveform with the model in~\autoref{eq:tau_model}. More details are given in the text.}
    \label{fig:waveform}
\end{figure*}

\begin{figure}[tbp]
    \centering
    \includegraphics[width=\linewidth]{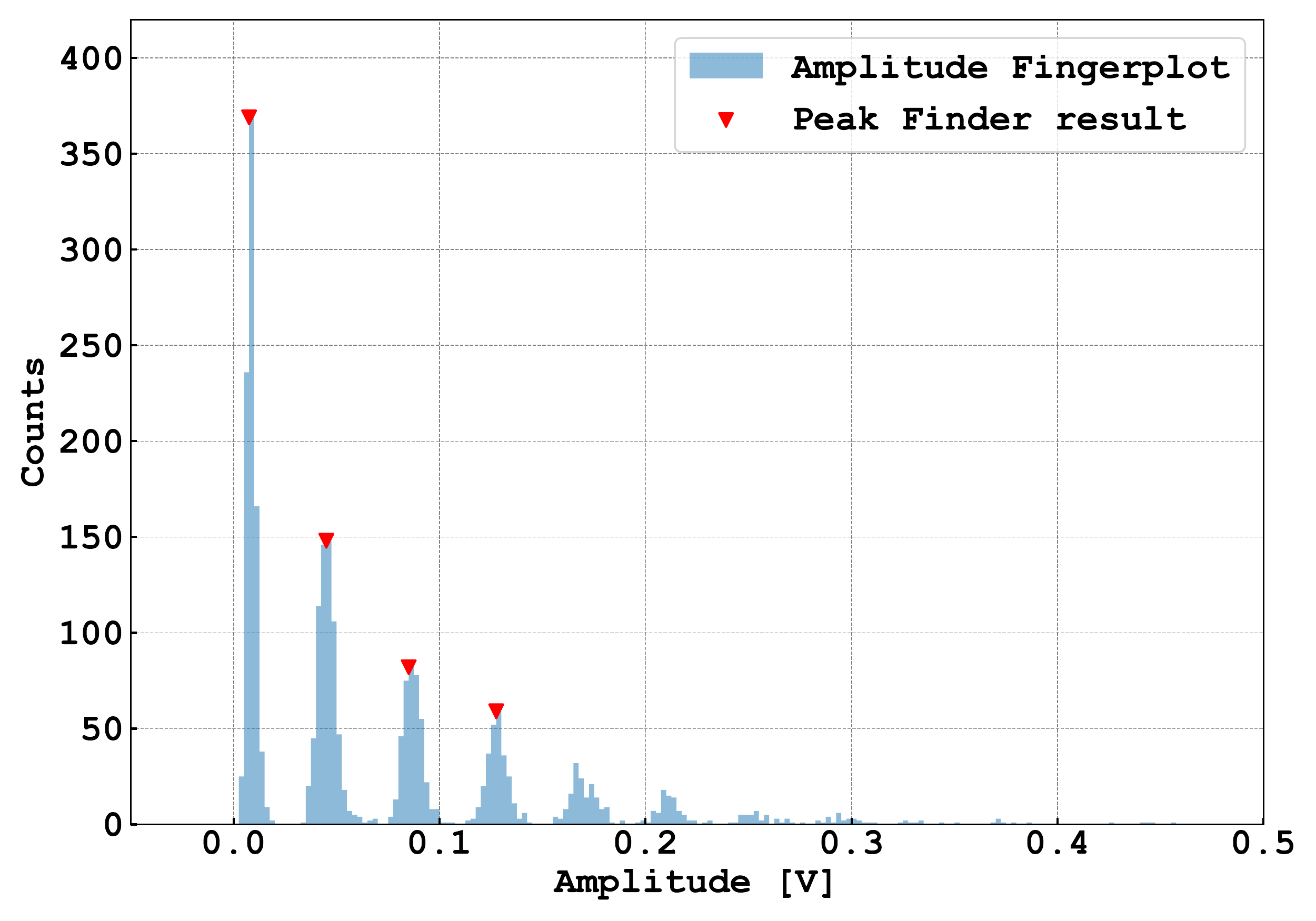}
    \includegraphics[width=\linewidth]{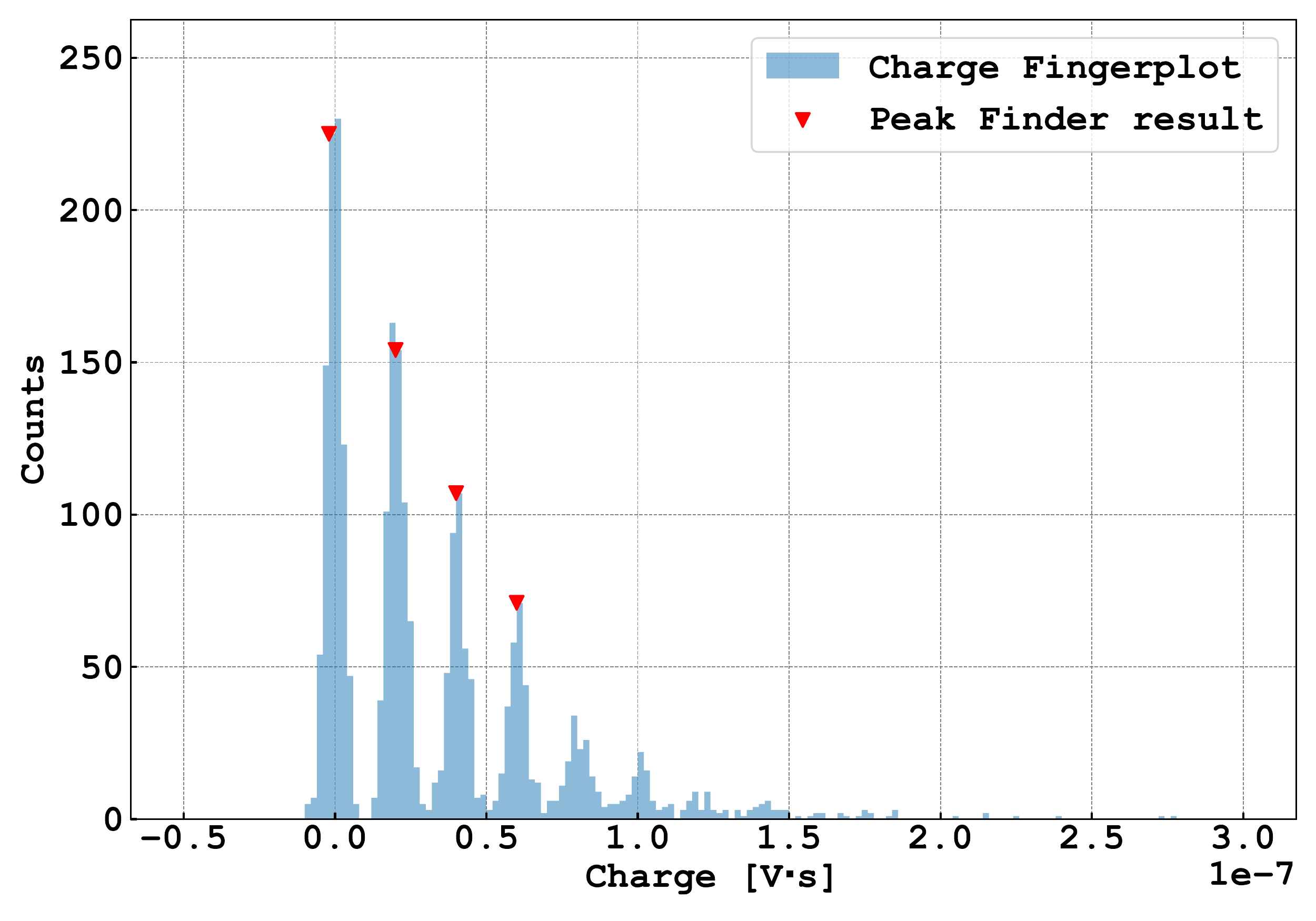}
    \caption{Example of an amplitude and charge fingerplot acquired at \SI{7}{\vov}. The peak finder algorithm correctly determines the abscissa of the peaks.}
    \label{fig:finger-plot-amplitude}
\end{figure}

\begin{figure}[tbp]
    \centering
    \includegraphics[width=\linewidth]{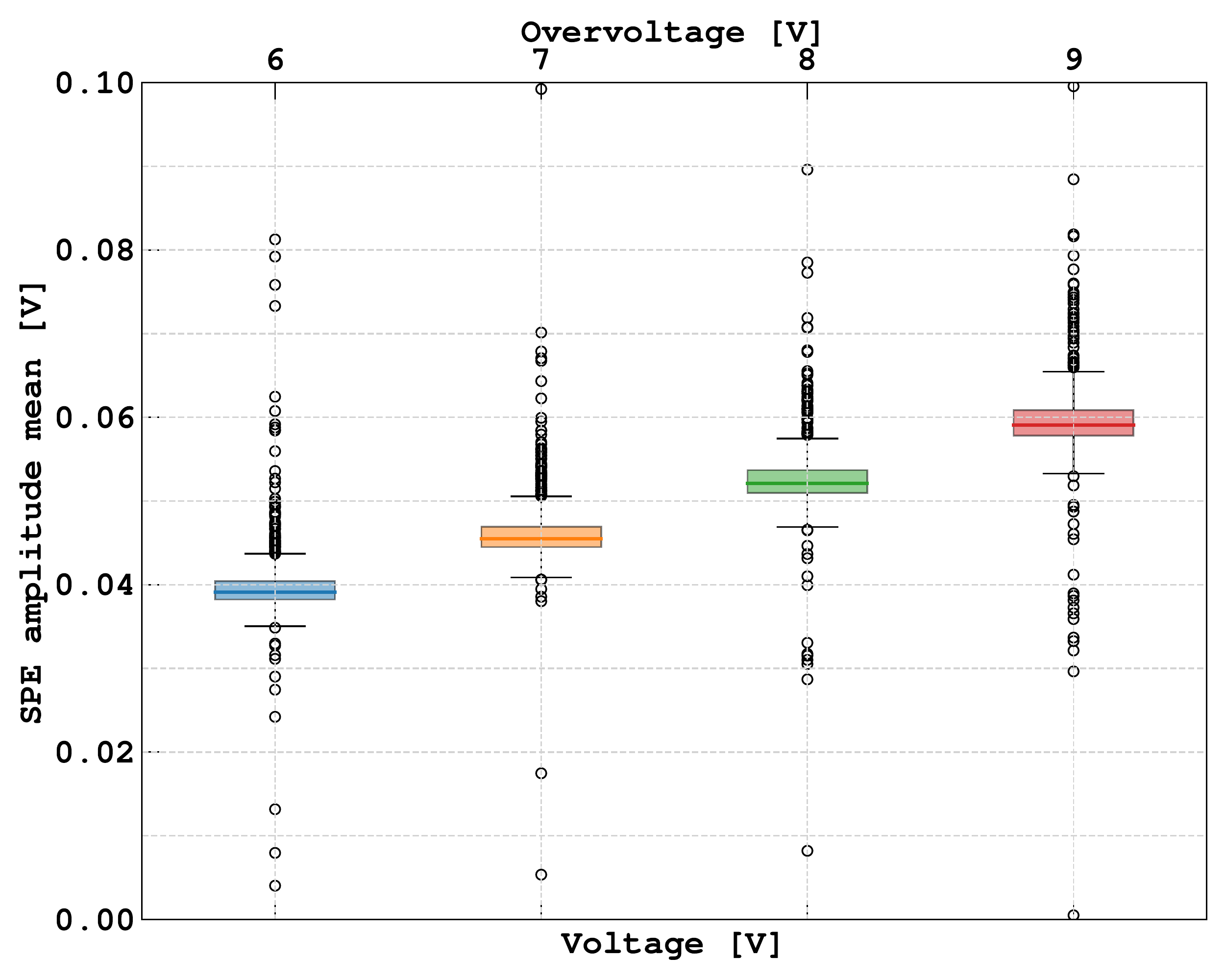}
    \includegraphics[width=\linewidth]{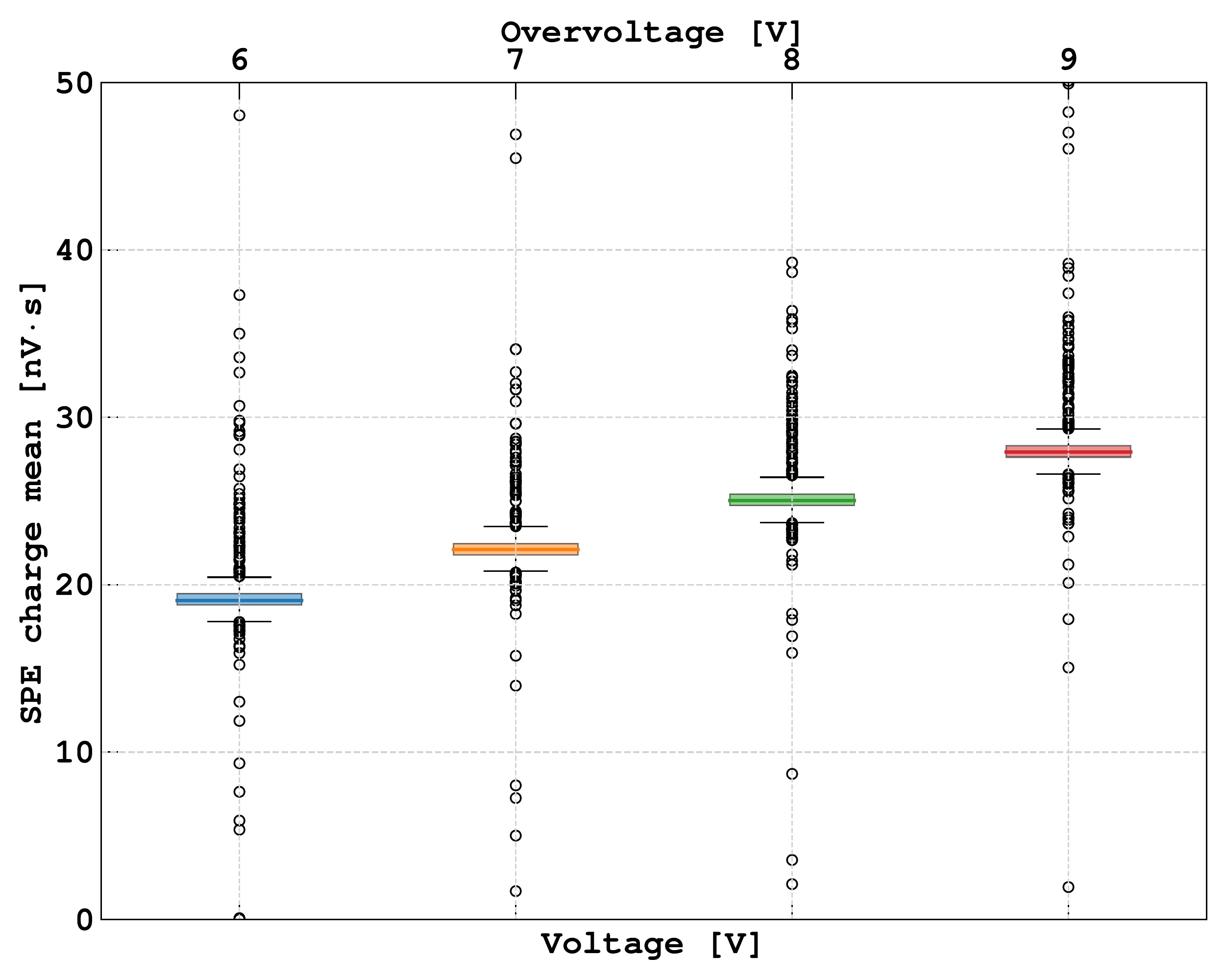}
    \caption{Single PhotoElectron (SPE) amplitude boxplot (Top) and charge (Bottom) as a function of the bias voltage and overvoltage. Boxes represent data between the first and third quartiles. Dots represent outliers more than 1.5 InterQuartile Range (IQR) away from the top or bottom of each box.}
    \label{fig:amp-charge-vs-vbias}
\end{figure}

\begin{figure}[tbp]
    \centering
    \includegraphics[width=\linewidth]{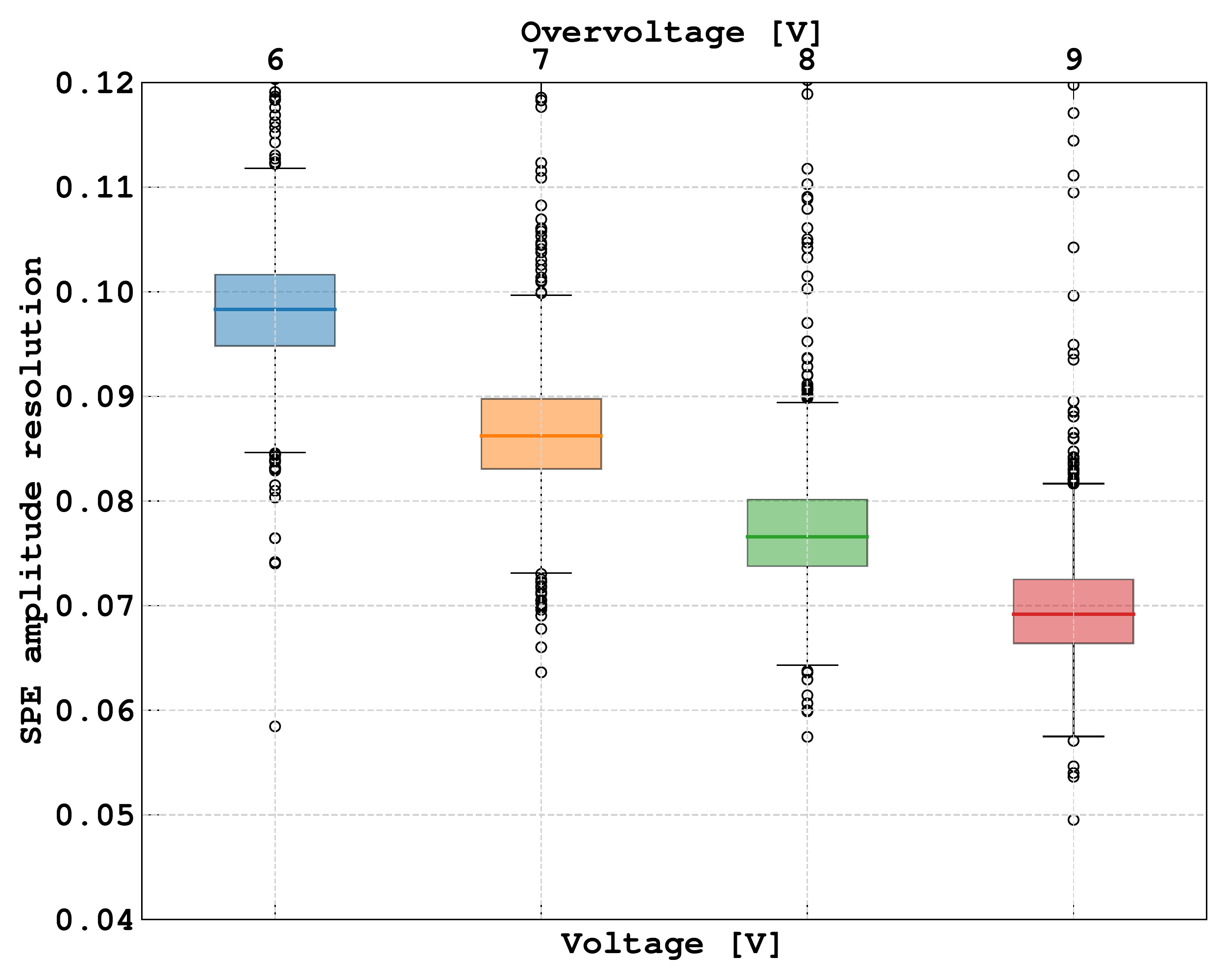}
    \includegraphics[width=\linewidth]{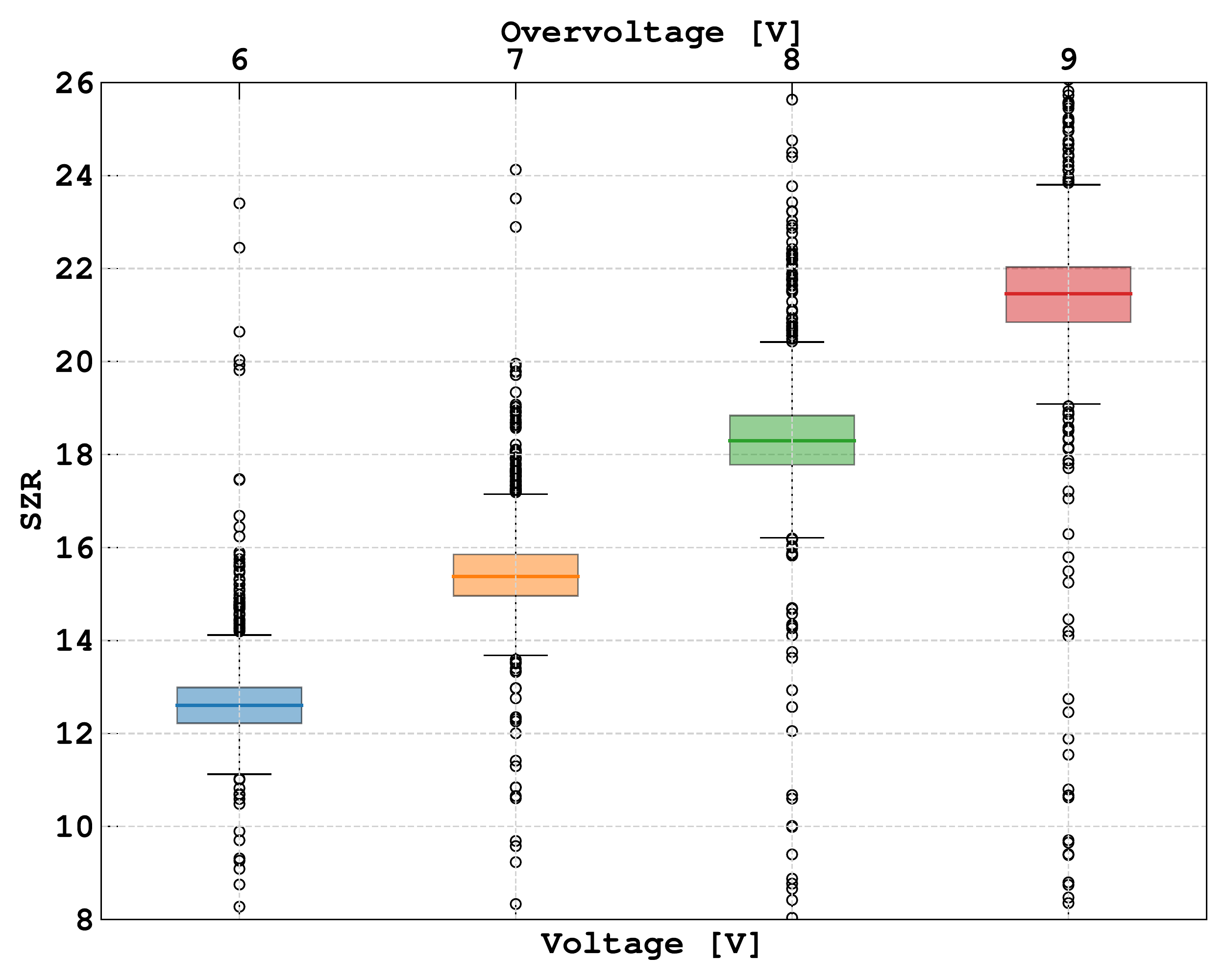}

    \caption{Single PhotoElectron (SPE) resolution boxplot (Top) and Single-to-Zero Ratio (SZR) (Bottom) as a function of the bias voltage and overvoltage. Boxes represent data between the first and third quartiles. Dots represent outliers more than 1.5 InterQuartile Range (IQR) away from the top or bottom of each box.}
    \label{fig:szr-res-vs-vbias}
\end{figure}


The characterisation of the \tile\ pulse output signals is performed to asses the quality of the \tiles\ when exposed to a pulsed light source. In this procedure, the \tile\ is submerged in liquid nitrogen and illuminated by a \tileLaser\ picosecond laser of \SI{400}{\nano \meter} wavelength, operating at \SI{500}{\hertz}. The laser's intensity is attenuated to ensure that the \tile\ works within the few-photon range.

The \tile\ output signal is acquired using the oscilloscope, which is triggered by the sync signal from the Hamamatsu laser module. 
The oscilloscope is configured to acquire waveforms that are \SI{5}{\micro \second} in duration, sampled at \SI{250}{\mega Sample/\second}. The signal starts approximately \SI{1}{\micro \second} after the waveform begins.
The vertical range of the oscilloscope is adjusted to \SI{0.8}{\volt} to account for potential DC offset shifts and large signal pulses.
 
The acquisition is repeated with the \tile\ supplied at \(4\) different bias voltages: \SI{132}{\volt}, \SI{136}{\volt}, \SI{140}{\volt}, and \SI{144}{\volt}, which correspond to about \(6\), \(7\), \(8\), and \(9\) Volt over voltage (\vov) for \sipms\ with nominal breakdown voltage of \vbdsipm.
For each bias voltage, \(10000\) oscilloscope triggers are acquired (the number of triggers was \(2500\) before the fifth month of production). An example of a raw SPE waveform collected at \SI{7}{\vov} is displayed in~\autoref{fig:waveform}.
The collected raw waveforms are saved in binary format and securely stored on a \ds\ server at LNGS. This method facilitates reanalysis of all the \tile\ pulse data, should there be any modification in the analysis procedure.

Before acquiring all triggers over the \(4\) different bias voltages, the LabVIEW application conducts an initial data acquisition at the highest overvoltage and displays a histogram of the amplitude distribution to the user. These histograms, commonly referred to as \emph{finger plots}, typically display \(4\) to \(5\) peaks resembling fingers. It is mandatory for the user to ensure the initial two peaks are discernible and that the laser intensity is set so that the peak with the most entries falls in either the first or second position. The user can continue to adjust the laser intensity and view additional preliminary acquisitions until this condition is satisfied.

The data analysis starts by estimating the pedestal, signal amplitude, and signal charge for each waveform gathered.
The pedestal is determined by averaging the waveform values from its start to \SI{0.5}{\micro \second}.
This calculated pedestal is then subtracted from the waveform.
The signal amplitude is identified as the highest value of the waveform after pedestal subtraction within the \SI{1}{\micro \second} to \SI{3}{\micro \second} interval.
The signal charge is obtained by integrating the pedestal-subtracted waveform over this same interval, employing the trapezoidal method for integration.

Upon processing each of the \(10000\) waveforms per bias voltage, we populate two histograms: one for the amplitude distribution and another for the charge distribution.
An example of these distributions is provided in \autoref{fig:finger-plot-amplitude}.
The amplitude histograms are divided into \(200\) bins ranging from \SI{0}{\volt} to \SI{0.5}{\volt}, while the charge histograms consist of 150 bins spanning from \SI{-20}{\nano \volt \second} to \SI{280}{\nano \volt \second}.

Once the finger plots are complete, a peak detection algorithm identifies the peak positions in the histograms. Two types of peak detection algorithms were compared and adjusted to yield reliable results. The first is the Peak Detector from LabVIEW's Signal Processing library, utilising an algorithm that fits a quadratic polynomial to sequential groups of data points. The second employs the TSpectrum module from the ROOT library, which instead fits normal distributions. Post-tuning, both algorithms pinpoint the abscissa of the peaks with discrepancies under a few per cent, provided there is sufficient data (several hundred events per peak).

Peak positions are solely employed to delineate the zero PhotoElectron (0-PE) and Single PhotoElectron (SPE) regions within the finger plots. The demarcation between the 0-PE and SPE regions is established by calculating the midpoint between the abscissae of the first and second peaks. The 0-PE region extends from the beginning of the histogram to this demarcation point. The SPE region, equal in width, begins at the end of the 0-PE region and encompasses the SPE peak.

After pinpointing the SPE region in the finger plot, its entire content undergoes fitting with a Gaussian distribution. The term \emph{SPE amplitude} refers to the mean obtained from fitting the SPE region in the amplitude finger plot. Similarly, in the charge finger plot, the \emph{SPE charge} is determined using the same method. \autoref{fig:amp-charge-vs-vbias} illustrates the median of the distribution of SPE means as a function of bias voltage and overvoltage, highlighting the range of the distribution (boxes) and its outliers. Dots represent outliers more than 1.5 InterQuartile Range (IQR) away from the top or bottom of each box. The median of the distribution indicates an approximately linear upward trend.

The \emph{SPE resolution} is defined as the ratio of the fitted $\sigma$ to the mean within the SPE region. By fitting across the entire SPE region, we can include both longer tails and false peaks as increases in the Gaussian's $\sigma$. This allows us to identify \tiles\ with less-than-ideal peaks. \autoref{fig:szr-res-vs-vbias} (Top) presents how the SPE resolution (derived from the amplitude finger plot) varies with bias voltage and overvoltage, by displaying the median of the SPE resolution distribution, its range (boxes) and its outliers. The resolution tends to improve as the overvoltage is raised.

A Gaussian fit is also used to model the content of the whole 0-PE region. The mean and the $\sigma$ of the Gaussian in this region are used to define the \emph{Single-to-Zero ratio} (SZR):
\begin{equation}
\text{SZR}=\frac{\mu_1-\mu_0}{\sigma_0} \, ,
\label{eq:snr_model}
\end{equation}
where $\mu_1$ ($\mu_0$) is the mean of the SPE (0-PE) peak and $\sigma_0$ is the 0-PE standard deviation.
The SZR parameter quantifies the single photon detectability over baseline fluctuations, an extremely important feature for the performance of the \ds\ experiment\footnote{
The SZR is akin to a Signal-to-Noise ratio (SNR) that is tuned for reliability and automation during the quality assessment in NOA.
However, since slightly different versions of SNRs have been used in \ds\ publications~\cite{Aalseth2018,GlobalArgon2021}, we have chosen to give a specific name to the one used in this context.}.
\autoref{fig:szr-res-vs-vbias} (Bottom) illustrates the SZR distribution in the form of a boxplot (from the amplitude finger plot) as a function of the bias voltage and overvoltage. Dots represent outliers more than 1.5 IQR away from the top or bottom of each box. As expected, the SZR increases with the overvoltage.

A final quality check assesses the conformity of the pulse shape of the SPE average waveform.
The latter is computed by averaging all waveforms whose charge is in the interval $\pm 2 \, \sigma$ centred on the SPE charge mean.
The SPE average waveform is modelled with the function
\begin{equation}
V(t; A, \tau_\text{rise}, \tau_\text{fall}) = A\left(e^{-t/\tau_\text{fall}} - e^{-t/\tau_\text{rise}}\right) \, .
\label{eq:tau_model}
\end{equation}
The values $A$, \taurise, and \taufall\ are obtained through a $\chi^2$ fit, where the uncertainties are set to the standard deviation of the average waveform in the pedestal region.
The parameters are initialised at their nominal values (\taurise\ at \SI{80}{\nano \second}, \taufall\ at \SI{350}{\nano \second}).
The fit is performed in an interval of \SI{2}{\micro \second}, with the start time (corresponding to $t = 0$ in the model in~\autoref{eq:tau_model}) assigned through a constant fraction discriminator set at \(5\)\perc\ of the average SPE waveform peak.
\autoref{fig:waveform} shows an example of an average SPE waveform along with the fitting function $V(t)$.

\autoref{tab:pulses-params} lists the quality parameters for the pulse assessment. The quality control allows us to discard \tiles\ with SPE means at unexpected values, poor resolution, or low SZR.
Very problematic \tiles\ with unusual finger plots, strictly not usable in \ds, typically fail all of the pulse checks.
The requirements have been fine-tuned to exclude the outliers in the distributions and have shown to provide PDUs with good performance\footnote{A paper containing details of the PDU performance in liquid nitrogen is in preparation.}. We perform the quality control for the measurements taken at two of the four bias voltages used, corresponding to \SI{7}{\volt} and \SI{9}{\volt} over voltage, as the other two have been shown to be redundant.

Starting in the fifth month of production (April 2025), the quality control pipeline was updated with a measurement of the Pulse Count Rate (PCR) using the waveforms acquired during the pulse counting assessment. 
The PCR is defined as the number of PE pulses in the pre-trigger region over the total acquisition time. Tails of large pulses from the preceding time window are excluded from the computation. The acquisition time is defined as the number of triggers (\(10000\)) multiplied by the width of the pre-trigger region (\SI{1}{\micro \second}). The result is expressed in \SI{}{\hertz/\milli \meter^2}. The sensitivity of this method is limited to 1 observed count, which corresponds to about \SI{0.05}{\hertz/\milli\meter^2}.

 An additional setup with dark \tile\ holders is dedicated to the measurement of the Dark Count Rate (DCR) and will be described in a future paper.

\section{Results}
\label{sec:results}

The \qaqc\ procedures implemented at NOA have enabled a systematic evaluation of the \tiles' performance, ensuring compliance with the requirements for integration into the PDUs of the \ds\ \tpc. 

A pre-production phase was successfully conducted from January to October 2024 to asses process capability, throughput, and failure rates. During this phase, \tiles\ were assembled using a first batch of wafers meant to confirm the quality of the \sipm, allowing for the refinement of quality assurance procedures until they reached the standards described in this paper.

Based on the production yield achieved to date, we expect that the complete production of the \tiles\ for the \tpc\ of \ds\ will be finalised in less than \(30\) months, with a throughput of \tilesthroughput. This production process encompasses the cryogenic characterisation of the total \nwafers\ silicon wafers (equivalent to \ndies\ individual dice), testing of \npcbtotnoa\ \pcbs, the packaging and qualification of \ntiletotnoa\ \tiles\ (including spares), and the assembly of \npduds\ functional TPC PDUs.

The rest of this section highlights some of the results of the \pcb\ and  \tile\ production and quality control, focusing on uniformity, and production yield achieved during mass testing.

\subsection{\pcb\ assessment performance and yield}
\label{subsec:pcb_perf_yield}

\begin{figure}[tpb]
    \centering
    \includegraphics[width=\linewidth]{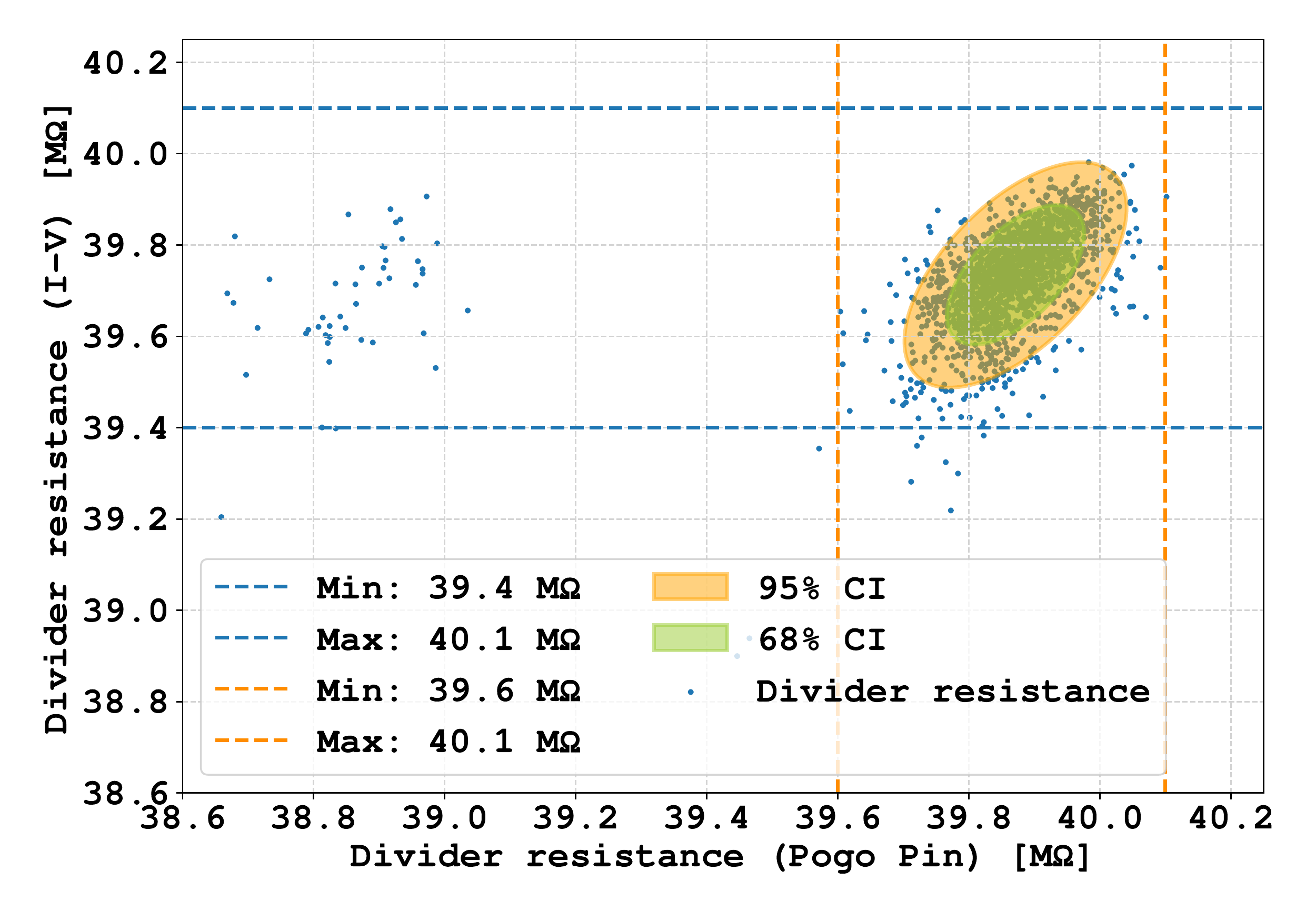}
    \caption{Measurements of the \tile’s divider resistance \rd\ with the \pcb\ test setup. The vertical axis refers to measurements derived from the I-V curve. The horizontal axis refers to the measurement performed with the pogo pin at \SI{40}{\volt} bias voltage. The quality requirements are shown as dashed lines. The elliptical contours are derived from the data that passed the quality requirements, assuming Gaussian confidence intervals (CI). The origin of the set of points in the left part of the plot is currently under investigation.}
    \label{fig:pcb-rd}
\end{figure}


The \pcb\ assessment demonstrated percent-level uniformity in the discrete components of the tested sample, confirming the stability of the component population process at EltHUB.

\ntestedpcbs\ \tile\ \pcbs\ were tested in \(2024-2025\) including \npreproductionpcbs\ from the pre-production batch and \nproductionpcbs\ from the production phase. About 10\perc\ failed the test, primarily due to poor soldering of a resistor or surface contamination. 

\pcbs\ that failed the test due to improper mounting of electronic components were reworked and retested, with only \(4\) units failing the second test, resulting in a final yield of 99.1\perc\ after rework.

\autoref{fig:pcb-rd} compares the two independent estimates of the divider resistance: the measurement using the pogo pin at \SI{40}{\volt} bias voltage (horizontal axis) and the value retrieved from the slope of the \iv\ curve (vertical axis). 
The quality requirements are also shown. Elliptical contours are derived from the data that passed the quality requirements, assuming Gaussian confidence intervals (CI).
The strong correlation between these methods highlights the reliability of the quality assurance.
The \pcb\ quality assurance significantly improves the \tile\ test yield, as it restricts the \tile\ test failures mainly to \sipm-related issues.

\subsection{\tile\ performance and yield}
\label{subsec:tile_perf_yield}

\begin{table}[tpb]
    \centering
    \begin{tabular}{c|c|c}
        \hline
        Warm test & Cold test &  \tiles\ percentage \\
        \hline
        Passed & Passed & \warmcoldpass\ \\
        Passed  & Failed  & \warmpasscoldfail\ \\
        Failed  &Passed   & \warmfailcoldpass\ \\
        Failed  & Failed & \warmcoldfail\ \\
        \hline
    \end{tabular}
    \caption{Results of the warm and cold \tile\ test with percentages associated with the nature of the failed test. More details on the failure modes are given in the text. A \tile\ must pass both the test at room temperature and in liquid nitrogen to be assembled in a PDU.}
    \label{tab:percentage-qaqc}
\end{table}

\begin{figure*}[tpb]
    \centering
    \includegraphics[width=0.98\textwidth]{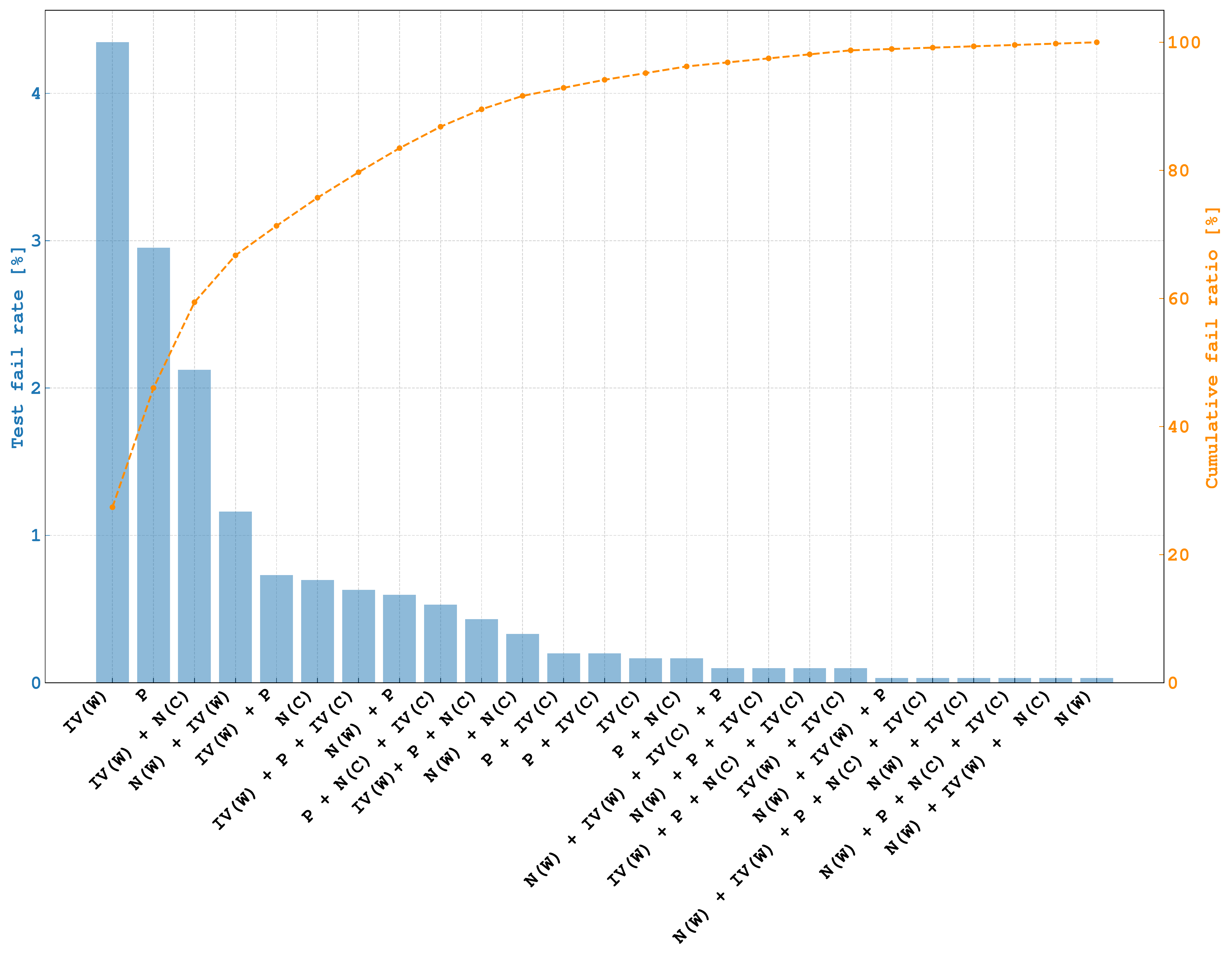}
    \vspace{0.5cm} 

     \noindent
    \begin{minipage}{0.48\textwidth}
        \centering
        \includegraphics[width=\textwidth]{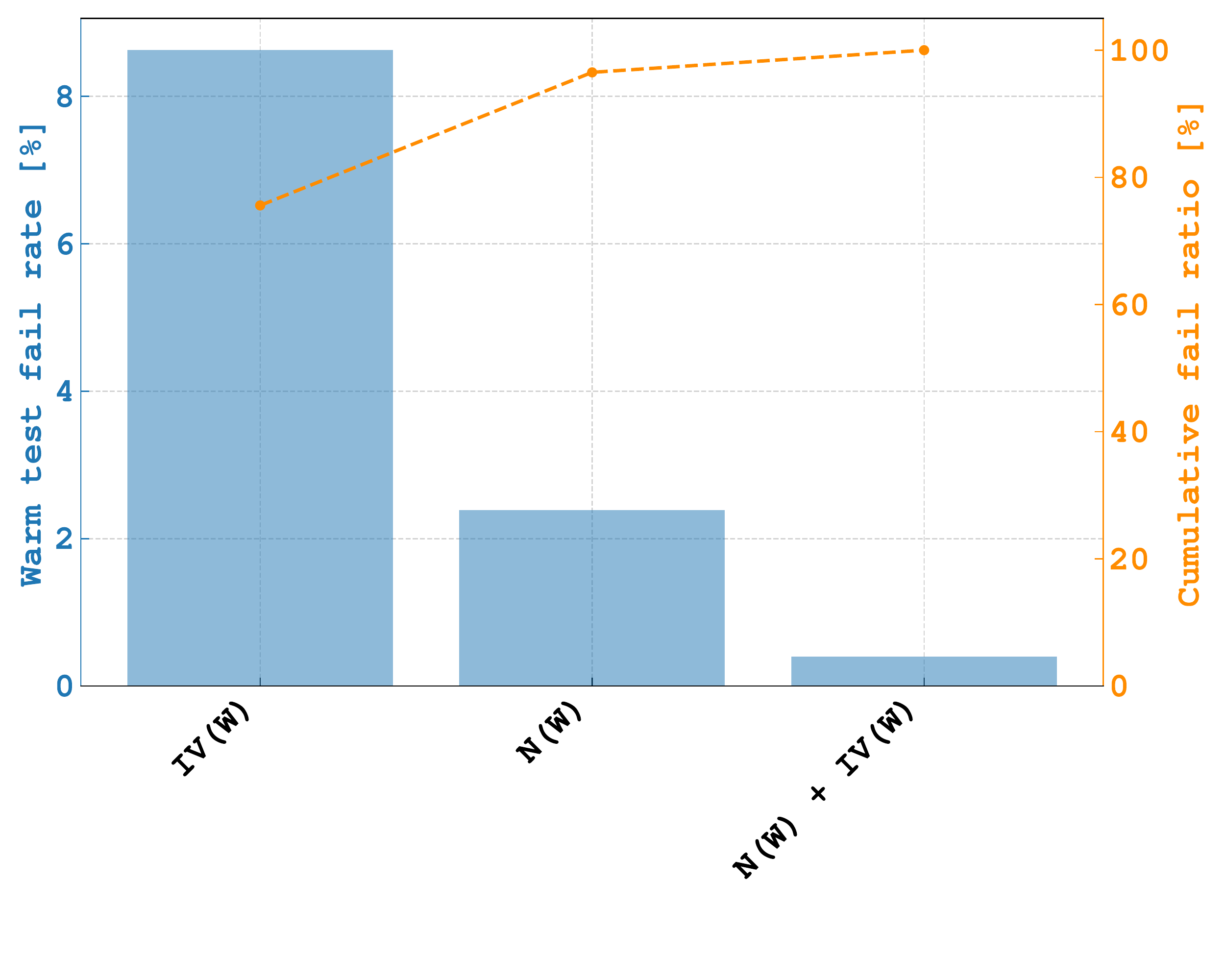}
    \end{minipage}
    \hfill
    \begin{minipage}{0.48\textwidth}
        \centering
        \includegraphics[width=\textwidth]{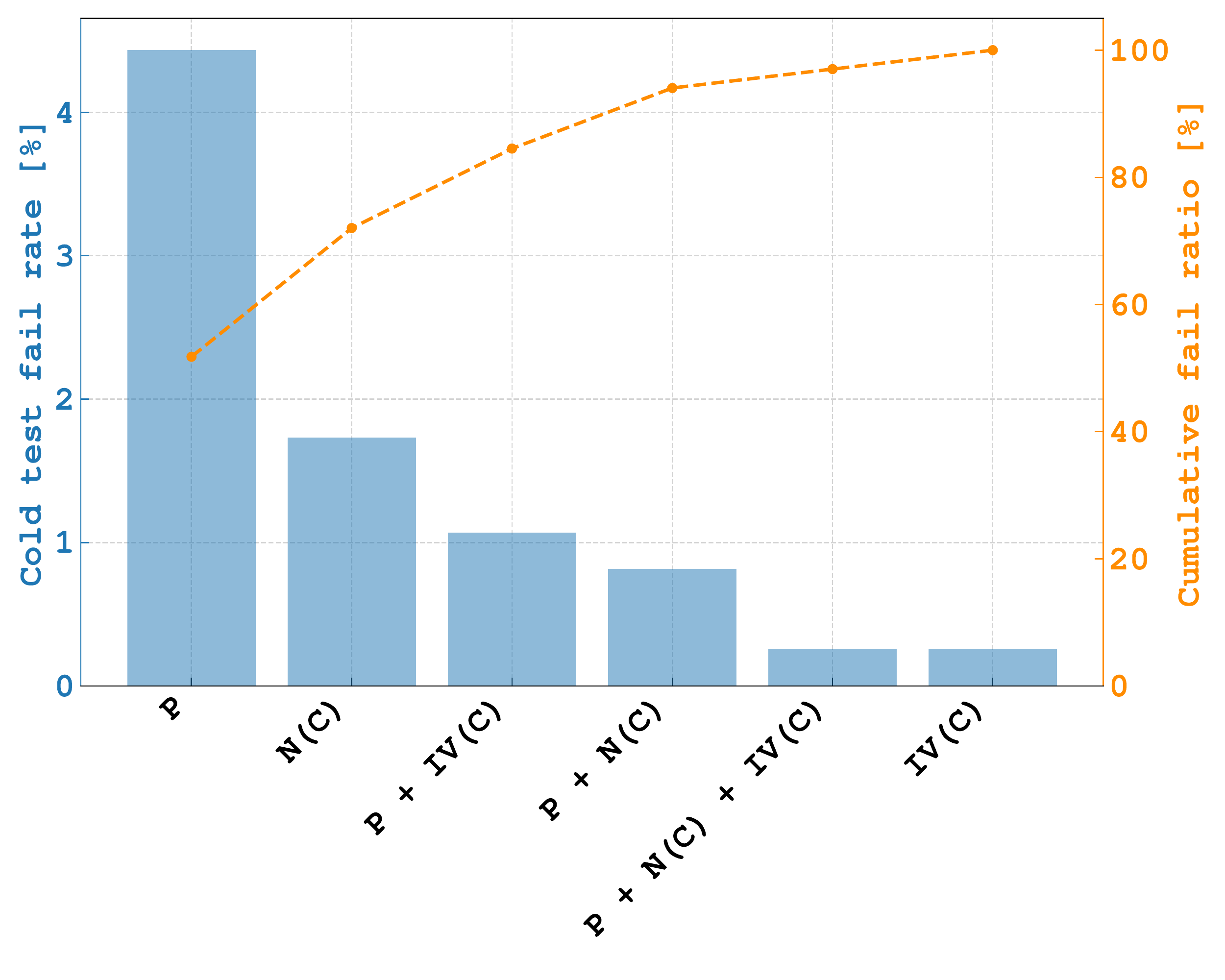}
    \end{minipage}

    \caption{Pareto charts illustrating the failure modes in the \tile\ quality assessment process. 
    The top figure presents the overall failure distribution. 
    The bottom row differentiates between tests conducted at room temperature (left) and in liquid nitrogen (right). 
    On the horizontal axes, \textbf{IV} indicates failures associated with the \iv\ curve, \textbf{N} denotes noise-related failures, and \textbf{P} pertains to issues within the pulse quality assessment. 
    The suffixes \textbf{W} and \textbf{C} inside parentheses distinguish between tests conducted at room temperature (warm) and those conducted in liquid nitrogen (cold).
    The most common failure is associated with the \iv\ curve at room temperature, accounting for \ivwarmfail\ of the overall tests, followed by pulse quality assessment in liquid nitrogen at \pulsefail, and the simultaneous failure of the \iv\ test at room temperature and the noise test in liquid nitrogen at \ivwarmnoisecoldfail. 
    Together, these three failure modes constitute approximately \(60\)\perc\ of the cumulative fail ratio.
    In the warm test, the most common failure involves the \iv\ curve, accounting for \ivwarmfailoverfailed\ of the total failures at room temperature. 
    In the cold test, the most frequent failure is the pulse quality, which represents 53\% of failures in defective \tiles\ in liquid nitrogen.}
    
    \label{fig:pareto-plots-combined}
\end{figure*}

\begin{figure}[tpb]
    \centering
\includegraphics[width=\linewidth]{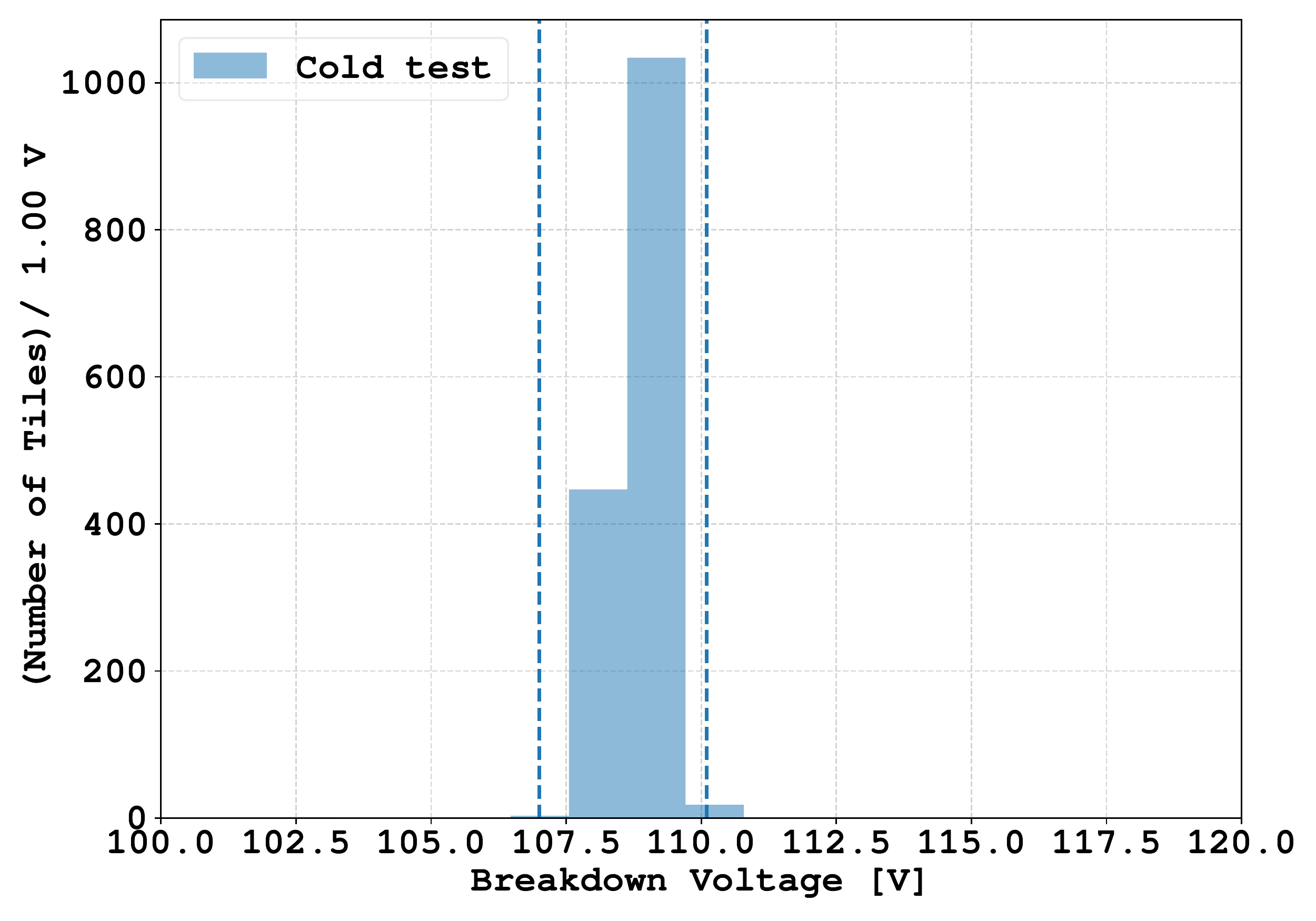}
    \caption{Distribution of the measured breakdown voltage, \vbd, in liquid nitrogen. The QC requirements are shown as dashed lines.}
    \label{fig:vbd-dist}
\end{figure}

\begin{figure}[tpb]
    \centering
\includegraphics[width=\linewidth]{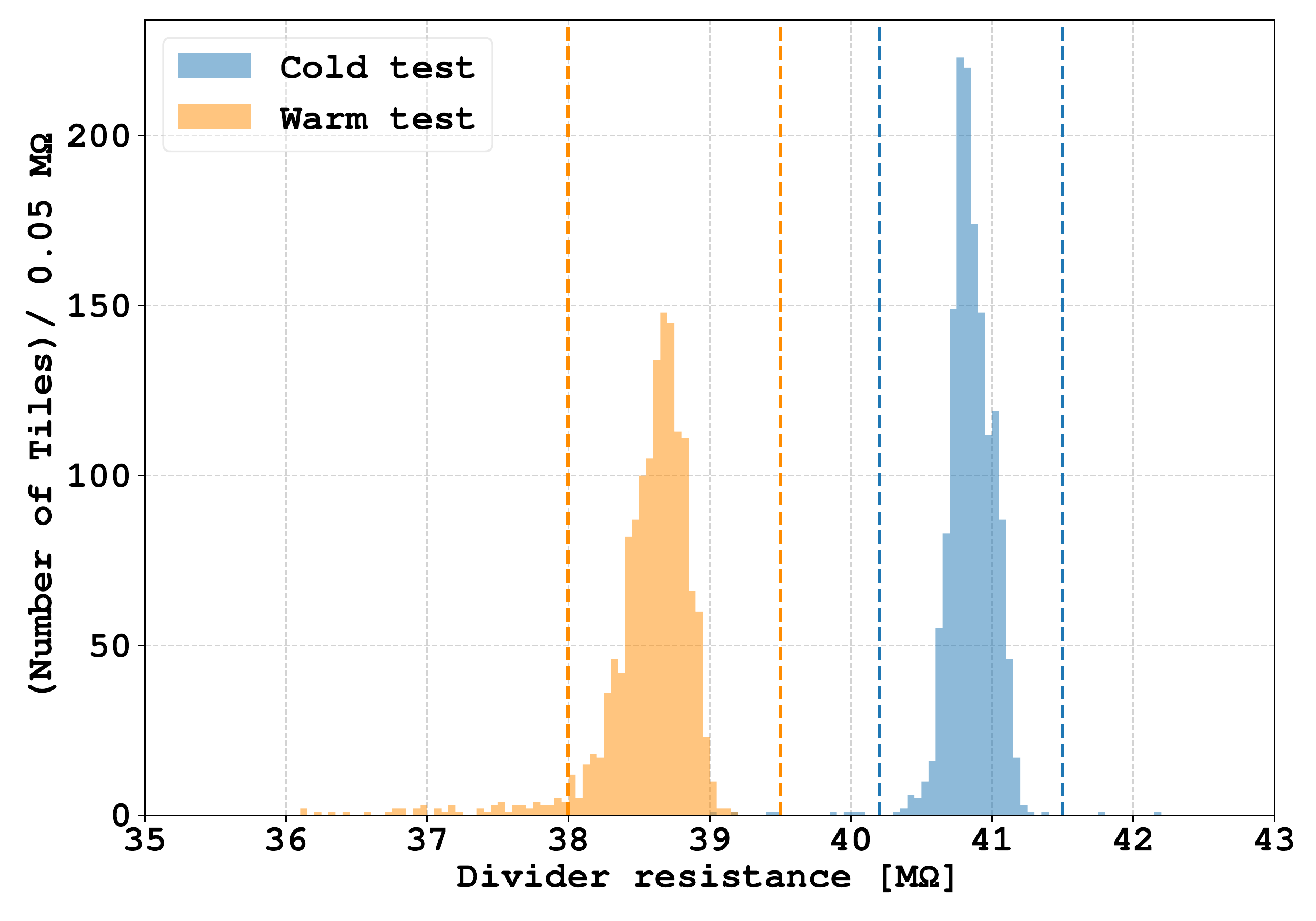}
    \caption{Distribution of the measured divider resistance, \rd, at room temperature (Orange) and in liquid nitrogen (Blue). The quality requirements are shown as dashed lines.}
    \label{fig:rd-dist}
\end{figure}

\begin{figure}[tpb]
    \centering
\includegraphics[width=\linewidth]{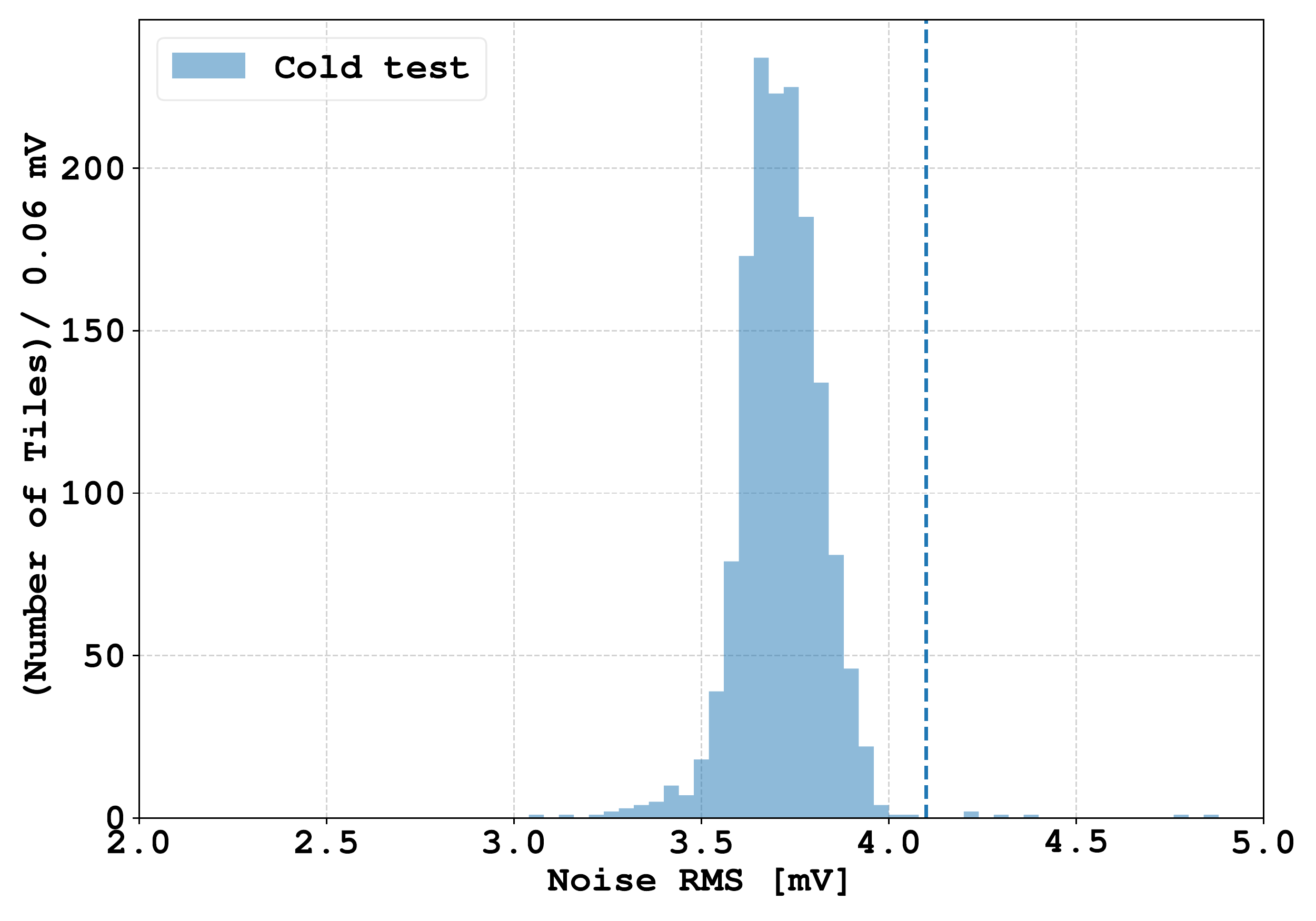}
    \caption{Distribution of the Noise RMS, as measured by the oscilloscope in liquid nitrogen with a \SI{40}{\volt} bias voltage (Blue). The quality requirement is shown as a dashed line.}
    \label{fig:noise-rms-std-dist}
\end{figure}

\begin{figure}[tpb]
    \centering
\includegraphics[width=\linewidth]{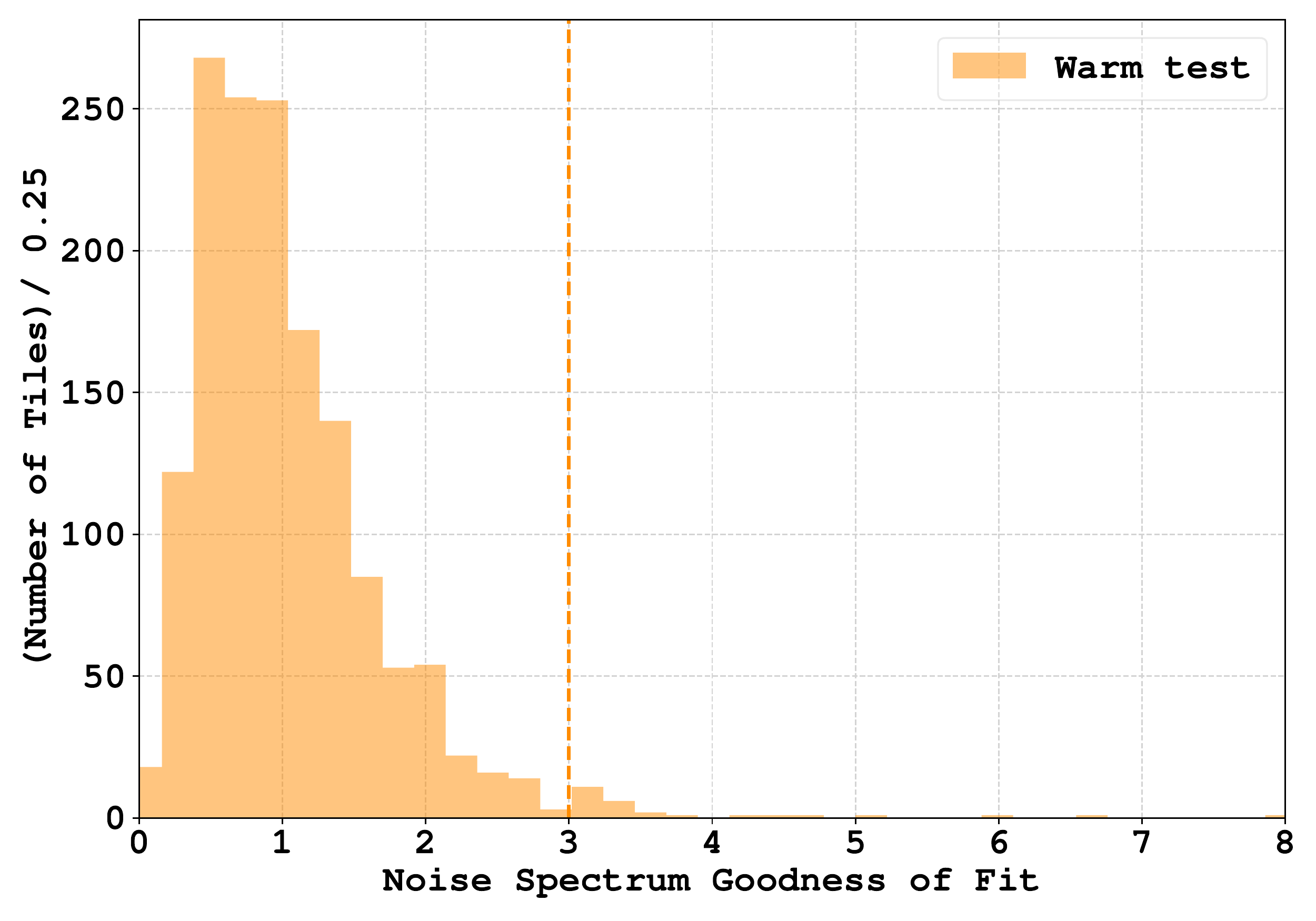}
    \caption{Distribution of the measured Goodness of Fit parameter (\gof) at room temperature. The quality requirement is shown as a dashed line.}
    \label{fig:noise-gof-dist}
\end{figure}

\begin{figure}[tpb]
    \centering
\includegraphics[width=\linewidth]{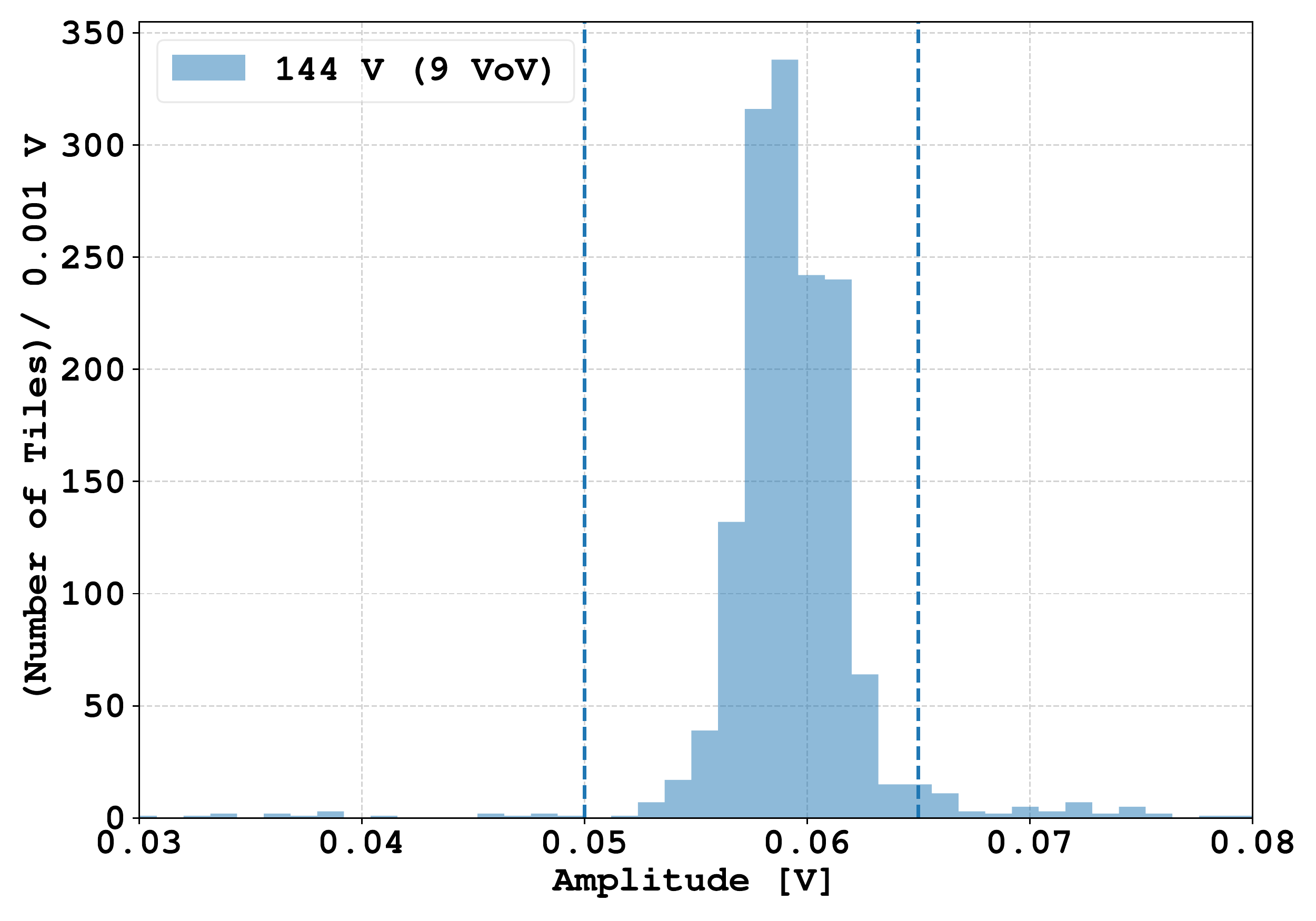}
\includegraphics[width=\linewidth]{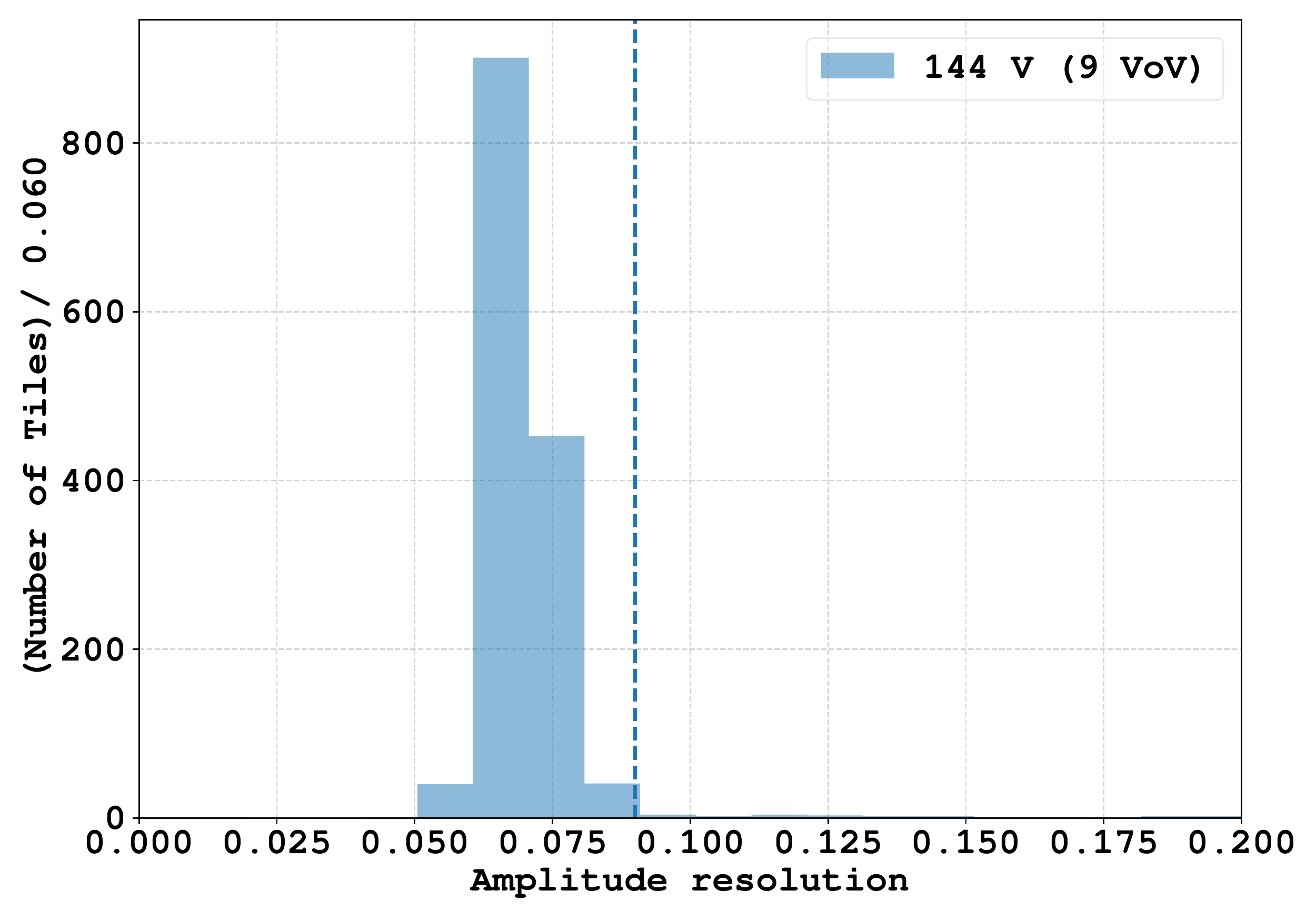}
\includegraphics[width=\linewidth]{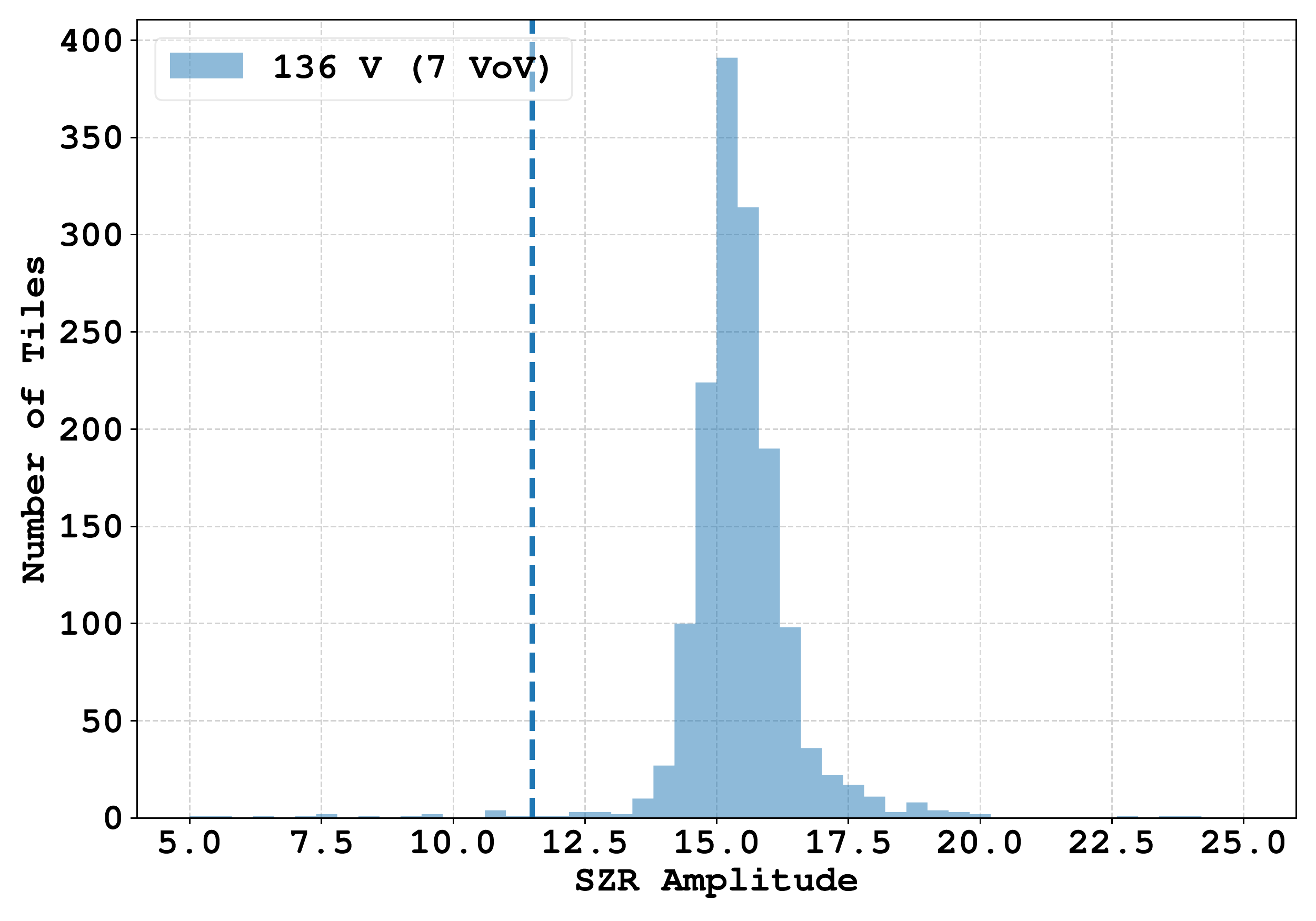}
    \caption{Distributions of the mean of the Single PhotoElectron (SPE) amplitude (first from the top) and its resolution (second from the top) at \SI{144}{\volt} (\SI{9}{\vov}) along with the Single to Zero Ratio (SZR, third from the top). The quality requirements are shown as dashed lines.}
    \label{fig:spe-amp-dist}
\end{figure}

\begin{figure*}[tpb]
\centering
\includegraphics[width=\linewidth]{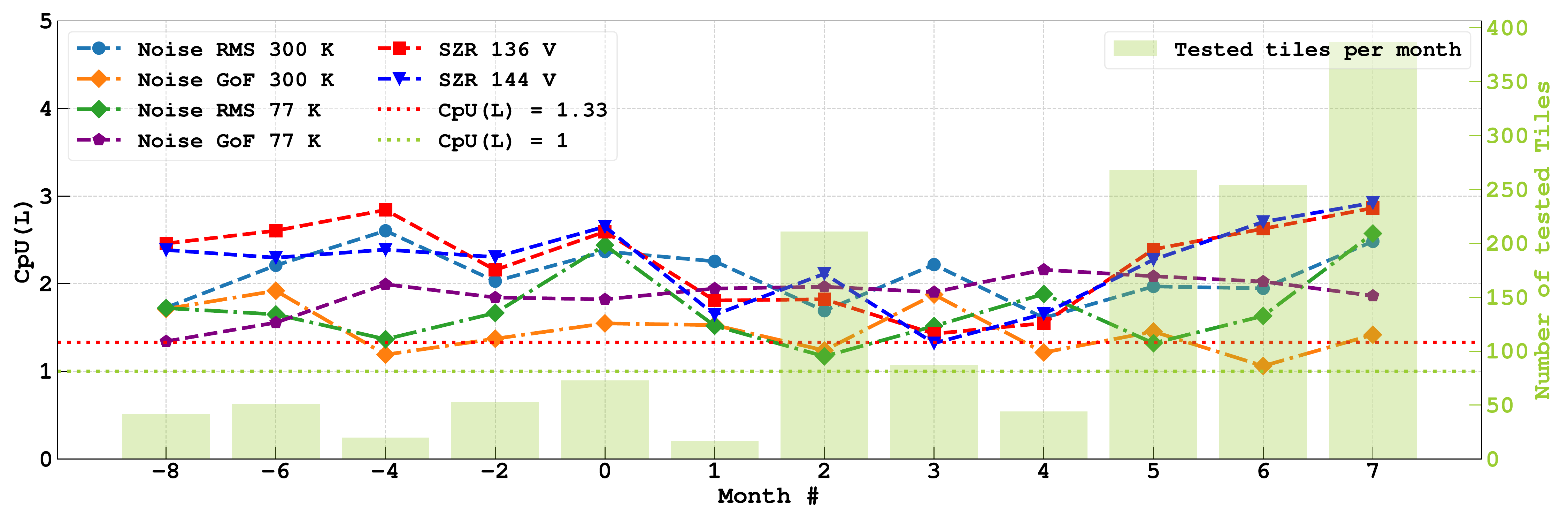}
\includegraphics[width=\linewidth]{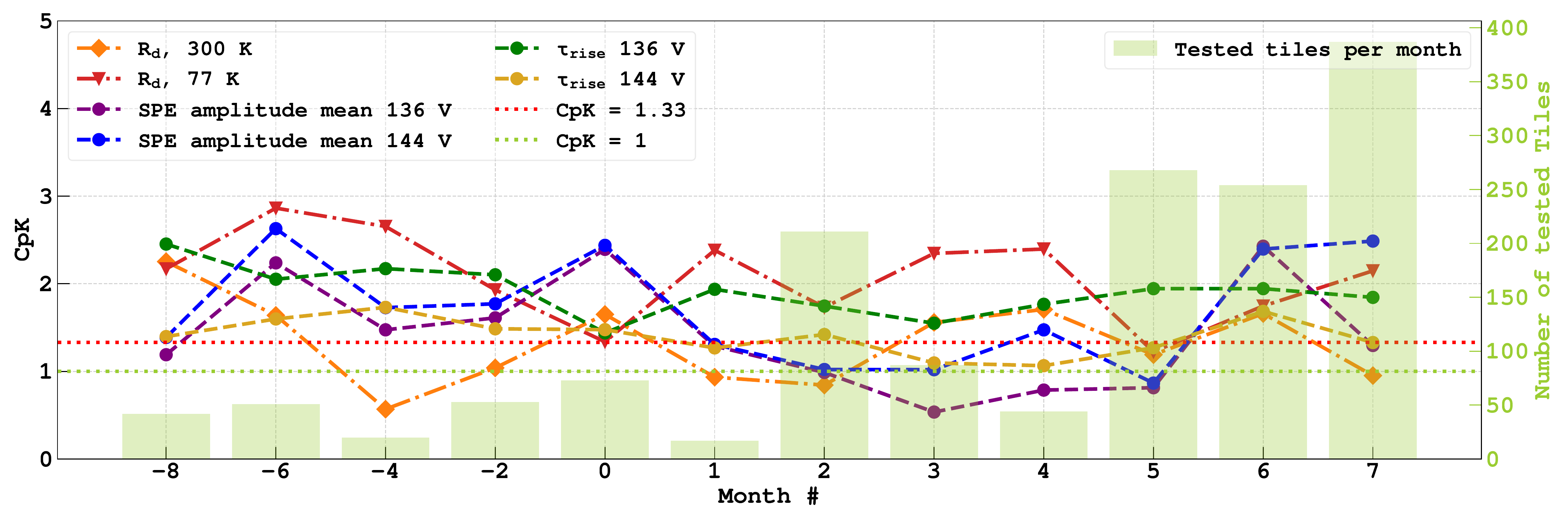}
\caption{The top figure shows the monthly CpU(L) of the \tiles\ noise RMS, Goodness of Fit, and SZR at room temperature and \cold, while the bottom figure displays the monthly CpK of the \tiles\ breakdown voltage \vbd, divider resistance \rd\ at room temperature and \cold\, the Single PhotoElectron (SPE) mean in amplitude and \taurise\ at \SI{136}{\volt} and \SI{144}{\volt}. Pre-production data is labelled with negative numbers from the beginning of the production and grouped in pairs.
The red horizontal dashed line at \(1.33\) is the conventional industry standard, indicating a \(4\sigma_X\) difference between the mean of the measured
quantity and its upper specification limit, while the green horizontal dashed line at \(1\) indicates the conventional acceptance limit. The number of \tiles\ tested per month is displayed as well. Fluctuations of this quantity result from interruptions due to process adjustments, which decrease the monthly \tiles\ throughput.}
    \label{fig:cpucpk}
\end{figure*}

The \tile\ test demonstrated good uniformity across the \ntestedtiles\ tested devices, including \npreproductiontiles\ pre-production and \nproductiontiles\ production \tiles.
Overall, the distributions confirm the high uniformity of the \tiles, with \warmcoldpass\ meeting all of the quality requirements. \autoref{tab:percentage-qaqc} summarises the \tile\ percentage that failed both the warm and cold tests, only one of them or none of them.

The overall performance of the \tile\ test, categorised by failure type, is illustrated in the Pareto charts in \autoref{fig:pareto-plots-combined}.
The most common failure is associated with the \iv\ curve at room temperature, accounting for \ivwarmfail, followed by pulse quality assessment in liquid nitrogen at \pulsefail, and the simultaneous failure of the \iv\ test at room temperature and the noise test in liquid nitrogen at \ivwarmnoisecoldfail. Together, these three failure modes constitute \(60\)\perc\ of the total failure rate.
An \iv\ failure at room temperature is a symptom of severe misbehaviour of the \tile.
\tiles\ may still pass the quality tests in liquid nitrogen but fail at room temperature. When retested in liquid nitrogen after some time, these \tiles\ often exhibit a worsened performance.\footnote{A long-term test in liquid nitrogen of PDUs built out of \tiles\ is carried out in the \ds\ Naples Photosensor Test Facility (PTF). The performance of the PDUs is the main topic of a paper in preparation.}

The majority of the room temperature test failures (refer to \autoref{fig:pareto-plots-combined} left) is due to \iv\ curves falling outside specifications, accounting for \ivwarmfailoverfailed\ of failed \tiles\ at room temperature,  (approximately \(8.7\)\% of the total sample).

The most common failure mode in the test in liquid nitrogen (refer to \autoref{fig:pareto-plots-combined} right) is related to the pulse quality control, representing \pulsefailoverfailed\ of failed \tiles\ in liquid nitrogen (around \(5.9\)\% of the full sample).
The pulse counting test remains the most frequent failure mode even for \tiles\ that pass the test at room temperature.

Regarding the \iv\ curve assessment,~\autoref{fig:vbd-dist} and~\autoref{fig:rd-dist} show the distribution of the breakdown voltage, \vbd, and the divider resistance, \rd, respectively, over all the \tiles\ tested in NOA at room temperature and in liquid nitrogen. Quality specifications are displayed as dashed lines.

The breakdown voltage distribution in liquid nitrogen peaks at \SI{108}{\volt}, consistent with the expected value for a configuration of four \sipms\ in series, each having a \vbd\ of \SI{27}{\volt}~\cite{acerbi2024qualityassurancequalitycontrol}.
Maintaining a narrow distribution of the breakdown voltage is essential when grouping \tiles\ in PDUs, since the \tiles\ within a PDU are operated at a common bias voltage. Significant variations in the breakdown voltage of the \tiles\ would cause gain mismatches, degrading the SPE resolution. Therefore, PDUs are constructed using \tiles\ that have a breakdown voltage consistency within a margin of  \SI{1}{\volt}.

The divider resistance distributions at room temperature and in liquid nitrogen exhibit the expected trend, with \rd\ increasing as temperature decreases.
There is a small tail of \tiles\ with \rd\ out of specification despite the corresponding \pcb\ meeting the specification of the quality control. The tail is mostly due to the double slope phenomenon introduced in \autoref{sec:iv}, where the \iv\ curves show an additional, though less pronounced, departure from linearity below \vbd.

The primary reasons for failures in noise assessment are broken TIAs or the detachment of one or more \sipm\ branches. In either situation, the noise quality parameters fall outside the acceptance range, especially the Goodness of Fit (GoF), which is highly sensitive to these types of defects.
Some examples of quality parameters distributions are shown in \autoref{fig:noise-rms-std-dist} and \autoref{fig:noise-gof-dist} showing respectively the RMS noise in liquid nitrogen (at \SI{40}{\volt} bias voltage), measured with the oscilloscope, and the GoF parameter at room temperature.
The mean GoF is consistent with one, meaning that the measured spectra are in excellent agreement with the reference spectrum. The tail of the distribution extends up to \(40\), with 3\% of the \tiles\ exceeding the requirements.

The results of the pulse counting test in liquid nitrogen are exemplified in \autoref{fig:spe-amp-dist} by the distributions of the SPE mean, resolution, and SZR in amplitude with bias voltage \SI{144}{\volt} (\SI{9}{\vov}).
All of the distributions are sharply peaked, with only a small fraction of entries deviating significantly from the bulk of the distribution, 10\% for the SPE amplitude mean, 6\% for its resolution, and 2\% for the SZR.
The \tiles\ exhibiting SPE values that fall outside the expected range often have this issue because of secondary peaks located between the 0-PE and the SPE peak. This occurrence is typical in \tiles\ displaying an \iv\ characteristic with a double slope. Abnormal SPE resolution values arise from a broadening of the SPE peak.
In finger plots with significant peak overlap, the SPE resolution surpasses the upper range of the histogram shown in \autoref{fig:spe-amp-dist} (second from the top).

Since April 2025, \tiles\ with a PCR greater than about \SI{1}{\hertz/\milli\meter^2} are not mounted on PDUs and transferred to the DCR testing setup for further investigation.
With this selection, about 6\% of the \tiles\ meeting all the aforementioned requirements are discarded.

As in~\cite{acerbi2024qualityassurancequalitycontrol}, process capability indices are used to measure the ability of the \tile\ manufacturing process to produce
\tiles\ within the specification limits.
Three process Capability (Cp) indices are monitored on a monthly basis: CpL, CpU and CpK. CpL(U) is a measurement of the \tile\ manufacturing process based on its Lower (Upper) Specification Limit LSL (USL). 
CpL and CpU are defined in~\autoref{eq:cpl} and~\autoref{eq:cpu}. They represent the ratio of the difference between the mean of the measured quantity \(X\) and the lower (upper) specification limit, LSL (USL), to the process's standard deviation, specifically the \(3\sigma_X\) variation.

\begin{equation}
\label{eq:cpl}
    \text{CpL}=\frac{\overline{X}-\text{LSL}}{3\sigma_X}
\end{equation}

\begin{equation}
\label{eq:cpu}
    \text{CpU}=\frac{\text{USL}-\overline{X}}{3\sigma_X}
\end{equation}
The CpK metric is essential for assessing process compliance when both upper and lower specification limits are required. It is defined as the ratio of the difference between the average of the measured quantity \(X\) and the nearest specification limit (either USL or LSL) to the standard deviation of the process (\(3\sigma_X\) variation), as shown in~\autoref{eq:cpk}.
\begin{equation}
\label{eq:cpk}
    \text{CpK}=\text{min}[\text{CpU}, \text{CpL}]
\end{equation}

In both cases, the mean and standard deviation are computed by removing the outliers, defined as \tiles\ with more than 1.5 InterQuartile Range (IQR) above the upper (lower) quartile.

As an example, \autoref{fig:cpucpk} (Top) shows the monthly CpU(L) of the \tiles' noise parameters (RMS, GoF, and SZR) at room temperature and in liquid nitrogen and the monthly CpK of the \tiles' \rd, SPE mean in amplitude,  and \taurise\ at \SI{136}{\volt} and \SI{144}{\volt} bias voltage (Bottom). Pre-production data is labelled with negative numbers from the beginning of the production and grouped in pairs.
The red horizontal dashed line at \(1.33\) is the conventional industry standard~\cite{Heckert2002}, indicating a \(4\sigma_X\) difference between the mean of the measured
quantity and its upper specification limit, while the green horizontal dashed line at \(1\) indicates the conventional acceptance limit. The number of \tiles\ tested per month is displayed as well. Fluctuations of this quantity result from interruptions due to process adjustments, which decrease the monthly \tiles\ throughput.

Decreasing trends in process capability indices can be attributed to either insufficient data, which may occur due to process interruptions, or production-related issues.
The decline seen in \autoref{fig:cpucpk} around month-\(4\) was caused by alterations in the Flip-Chip process, resulting in some \sipms\ with scratches that produced irregular I-V curves in \tiles.
The slight decline observed around month \(3\) was due to the use of wafers exhibiting a marginally higher \sipm\ breakdown voltage. Besides these identified issues, the process capability reflects a consistently stable production campaign.

\section{Conclusions}
\label{sec:conclusions}

The \ds\ experiment aims to achieve unprecedented sensitivity in WIMP dark matter direct detection using a double phase UAr-TPC instrumented with a novel \sipm-based photodetection system. The underlying photodetector module, the \tile, has been developed and demonstrated readiness for mass production at the Nuova Officina Assergi (NOA) facility at LNGS. This paper reported the quality assurance and quality control procedures implemented to ensure the performance, reliability, and uniformity of the \tiles\ throughout the production process.  

A comprehensive \qaqc\ protocol was developed to asses the electrical and optical properties of each \tile\ at both room temperature and in liquid nitrogen. The characterisation included \iv\ curve measurements, noise spectrum analysis, and photon response testing. The results demonstrate a high level of uniformity across the \ntestedtiles\ tested \tiles. Key performance parameters, such as breakdown voltage, divider resistance, and single-photon response, have shown consistent and reproducible behaviour over the production time.  

A well-optimised testing workflow enabled an efficient validation process, significantly improving production yield. The \pcb\ and \tile\ test assessments at room temperature successfully identify major electronic faults early in the process. The quality check in liquid nitrogen ensures that the \tiles\ work as expected. The overall production yield exceeded \warmcoldpass, with failures primarily attributed to \iv\ curve deviations and inconsistencies in Single PhotoElectron measurements. These results validate the robustness of the \tile\ design and its suitability for operation in a cryogenic environment.  

The full production of \ntiletotnoa\ \tiles, corresponding to an overall instrumented area of \opticalplanes, is expected to end in less than \(30\) months with the completion of the cryogenic characterisation of the overall amount of \nwafers\ silicon wafers (equivalent to \ndies\ single dice), the test of \npcbtotnoa\ bare \pcbs, the packaging and qualification of \ntiletotnoa\ \tiles\ and the building of \npduds\ functional TPC PDUs.

 Continuous refinements in the assembly and testing processes will further enhance production efficiency and ensure that only high-quality photodetectors are integrated into the experiment. The success of this effort establishes a strong foundation for deploying large-area \sipm-based photodetection systems in future low-background physics experiments.

\vspace{1cm}
\section{Acknowledgment}
\small{
This paper is based upon work supported by the U. S. National Science Foundation (NSF) (Grants No. PHY-0919363, No. PHY-1004054, No. PHY-1004072, No. PHY-1242585, No. PHY-1314483, No. PHY-1314507, No. PHY-1622337, No. PHY-1812482, No. PHY-1812547, No. PHY-2310091, No. PHY-2310046, associated collaborative grants, No. PHY-1211308, No. PHY-1314501, No. PHY1455351 and No. PHY-1606912, as well as Major Research Instrumentation Grant No. MRI-1429544), the Italian Istituto Nazionale di Fisica Nucleare (Grants from Italian Ministero dell’Istruzione, Università, e Ricerca Progetto Premiale 2013 and Commissione Scientific Nazionale II), the Natural Sciences and Engineering Research Council of Canada, SNOLAB, and the Arthur B. McDonald Canadian Astroparticle Physics Research Institute. This work received support from the French government under the France 2030 investment plan, as part of the Excellence Initiative of Aix-Marseille University — A*MIDEX (AMX-19-IET-008 — IPhU). We also acknowledge the financial support by LabEx UnivEarthS (ANR-10-LABX-0023 and ANR18-IDEX-0001), Chinese Academy of Sciences (113111KYSB20210030) and National Natural Science Foundation of China (12020101004). This work has been supported by the São Paulo Research Foundation (FAPESP) grant 2021/11489-7 and by the National Council for Scientific and Technological Development (CNPq). Support is acknowledged by the Deutsche Forschungsgemeinschaft (DFG, German Research Foundation) under Germany’s Excellence Strategy — EXC 2121: Quantum Universe — 390833306. The authors were also supported by the Spanish Ministry of Science and Innovation (MICINN) through the grant PID2022-138357NB-C22, the “Atraccion de Talento” grant 2018-T2/TIC-10494, the Polish NCN (Grants No. UMO-2022/47/B/ST2/02015 and UMO-2023/51/B/ST2/02099), the Polish Ministry of Science and Higher Education (MNiSW,grant number 6811/IA/SP/2018), the International Research Agenda Programme AstroCeNT (Grant No. MAB/2018/7) funded by the Foundation for Polish Science from the European Regional Development Fund, the European Union’s Horizon 2020 research and innovation program under grant agreement No 952480 (DarkWave), the Science and Technology Facilities Council, part of the United Kingdom Research and Innovation, and The Royal Society (United Kingdom), and IN2P3-COPIN consortium (Grant No. 20-152). We also wish to acknowledge the support from Pacific Northwest National Laboratory, which is operated by Battelle for the U. S. Department of Energy under Contract No. DE-AC05-76RL01830. This research was supported by the Fermi National Accelerator Laboratory (Fermilab), a U. S. Department of Energy, Office of Science, HEP User Facility. At the time of this work, Fermilab was managed by Fermi Research Alliance, LLC (FRA), acting under contract No. DE-AC02-07CH11359. This work was supported (in part) by the PRIN2020 project of the Italian Ministry of Research (MUR) (Grant No. PRIN 20208XN9TZ).

}

\printbibliography
\newcommand{\notds}{\nolinebreak\footnotemark\nolinebreak}
\renewcommand{\thefootnote}{$*$}
{
\onecolumn
\textbf{The DarkSide-20k Collaboration}

F.~Acerbi\thanksref{l_TNFBK}\nolinebreak,
P.~Adhikari\thanksref{l_Carleton}\nolinebreak,
P.~Agnes\thanksref{l_AQGSSI}\textsuperscript{,}\thanksref{l_AQLNGS}\nolinebreak,
I.~Ahmad\thanksref{l_AstroCeNT}\nolinebreak,
S.~Albergo\thanksref{l_CTUNI}\textsuperscript{,}\thanksref{l_CTINFN}\nolinebreak,
I.~F.~Albuquerque\thanksref{l_USP}\nolinebreak,
T.~Alexander\thanksref{l_PNNL}\nolinebreak,
A.~K.~Alton\thanksref{l_Augustana}\nolinebreak,
P.~Amaudruz\thanksref{l_TRIUMF}\nolinebreak,
M.~Angiolilli\thanksref{l_AQGSSI}\textsuperscript{,}\thanksref{l_AQLNGS}\nolinebreak
% ...

\begin{enumerate}
\item{\TNFBK\label{l_TNFBK}}
\item{\Carleton\label{l_Carleton}}
\item{\label{l_AQGSSI}\AQGSSI}
\item{\label{l_AQLNGS}\AQLNGS}
\item{\AstroCeNT\label{l_AstroCeNT}}
\item{\CTINFN\label{l_CTINFN}}
\item{\CTUNI\label{l_CTUNI}}
\item{\USP\label{l_USP}}
\item{\PNNL\label{l_PNNL}}
\item{\Augustana\label{l_Augustana}}
\item{\TRIUMF\label{l_TRIUMF}}
\item{\Columbia\label{l_Columbia}}
\item{\CAINFN\label{l_CAINFN}}
\item{\CAUniPHY\label{l_CAUniPHY}}
\item{\Alberta\label{l_Alberta}}
\item{\MendeleevUniverisity\label{l_MendeleevUniverisity}}
\item{\LNLINFN\label{l_LNLINFN}}
\item{\SNL\label{l_SNL}}
\item{\RHUL\label{l_RHUL}}
\item{\CIEMAT\label{l_CIEMAT}}
\item{\CPPM\label{l_CPPM}}
\item{\PIINFN\label{l_PIINFN}}
\item{\PIUniPHY\label{l_PIUniPHY}}
\item{\Oxford\label{l_Oxford}}
\item{\TOINFN\label{l_TOINFN} }
\item{\TOPoli\label{l_TOPoli}}
\item{\RMUnoINFN\label{l_RMUnoINFN}}
\item{\GEUni\label{l_GEUni}}
\item{\GEINFN\label{l_GEINFN}}
\item{\MIPoliICA\label{l_MIPoliICA}}
\item{\MIINFN\label{l_MIINFN}}
\item{\WUT\label{l_WUT} }
\item{\RMTreINFN\label{l_RMTreINFN} }
\item{\NAINFN\label{l_NAINFN} }
\item{\VTech\label{l_VTech} }
\item{\CAUniEEE\label{l_CAUniEEE} }
\item{\Zaragoza\label{l_Zaragoza} }
\item{\MSU\label{l_MSU} }
\item{\BOINFN\label{l_BOINFN} }
\item{\BOUniPHY\label{l_BOUniPHY} }
\item{\Laurentian\label{l_Laurentian} }
\item{\SNOLAB\label{l_SNOLAB} }
\item{\AQUni\label{l_UnivAQ} }
\item{\Krakow\label{l_Krakow} }
\item{\NAUniDIST\label{l_NAUniDIST} }
\item{\MIUni\label{l_MIUni} }
\item{\RMUnoUni\label{l_RMUnoUni} }
\item{\MIPoliCHE\label{l_MIPoliCHE} }
\item{\Petersburg\label{l_Petersburg} }
\item{\NAUniPHY\label{l_NAUniPHY} }
\item{\Queens\label{l_Queens} }
\item{\Princeton\label{l_Princeton} }
\item{\Kurchatov\label{l_Kurchatov} }
\item{\Belgorod\label{l_Belgorod} }
\item{\UCDavis\label{l_UCDavis} }
\item{\Lancaster\label{l_Lancaster} }
\item{\APC\label{l_APC} }
\item{\Liverpool\label{l_Liverpool} }
\item{\BINP\label{l_BINP} }
\item{\BOCentroFermi\label{l_BOCentroFermi} }
\item{\Manchester\label{l_Manchester} }
\item{\ETHZ\label{l_ETHZ} }
\item{\WilliamsCollege\label{l_WilliamsCollege} }
\item{\Hawaii\label{l_Hawaii} }
\item{\Chicago\label{l_Chicago} }
\item{\CTLNS\label{l_CTLNS} }
\item{\IHEP\label{l_IHEP} }
\item{\UniversityofEdinburgh\label{l_UniversityofEdinburgh} }
\item{\Houston\label{l_Houston} }
\item{\NAUniPHARM\label{l_NAUniPHARM} }
\item{\UCRiverside\label{l_UCRiverside} }
\item{\JINR\label{l_JINR} }
\item{\MEPhI\label{l_MEPhI} }
\item{\Birmingham\label{l_Birmingham} }
\item{\UniHAM\label{l_UniHAM} }
\item{\Temple\label{l_Temple} }
\item{\Warwick\label{l_Warwick} }
\item{\INFN\label{l_INFN} }
\item{\UMass\label{l_UMass} }
\item{\NAUniCHE\label{l_NAUniCHE} }
\item{\CAUniCHE\label{l_CAUniCHE} }
\item{\UCAS\label{l_UCAS} }
\item{\UCLA\label{l_UCLA} }
\item{\FortLewis\label{l_FortLewis} }
\item{\Lodz\label{l_Lodz}}
\item{\STFCInterconnect\label{l_STFCInterconnect}}
\item{\ENUniCEE\label{l_ENUniCEE}}
\item{\STFCppd\label{l_STFCppd}}
\item{\RMTreUni\label{l_RMTreUni}}
\item{\Washington\label{l_Washington}}
\end{enumerate}

}

\end{document}